\documentclass[a4paper,fleqn,usenatbib]{mnras}
\usepackage[dvipsnames,usenames]{color}
\usepackage{colortbl}
\usepackage{mathptmx}
\usepackage{etoolbox}
\usepackage{graphicx}
\usepackage{amsmath}
\usepackage{natbib}
\newcommand{\etal}{{et al}\/.}
\pubyear{2016}
\begin{document}
\title[LOFAR survey of H-ATLAS NGP]{LOFAR/H-ATLAS: A deep
  low-frequency survey of the {\it Herschel}-ATLAS North Galactic Pole
field}
\author[M.J.\ Hardcastle \etal]
{M.J.\ Hardcastle$^1$, G.\ G\"urkan$^1$, R.J. van Weeren$^2$,
  W.L.\ Williams$^1$, P.N.\ Best$^3$,\newauthor
  F.\ de Gasperin$^4$, D.A. Rafferty$^5$, S.C.\ Read$^1$,
  J.\ Sabater$^3$, T.W.\ Shimwell$^4$,\newauthor
  D.J.B.\ Smith$^1$, C.\ Tasse$^6$, N.\ Bourne$^3$, M.\ Brienza$^{7,8}$,
  M.\ Br\"uggen$^5$, G. Brunetti$^{9}$,\newauthor K.T.\  Chy\.zy$^{10}$,
  J.\ Conway$^{11}$, L.\ Dunne$^{3,12}$, S.A.\ Eales$^{10}$,
  S.J.\ Maddox$^{3,12}$,\newauthor M.J.\ Jarvis$^{13,14}$,
  E.K.\ Mahony$^{15,16}$,\ R.\ Morganti$^{7,8}$, I.\ Prandoni$^{9}$,
  H.J.A.\ R\"ottgering$^4$,\newauthor E.\ Valiante$^{12}$ and G.J.\ White$^{17,18}$\\
$^1$ Centre for Astrophyics Research, School of Physics, Astronomy and Mathematics, University of
  Hertfordshire, College Lane, Hatfield AL10 9AB, UK\\
$^2$ Harvard-Smithsonian Center for Astrophysics, 60 Garden Street,
  Cambridge, MA 02138, USA\\
$^3$ SUPA, Institute for Astronomy, Royal Observatory, Blackford Hill, Edinburgh, EH9 3HJ, UK\\
$^4$ Leiden Observatory, Leiden University, PO Box 9513, 2300 RA Leiden, the Netherlands\\
$^5$ University of Hamburg, Hamburger Sternwarte, Gojenbergsweg 112,
  21029 Hamburg, Germany\\
$^6$ GEPI, Observatoire de Paris, CNRS, Universit\'e Paris Diderot, 5
  place Jules Janssen, 92190 Meudon, France  \\
  $^{7}$ ASTRON, the Netherlands Institute for Radio Astronomy, Postbus 2, 7990 AA,
  Dwingeloo, The Netherlands\\
  $^{8}$ Kapteyn Astronomical Institute, University of Groningen, P.O. Box 800,
9700 AV Groningen, The Netherlands\\
  $^{9}$ INAF-Institute of Radioastronomy, Via P. Gobetti 101, 40129, Bologna, Italy\\
$^{10}$ Astronomical Observatory, Jagiellonian University, ul. Orla 171,
  30-244 Krak\'ow, Poland\\
$^{11}$ Chalmers University of Technology, Onsala Space Observatory, S-43992, Sweden\\
$^{12}$  School of Physics and Astronomy, Cardiff University, Queen's Buildings, The Parade, Cardiff CF24 3AA, UK\\
  $^{13}$ Oxford Astrophysics, Denys Wilkinson Building, Keble Road,
  Oxford OX1 3RH, UK\\
  $^{14}$ Physics Department, University of the Western Cape, Private Bag
  X17, Bellville 7535, South Africa\\
  $^{15}$ Sydney Institute for Astronomy, School of Physics A28, The
  University of Sydney, NSW 2006, Australia\\
  $^{16}$ ARC Centre of Excellence for All-Sky Astrophysics (CAASTRO)\\
  $^{17}$ Department of Physical Sciences, The Open University, Milton
  Keynes MK7 6AA, UK\\
  $^{18}$ RAL Space, The Rutherford Appleton Laboratory, Chilton, Didcot, Oxfordshire OX11 0NL, UK\\ 
}
\maketitle
\begin{abstract}
  We present LOFAR High-Band Array (HBA) observations of the {\it
    Herschel}-ATLAS North Galactic Pole survey area. The survey we
  have carried out, consisting of four pointings covering around 142
  square degrees of sky in the frequency range 126--173 MHz, does not
  provide uniform noise coverage but otherwise is representative of
  the quality of data to be expected in the planned LOFAR wide-area
  surveys, and has been reduced using recently developed `facet
  calibration' methods at a resolution approaching the full resolution
  of the datasets ($\sim 10 \times 6$ arcsec) and an rms off-source
  noise that ranges from 100 $\mu$Jy beam$^{-1}$ in the centre of the
  best fields to around 2 mJy beam$^{-1}$ at the furthest extent of
  our imaging. We describe the imaging, cataloguing and source
  identification processes, and present some initial science results
  based on a 5-$\sigma$ source catalogue. These include (i) an initial
  look at the radio/far-infrared correlation at 150 MHz, showing that
  many {\it Herschel} sources are not yet detected by LOFAR; (ii)
  number counts at 150 MHz, including, for the first time,
  observational constraints on the numbers of star-forming galaxies;
  (iii) the 150-MHz luminosity functions for active and star-forming
  galaxies, which agree well with determinations at higher frequencies
  at low redshift, and show strong redshift evolution of the
  star-forming population; and (iv) some discussion of the
  implications of our observations for studies of radio galaxy life
  cycles.
\end{abstract}

\section{Introduction}
\label{sec:intro}

Low-frequency continuum radio emission from galaxies originates in the
synchrotron process, with the two sources of energy for the required
high-energy electrons and positrons being supernovae and their
  remnants (in star-forming galaxies) or the activity of radio-loud
active galactic nuclei, which drive relativistic jets of magnetized
plasma into the external medium. In principle, both of these processes
provide us with information that cannot be accessed in any other way.
The inferred cosmic-ray population of star-forming galaxies stores
some fraction of the energy deposited by supernova and supernova
  remnant activity, albeit
with an integration timescale that depends on radiative losses and
transport processes in the host galaxy, and thus depends on the
time-integrated star-formation rate, giving rise to the well-known
radio/far-infrared correlation \citep{vanderKruit71, deJong+85,
  Helou+85, Yun+01, Ibar+08, Murphy09, Jarvis+10,
  Ivison+10,Ivison+10b, Lacki+10, Smith+14}. The luminosity and other
properties (structure, spectrum, and polarization) of radio emission
from radio-loud AGN offer us the only method, in the absence of deep
X-ray observations for every target, of assessing the kinetic
luminosity produced by the AGN, or jet power, and the radio luminosity
alone is widely used for this purpose \citep{Willott+99} although
there are serious uncertainties in applying this method to individual
objects \citep{Hardcastle+Krause13}. In the local Universe, there is a
large population of radio-loud AGN which exhibit no signatures of
conventional thin-disc accretion, generally referred to as
low-excitation radio galaxies or jet-mode objects
\citep{Hardcastle+07,Hardcastle+09}: radio observations represent the
only way to study the accretion onto the central supermassive black
hole in these objects and by far the most efficient way (in the
absence of sensitive X-ray observations for large samples) to
constrain their effects on the external medium, the so-called feedback
process thought to be responsible for preventing massive star
formation from the hot phase of the intergalactic medium
\citep{Croton+06}.

Wide-area, sensitive radio surveys, in conjunction with wide-area
optical photometric and spectroscopic surveys such as the Sloan
Digital Sky Survey (SDSS: \citealt{Eisenstein+11}) provide us with the
ideal way to study both these processes in a statistical way in the
local Universe. Sensitive surveys are required to detect the radio
emission expected from low-level star formation, which can be faint;
star-forming objects start to dominate the radio-emitting population
at luminosities below about $10^{23}$ W Hz$^{-1}$ at 1.4 GHz
\citep{Mauch+Sadler07}, corresponding to 4 mJy for a source redshift
of 0.1 and 0.3 mJy at $z=0.3$. Wide-area surveys are required in order
to find statistically meaningful samples of powerful AGN that are
close enough to be optically identified and have their redshifts
determined using available optical data. A key problem, however, is
{\it distinguishing} between radio emission driven by low-level
star-formation activity and that powered by low-luminosity AGN \citep{Ibar+09}. In an
era where radio survey capabilities are expected to become vastly more
powerful, it is important to develop diagnostics that will help us to
understand this problem, or at least to understand its true extent. To
do this we need to calibrate the radio properties of identified radio
sources against their instantaneous star-formation rates and
star-formation histories obtained by other means. This motivates radio
observations of wide regions of the sky with good constraints on
star-formation activity.

One widely used diagnostic of star formation is the luminosity and
temperature of cool dust, heated by young stars. The ability of the
{\it Herschel} satellite \citep{Pilbratt+10} to make sensitive
far-infrared observations over a broad bandwidth made it exquisitely
sensitive to this particular tracer of star-formation activity. The
{\it Herschel}-ATLAS survey (H-ATLAS: \citealt{Eales+10}) carried out
wide-area surveys of several large areas of the sky in northern,
equatorial and southern fields using the PACS \citep{Poglitsch+10} and
SPIRE \citep{Griffin+10} instruments (at wavelengths of 100, 160, 250,
350 and 500 $\mu$m), allowing investigations of the relationship
between star formation and radio emission \citep{Jarvis+10} and
between star formation and AGN activity of various types
\citep{Serjeant+10,Hardcastle+10b,Bonfield+11,Hardcastle+13,Kalfountzou+14,Gurkan+15}.
However, these studies have been limited by the availability of
high-quality radio data, as they rely on the 1.4-GHz VLA surveys Faint
Images of the Radio Sky at Twenty-cm \citep[FIRST,][]{Becker+95} and
NRAO VLA Sky Survey \citep[NVSS;][]{Condon+98}, and, while these
surveys have proved extremely valuable, they have inherent weaknesses
when it comes to studying faint star formation and distant AGN. NVSS
is sensitive to all the radio emission from sources extended on scales
of arcminutes, but its resolution and sensitivity are low (resolution
of 45 arcsec; rms noise level $\sim 0.5$ mJy beam$^{-1}$) which means that
it can only detect luminous or nearby objects, and has difficulty
identifying them with optical counterparts. FIRST is higher-resolution
(5 arcsec) and more sensitive ($\sim 0.15$ mJy beam$^{-1}$) but its
lack of short baselines means that it resolves out extended emission
on arcmin scales, often present in nearby radio-loud AGN. Constructing
samples of radio-loud AGN from these surveys with reliable
identifications and luminosities involves a painstaking process of
combining the two VLA surveys \citep[e.g.][]{Best+05,Virdee+13}, and
good imaging of the sources is often not possible.

The Low-Frequency Array (LOFAR: \citealt{vanHaarlem+13}) offers the
opportunity to make sensitive surveys of large areas of the northern
sky with high rates of optical counterpart detection because of its
combination of collecting area, resolution (up to 5 arcsec with the
full Dutch array) and field of view. The LOFAR Surveys Key Science
Project \citep{Rottgering+06} aims to conduct a survey (the `Tier 1'
High Band Array survey, hereafter referred to as `Tier 1': Shimwell
\etal\ in prep.) of the northern sky at 5-arcsec resolution to an rms
noise at 150 MHz of $\sim 100$ $\mu$Jy beam$^{-1}$, which for a
typical extragalactic source with spectral index\footnote{Here and
  throughout the paper spectral index is defined in the sense $S_\nu
  \propto \nu^{-\alpha}$.} $\alpha = 0.7$ implies a depth 7 times
greater than FIRST's for the same angular resolution. Crucially, LOFAR
has good $uv$ plane coverage on both long and short baselines, and so
is able to image all but the very largest sources at high resolution
without any loss of flux density, limited only by surface brightness
sensitivity. Deep observations at these low frequencies are rare, and
the previous best large-area survey at frequencies around those of the
LOFAR High Band Array (HBA) is the TIFR GMRT Sky
Survey\footnote{\url{http://tgss.ncra.tifr.res.in/}} (TGSS), full data
from which were recently released \citep{Intema+16}: however, this has
a best resolution around 20 arcsec, which is substantially lower
  than the $\sim 5$ arcsec that LOFAR can achieve, and, with an rms
  noise of $\sim 5$ mJy beam$^{-1}$, significantly lower sensitivity
  than will be achieved for the LOFAR Tier 1 survey.

AGN selection at the lowest frequencies has long been recognised to
provide the most unbiased AGN samples, because the emission is
dominated by unbeamed radiation from the large-scale lobes, a fact
which has ensured the long-term usefulness of low-frequency-selected
samples of AGN such as 3CRR, selected at 178 MHz \citep{Laing+83} or
samples derived from the 151-MHz 6C and 7C surveys
\citep[e.g.][]{Eales85,Rawlings+01,Willott+02,Cruz+06}. These AGN
surveys, however, have had little or no bearing on star-formation
work, since the flux density limits of the surveys exclude all but a few
bright nearby star-forming objects. The relationship between star
formation and radio luminosity at low frequencies is essentially
unexplored. LOFAR observations of {\it Herschel}-ATLAS fields
therefore offer us the possibility both to accumulate large, unbiased,
well-imaged, samples of radio-loud AGN {\it and} to study the
radio/star-formation relation in both radio-loud and radio-quiet
galaxies in the nearby Universe.

In this paper we describe an exploratory LOFAR HBA observation, of the
H-ATLAS North Galactic Pole (NGP) field, a rectangular contiguous area
of sky in the SDSS sky area covering $\sim 170$ square degrees around
$\mathrm{RA}= 13.5$ h and $\mathrm{Dec} = 30^\circ$, and therefore
well positioned in the sky for LOFAR, with a substantial overlap with
the position of the Coma cluster at low $z$. Our survey prioritizes
sky coverage over uniform sensitivity but achieves depth comparable to
the eventual Tier 1 LOFAR survey. We describe the imaging, cataloguing
and source identification process and the tests carried out on the
resulting catalogues. We then present some first results on the
radio/far-infrared relation observed in the fields and the properties
of optically identified radio sources, together with number counts for
star-forming sources at 150 MHz and a first $z=0$ 150-MHz luminosity
function. A subsequent paper (G\"urkan \etal\ in prep.) will explore
the 150-MHz radio/star-formation relation derived from LOFAR and
H-ATLAS data and we expect to carry out further analysis of the bright
AGN population.

Throughout the paper we assume a cosmology in which $H_0 = 70$ km
s$^{-1}$, $\Omega_{\rm m} = 0.3$ and $\Omega_\Lambda = 0.7$.

\section{Observations}

\begin{table*}
  \caption{LOFAR observations of the NGP field}
  \label{obslist}
  \begin{tabular}{llllr}
    \hline
    Field name&RA&Dec&Start date/time&Duration (h)\\
    \hline
    Central&13h24m00s&+27d30m00s&2013-04-26 17:42:15&9.7\\
    NW&13h00m00s&+31d52m00s&2014-04-22 18:30:30&8.0\\
    SW&13h04m00s&+25d40m00s&2014-04-25 18:17:00&8.0\\
    NE&13h34m00s&+32d18m00s&2014-07-15 13:28:38&8.0\\ 
    \hline
  \end{tabular}
\end{table*}

The NGP field was observed in four separate pointings, chosen to
maximise sky covered, with the LOFAR HBA (Table \ref{obslist}) as part
of the Surveys Key Science project. Observations used the
HBA\_DUAL\_INNER mode, meaning that the station beams of core and
remote stations roughly matched each other and giving the widest
possible field of view. The first observation, which was made early on
in LOFAR operations, was of slightly longer duration ($\sim 10$ h)
than the others ($\sim 8$ h). International stations were included in
some of the observations in 2014 but were not used in any of our
analysis, which uses only the Dutch array.

In each case, the observations of the field were preceded and followed
by short, 10-minute observations of calibrator sources (3C\,196 at the
start of the run and 3C\,295 at the end). Each observation used the
full 72 MHz of bandwidth provided by the HBA on the target field,
spanning the frequency range 110 to 182 MHz. As LOFAR is a software
telescope, multiple beams can be formed on the sky, and the total
bandwidth that can be processed by the correlator exceeds the total
bandwidth available from the HBA: this allowed us to observe an
additional 24 MHz spread throughout this frequency range on an
in-field calibrator, the bright point source 3C\,287, which lies in
the SE corner of the NGP field. The original intention was to use this
calibrator pointing for determination of the clock offsets between the
core and remote stations, but this proved unnecessary, as we shall see
below. As data with non-contiguous frequency coverage could not easily
be analysed using the facet calibration method (see below) at the time
of our analysis, we do not consider the 3C\,287 observations further.

After observation, the data were averaged by the observatory to 4
channels per sub-band (an HBA sub-band has a bandwidth of 195.3 kHz) and
a 5-second integration time. No `demixing' of bright off-axis sources
was carried out -- this was deemed unnecessary
given the sky positions of bright objects like Cyg A and Cas A --
and all further processing was carried out by us using the University
of Hertfordshire high-performance computing facility.

\section{Data processing, imaging and cataloguing}
\label{sec:data}

\subsection{Facet calibration}

The data were processed using techniques which are described in detail
by \cite{vanWeeren+16} (hereafter vW16) and \cite{Williams+16}
(hereafter W16), implemented by us in a
way which was intended to maximise the data processing efficiency on
the Hertfordshire cluster\footnote{The facet calibration scripts may
  be found at \url{https://github.com/tammojan/facet-calibration}; the code
  for the implementation described in this paper is available at
  \url{https://github.com/mhardcastle/surveys-pipeline} .}. Here we give a
brief overview of the processes, highlighting steps in which our
approach differs from that of vW16 and W16.

Initial flagging using {\sc rficonsole} was done on each sub-band of
the target and calibrator observation (we used 3C\,196 as the primary
calibrator for all four observations) and we then solved for per
station for amplitude, phase and `rotation angle' -- a
  term that accounts for differential Faraday rotation within a
  sub-band, as described in
  section 4.3 of vW16 -- on the calibrator observations using the `Black
Board Self-Calibration' ({\sc bbs}) software \citep{Pandey+09}, making
use of a high-resolution model of 3C\,196 kindly supplied by
V.N.\ Pandey. Because each sub-band was treated independently, we were
able to efficiently run many of these steps in parallel. We then
combined the complex gain solutions on the calibrator for each
sub-band, using tools in the LoSoTo
package\footnote{\url{https://github.com/revoltek/losoto}}. Bad
stations or sub-bands could be identified at this point by looking for
large rms values or gain outliers: when one or more stations were
identified as bad, we flagged them throughout the observation and
re-ran the calibration.

With all bad data removed from the calibrator observations, we then
fitted the phase solutions at all frequencies with a model intended to
solve for the effects of clock offsets (which introduce a phase offset
which is linear in observing frequency $\nu$) and the differential
total ionospheric electron content, or TEC (which introduces phase offsets which
go as $\nu^{-1}$). This so-called clock-TEC separation can only be run on
the calibrator observations, because of their high signal-to-noise
ratio, and must be run over as broad a bandwidth as possible to
maximize the effectiveness of the fitting process. The result was a
set of per-station clock offset values which we transferred, along
with the gain amplitudes, to the data for the target field, again for
each sub-band. (The ionospheric TEC values are not transferred to the
target because the ionosphere toward the target will be different, and
in general variable: the clock offsets, on the other hand, are not
expected to vary significantly over the observation.)

The clock-corrected sub-bands on the target field were now
concatenated into `bands' of ten sub-bands each: such a band has a
bandwidth of just under 2 MHz. This concatenation gives sufficient
signal-to-noise on the target field for per-band phase calibration.
The next stage was to generate a model to allow us to calibrate the
fields. To do this we phase calibrated band 20 (data at frequencies of
150-152 MHz, around the frequency range where the HBA is most
sensitive) using the observatory-provided global sky model, which is
based on low-resolution observations from other telescopes. We then
imaged this dataset with {\sc awimager} \citep{Tasse+13} at $\sim 10$-arcsec
resolution, derived a list of Gaussians from the image with {\sc
  pybdsm}\footnote{\url{http://www.astron.nl/citt/pybdsm/}}
\citep{Mohan+Rafferty15} and cross-matched them with the FIRST
catalogue to obtain a list of sources in the field detected by both
LOFAR and FIRST, with their 151-MHz and 1.4-GHz flux densities. This
catalogue, while shallow, is virtually free from artefacts because of
the FIRST cross-matching and has excellent positional and structural
information from FIRST; this is achieved at the price of omitting only
a few bright steep-spectrum or resolved objects that are not seen in
FIRST.

For each band, the data were run through {\sc rficonsole} again (in
order to catch low-level RFI with the improved signal-to-noise of the
broader bandwidth) and then averaged by a factor 2 in time and
frequency, i.e. down to a 10-s integration time and 20 channels per
band. This averaging was selected for computational speed in the facet
calibration process, though it produces moderate bandwidth and
time-averaging smearing, increasing the apparent size of sources while
preserving their total flux density, beyond 2-3 degrees from the
pointing centre at the full LOFAR imaging resolution of $\sim 5$
arcsec. (In this paper we work at resolutions somewhat lower than this
full resolution, around $10 \times 6$ arcsec, so that smearing
effects are less important, but not negligible; we would expect a drop
in peak flux density by a factor 0.8 and a broadening by a factor 1.2
by 4 degrees from the pointing centre, which is more or less the
largest pointing centre offset considered in this work.) The
cross-matched catalogue was then used (scaling appropriately using the
LOFAR/FIRST spectral index) to provide the sky model for an initial
phase-only calibration for each band using {\sc bbs}, dividing the
thousand or so cross-matched sources in each field into about 100
discrete sky regions or patches to reduce computing time. As the
objective is only to provide phase calibration good enough to start
the facet calibration process, small defects in the sky model are not
important and in principle should not affect the final result. Again,
it was possible to apply this process to all bands in parallel.

With the initial phase-only calibration computed, we corrected the
data for the effect of the element beam and array factor using {\sc
  bbs}, and all further imaging work until the very end of the process
was then carried out using apparent flux densities (i.e. no primary beam
correction was carried out in the imaging). At this point we could
image each band [we now used {\sc wsclean} \citep{Offringa+14} for
  imaging, since we no longer required the primary beam correction
  abilities of {\sc awimager}] and subtract the detected sources in
the two-stage manner (first using `high-resolution' images with
approximately 30 arcsec beam size, then using a lower-resolution $\sim
100$-arcsec image) described by vW16. Once each band had had the
sources subtracted, and we had a per-band sky model describing the
subtraction that has been done, we were in a position to start the
facet calibration itself.

\begin{table}
  \caption{Spectral windows, frequencies and LOFAR band/sub-band
    numbers used}
  \label{tab:frequencies}
  \begin{tabular}{rrrr}
    \hline
    Spectral&Frequency range&Band&Sub-band\\
    window&(MHz)&numbers&numbers\\
    \hline
    1&126-134&8-11&80-119\\
    2&134-142&12-15&120-159\\
    3&142-150&16-19&160-169\\
    4&150-158&20-23&200-239\\
    5&158-166&24-27&240-279\\
    6&166-173&28-31&280-319\\
    \hline
  \end{tabular}
\end{table}

Our approach to facet calibration is slightly different from that of
vW16. We are interested in imaging in several separate frequency
ranges (which we refer to hereafter as `spectral windows'), since we
would like to be able to measure in-band spectral indices for detected
sources. In addition, facet calibrating in different spectral windows
can be done in parallel, speeding the process up considerably.
Accordingly, we chose to facet calibrate with six spectral windows,
each made up of four bands and thus containing about 8 MHz of
bandwidth\footnote{When fitting over these narrow bandwidths, fitting
  for differential TEC as described by vW16 becomes, effectively,
  fitting for phase tied together over all four datasets. However, the
  key point is the gain in signal to noise of a factor 2 derived from
  the joint fit to all four bands.} (Table \ref{tab:frequencies}). We
intentionally did not include in this spectral range the very lowest
frequencies in the data, below 126 MHz, as they have significantly
worse sensitivity than the higher HBA frequencies, and also discarded
frequencies above 173 MHz, which are badly affected by RFI even after
flagging: thus our final images contain about 48 MHz of bandwidth out
of a possible total 72~MHz, but would probably gain little in
sensitivity by including the missing data.

Facets were defined using our knowledge of the bright source
distribution from the `high-resolution' (30 arcsec) images used for
subtraction: we aimed for between 20 and 30 facets per pointing in
order to sample the ionosphere as well as possible, with the number
actually used being determined by the number of bright sources
available. Calibration positions in the facets were defined, as
described by vW16, by selecting square regions (of size less than $50
\times 50$ arcmin, and normally several times smaller than that)
containing sources with a total flux density normally greater than 0.4
(apparent) Jy at 150~MHz. The boundaries between each facet were set
using Voronoi tessellation. Each facet was then calibrated and imaged
in the manner described by vW16, runnning in parallel over all six
spectral windows. In practice, we always started the run for the data
at 150--158~MHz before the other five, so that problems with, for
example, the definition of the facets could be ironed out using a
single dataset; such problems generally showed up as a failure of
phase or amplitude self-calibration, resulting in poor solutions
and/or increased residuals after subtraction, and were dealt with by
increasing the size of the calibration region to include more sources
or changing the phase/amplitude solution intervals. The whole facet
calibration process, for the six spectral windows, took between one
and two weeks per field, depending on the number of facets and the
number of problems encountered; typically the self-calibration,
imaging and subtraction step for one facet in one spectral window took
4--5 hours on a 16-core node with 64 GB RAM and 2.2-GHz clock speed.
The run time per facet was shortened by a factor $\sim 2$ with respect
to earlier implementations of facet calibration by the use of {\sc
  wsclean} both for the final facet imaging and for the `prediction'
of the visibilities to subtract from the un-averaged data, by blanking
of the images given to {\sc pybdsm} for mask generation to reduce the
area that it searched for sources, and by some alterations to {\sc
  pybdsm} to improve efficiency, particularly in using the efficient
FFTW algorithm for the Fourier transforms involved in cataloguing
extended sources. The use of {\sc wsclean} rather than a {\sc bbs}
subtract meant that care was necessary with the placement of facets in
the final images to avoid aliasing of sources near the field edge in
the prediction step, resulting in negative `sources' outside the facet
after subtraction. A few negative sources generated in this way
propagate through into the final images, but have no significant
effect on the final science results as they are not detected by the
source finding algorithms.

Once the facet calibration and imaging process was complete, the
resulting images for each spectral window were mosaiced into a single
large image for each field, masking each facet to ensure that only the
good portion of the image was used. Previous imaging of each band with
{\sc awimager} was used to make images of the primary beam for each
band and we divided through by these images (regridding them with {\sc
  montage} to the scale of the facet calibration images and averaging
appropriately for each spectral window) to make maps of true rather
than apparent flux density. (To maximize the area covered by the survey we
image further down the primary beam than was attempted by e.g. W16:
primary beam corrections at the edge of the largest fields can
approach a factor 5.)

\subsection{Flux calibration}
\label{sec:fluxcal}

LOFAR flux density calibration is in practice somewhat problematic:
effectively, the gain normalizations transferred from the calibrator
to the source are only valid if the elevation of the source and the
calibrator are the same (implying the same effective telescope beam on
the sky), which can only be guaranteed for a snapshot observation and
is not generally true even then. This error manifests itself (after
facet calibration, which removes time-dependent effects) as a
frequency-dependent error in the flux scale. Currently the only
  method available to ensure consistency in the flux scale and
  correctness of in-band spectral index is to derive correction factors
as a function of frequency using flux densities from other
low-frequency surveys. Data available for our field include the VLSS
at 74 MHz, the TGSS, 7C and 6C surveys at 150 MHz, the WENSS survey at
327 MHz and the B2 survey at 408 MHz. Of these, we elected not to use
the 150-MHz data, although catalogues were available, in order to be
able to use them as comparison datasets later (see below). The VLSS
covers the whole survey area and should be properly calibrated on the
scale of \cite{Scaife+Heald12}, hereafter SH (we use the VLSSr
catalogue of \citealt{Lane+14}). The B2 survey also covers the whole
survey area, and SH report that it needs no systematic correction to
their flux scale; we assign errors to the flux densities based on the
recipes given by \cite{Colla+73}. WENSS is more problematic. Firstly,
the flux density of 3C\,286 in WENSS is a factor 1.23 above the SH
value. SH report an overall scaling factor for WENSS of 0.90, but in
order to ensure that 3C\,286's flux density is correct we scale all
WENSS flux densities for NVSS objects down by a factor 0.81. Secondly,
WENSS only goes down to a declination of $30^\circ$, leaving us with
almost no coverage for the SW field. We therefore supplemented our
catalogues with one derived using {\sc pybdsm} from the 350-MHz WSRT
survey\footnote{Data kindly provided by Shea Brown.} of the Coma
cluster by \cite{Brown+Rudnick11}, having verified that these data
were on the SH scale by comparison to a less sensitive VLA map at 330
MHz reduced by us from the VLA archive. We refer to this survey as
WSRT-Coma in what follows.

Our corrections are based on an initial catalogue of LOFAR sources
which was generated by concatenating the six beam-corrected spectral
window images, convolved to the same resolution using {\sc miriad},
into a single data cube and then using the spectral index mode of {\sc
  pybdsm} to extract flux densities for each `channel' of the cube. We
used the stacked broad-band image without beam correction as the
detection image since this should have roughly uniform noise across
the field. Using the spectral index mode, rather than making
independent catalogues for each image, ensures that flux densities are
measured from matched apertures in each spectral window. The catalogue
returned at this point is of course expected to have incorrect flux
calibration, which affects both the total flux densities for each
source measured from the frequency-averaged images and the individual
channel flux densities.

We then filtered this catalogue to include only bright LOFAR sources
($>0.1$ Jy) and cross-matched positionally using {\sc stilts} with the
VLSS, B2, WENSS and WSRT-Coma catalogues, requiring that all LOFAR
sources should be detected in VLSS and at least one of the three
higher-frequency catalogues for further consideration. This gave us
50-70 sources per field with flux densities spanning the frequency
range between 74 and a few hundred MHz, all, of course, being
relatively bright and with high signal to noise in the LOFAR data: the
lowest flux densities at 74 MHz were around 0.4 Jy, corresponding to
around 0.2 Jy at LOFAR frequencies and 0.1 Jy by 300--400 MHz.
Integrated flux densities were used in all cases. We then used Markov-Chain
Monte Carlo methods, implemented in the {\it emcee} {\sc python}
package \citep{Foreman-Mackey+13}, to fit for the LOFAR flux correction factors for each
frequency. The likelihood function was calculated from the total
$\chi^2$ for a given selection of correction factors for power-law
fits\footnote{We investigated the use of models with spectral
  curvature, e.g. quadratics in log space, but the parameters of these
  were poorly constrained because the fit can trade off curvature
  against correction factors to some extent; power laws should be
  adequate for most sources over the less than one decade in frequency
  that we use here.} to all frequencies of the data (i.e. both the
unscaled VLSS and higher-frequency data and the scaled LOFAR
measurements), with the normalization and power-law index for each
source being free parameters which are determined with a standard
Levenberg-Marquart $\chi^2$ minimization. We used a Jeffreys prior for
the scale factors. In the first round of this fitting we included all
sources, but in the second round we removed sources that were outliers
in the $\chi^2$ distribution before re-fitting, which eliminates any
sources that might be erroneous matches or heavily affected by
resolution effects or might have spectra that are intrinsically poorly
fit by a power law over the band, e.g. because of spectral curvature
or variability; in practice this second round generally eliminated at
most around 10 per cent of sources and gave results very similar to
the first round. The final flux calibration factors obtained in this
way for each field are tabulated in Table \ref{tab:factors}. Nominal
MCMC-derived credible intervals (errors) on these correction factors
are small, $<1$ per cent in all cases; the dominant source of error is
of course the real deviation from a power-law model of the sources
selected as calibrators, which is hard to quantify.

\begin{table}
  \caption{Correction factors applied per field and spectral window,
    and mean calibrator and target elevations}
  \label{tab:factors}
  \begin{tabular}{lrrrrrrrr}
    \hline
    Field&\multicolumn{6}{c}{Spectral window
      number}&\multicolumn{2}{c}{Mean elevation}\\
    &1&2&3&4&5&6&3C\,196&Field\\
    \hline
    Central&1.02&1.08&1.17&1.29&1.42&1.52&85.3&51.6\\
    NW     &0.97&1.03&1.07&1.14&1.21&1.30&82.2&58.6\\
    SW     &0.93&1.00&1.09&1.17&1.29&1.34&82.4&53.4\\
    NE     &0.72&0.75&0.79&0.84&0.89&0.92&78.2&58.9\\
    \hline
  \end{tabular}
\end{table}

Once this correction had been applied to each spectral window, it was
possible to cross-check the flux scale between fields by comparing
sources in the overlap regions. By this method we established that the
flux scale for the Central, NW and SW fields was in good agreement.
The NE field appeared to have a systematic offset with respect to the
others, which we attribute to the fact that the correction factor fits
are strongly affected by 3C\,286, which is rather poorly
facet-calibrated in this field (that is, relatively large artefacts
remain around the source, so that its flux density is probably poorly
measured). The flux densities in the NE field were systematically
high. To correct for this we scaled all correction factors in the NE
field by a further factor of 0.8, determined by cross-checking the
flux densities with the overlapping fields, which brings the flux scales in line
across the four fields; this factor is included in the numbers presented in
Table \ref{tab:factors}. We comment on the effectiveness of these
corrections in allowing us to recover reliable flux densities and
in-band spectral indices in the next section.

These corrections were then applied to the pre-existing images, and
new combined images at an effective frequency of 150 MHz, again
convolved to a matching common resolution, were made by averaging the
images from the 6 spectral windows for each field. These images, for
each field, are our deepest view of the data, and attain an rms noise
of 100 $\mu$Jy beam$^{-1}$ in the centre of the best fields. Their
resolution (Table \ref{tab:field-results}) is determined by the
Gaussian fit by {\sc wsclean} to the $uv$ plane coverage in spectral
window 1, which is slightly different for each field, and particularly
different for the Central field, which was observed for a longer time
but without all the long baselines available in later years. As
convolving to a common resolution would reduce the resolution of all
the images, we elected to retain these slightly different resolutions
between fields.

We then generated source catalogues for the corrected, matched-resolution
data with {\sc pybdsm} in its spectral index mode in the same way as
described above. {\sc pybdsm}'s ability to detect sources on multiple
scales and to associate several Gaussian components as single sources
were enabled in this cataloguing step, and sources were only
catalogued if they are detected with a peak above the local $5\sigma$
value. We also make use of {\sc pybdsm}'s ability to generate a map of
the rms of the four 150-MHz images, referred to as the rms map in what
follows.

\subsection{Image quality}

In general facet calibration worked well in these fields, succeeding
in its two design aims of allowing greatly improved subtraction of
bright sources and of making high-fidelity imaging of the data at
close to the full resolution of LOFAR tractable. The resulting images
are not artefact-free, but this is a consequence of not being able to
amplitude self-calibrate every bright source: artefacts manifest
themselves as dynamic range limitations around bright sources
(particularly those that were {\it not} in the calibration sub-region
of the facet) and should have only a limited effect on the quality of
the final catalogue. Where facet calibration fails, it generally does
so by having poor signal to noise on the calibrator, leading to poor
phase and/or amplitude solutions and, generally, no improvement in
self-calibration: if these poor solutions are then applied to the data
we see a facet with much higher rms noise than would be expected. As
discussed above, this problem can sometimes be solved by increasing
the calibration region size or the interval for phase or amplitude
self-calibration, but in extreme cases the facet simply has to be
abandoned: we discarded a few facets at the edges of the NE and
Central fields, typically at $>3$ degrees and so substantially down
the primary beam, where it was not possible to obtain good
self-calibration solutions (in addition to the fact that the
calibrator sources are generally fainter when they are further down
the primary beam, bandwidth and time-averaging smearing also start to
affect the quality of the results). Other facets with poor solutions
remain in the images and give rise to high-noise regions in the rms
map.

\begin{table*}
  \caption{Basic field image and catalogue properties for the combined
  150-MHz images}
  \label{tab:field-results}
  \begin{tabular}{lrrrrr}
    \hline
    
    Field&Area covered&central rms&Median rms&Image resolution&Catalogued\\
    &(sq. deg.)&($\mu$Jy)&($\mu$Jy)&(arcsec)&sources\\
    \hline
    Central&44.5&223&782&$9.62 \times 7.02$&2,473\\
    NW&34.7&104&296&$10.01 \times 5.54$&5,335\\
    SW&36.1&111&319&$9.87 \times 5.49$&5,747\\
    NE&40.6&100&382&$11.02 \times 4.68$&4,172\\
    \hline
  \end{tabular}
\end{table*}

\begin{figure}
  \includegraphics[width=1.0\linewidth]{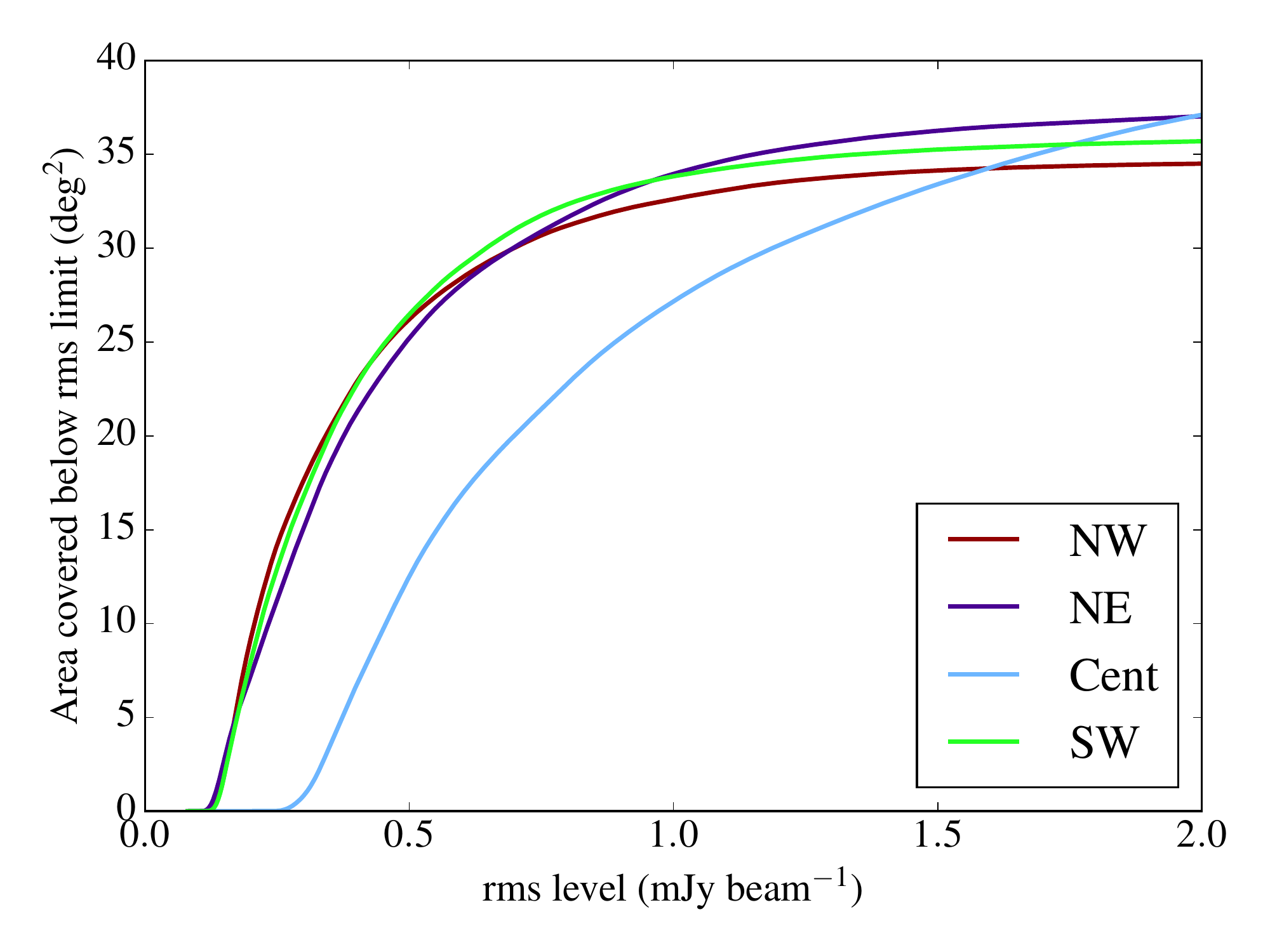}
\caption{Cumulative histogram of the area below a given rms value,
  calculated from the rms noise map derived from {\sc pybdsm} for the
  four fields. Note that, since the rms values are corrected for the
  telescope beam, the shape of this distribution depends on the
  placement of the facets as well as the intrinsic noise qualities of
  the image.}
\label{fig:rms}
\end{figure}

In Table \ref{tab:field-results} we tabulate the areal coverage,
central rms, and median rms of the four fields for the broad-band
150-MHz images, along with the final resolution achieved. The median rms is the rms below which the best half of
the field falls: this clearly depends on the placement of facets in
the beam as well as on the image quality. The area-rms distribution of
the fields is illustrated in Fig.\ \ref{fig:rms}. These rms values may
be compared to the FIRST and NVSS rms values converted to 150 MHz for
a source with $\alpha = 0.75$, which are 0.8 and 2.4 mJy beam$^{-1}$
respectively. Thus, purely considering rms levels, the LOFAR survey is
better than NVSS even far down the beam and always deeper than FIRST
in the central 50 per cent of each field. The LOFAR data are also
significantly better, in these terms, than the GMRT survey of the
equatorial H-ATLAS fields \citep{Mauch+13}, which has a best rms level
of 1 mJy beam$^{-1}$ at 325 MHz, corresponding to 1.8 mJy beam$^{-1}$
at 150 MHz. (We draw attention to the very different resolutions
  of these comparison surveys: FIRST has a resolution of 5 arcsec,
  NVSS 45 arcsec, and the GMRT survey between 14 and 24 arcsec. In
  terms of resolution, our data are most comparable to FIRST.)

The Central field, the first that was observed and the only one to be
taken in Cycle 0 with the original correlator, was by far the worst of
the fields in terms of noise despite its slightly longer observing
duration. In this field the initial subtraction was simply not very
good, suggesting large amplitude and/or phase errors in the original
calibration, although it was carried out in exactly the same way as
for the other fields. As a consequence, facet calibration did not
perform as well as in the other fields (presumably because the
residuals from poorly subtracted sources acted as additional noise in
the visibilities) and there are more bad facets than in any other
field, together with a higher rms noise even in the good ones. It
should be noted that this is also the worst field in terms of
positioning of bright sources, with 3C\,286, 3C\,287 and 3C\,284 all a
couple of degrees away from the pointing centre: we do not know
whether this, the fact that the observation was carried out early on
in the commissioning phase, poor ionospheric conditions, or a
combination of all of these, are responsible for the poor results. The
other three fields are of approximately equal quality, with rms noise
values of 100~$\mu$Jy beam$^{-1}$ in the centre of the field and each
providing around 31 square degrees of sky with rms noise below 0.8 mJy
beam$^{-1}$ after primary beam correction. The worst of these three
fields, the NE field, which has some facets where facet
calibration worked poorly or not at all, was observed partly in
daytime due to errors at the observatory, which we would expect would
lead to poorer ionospheric quality which may contribute to the lower
quality of the data.

\begin{figure*}
\includegraphics[width=1.0\linewidth]{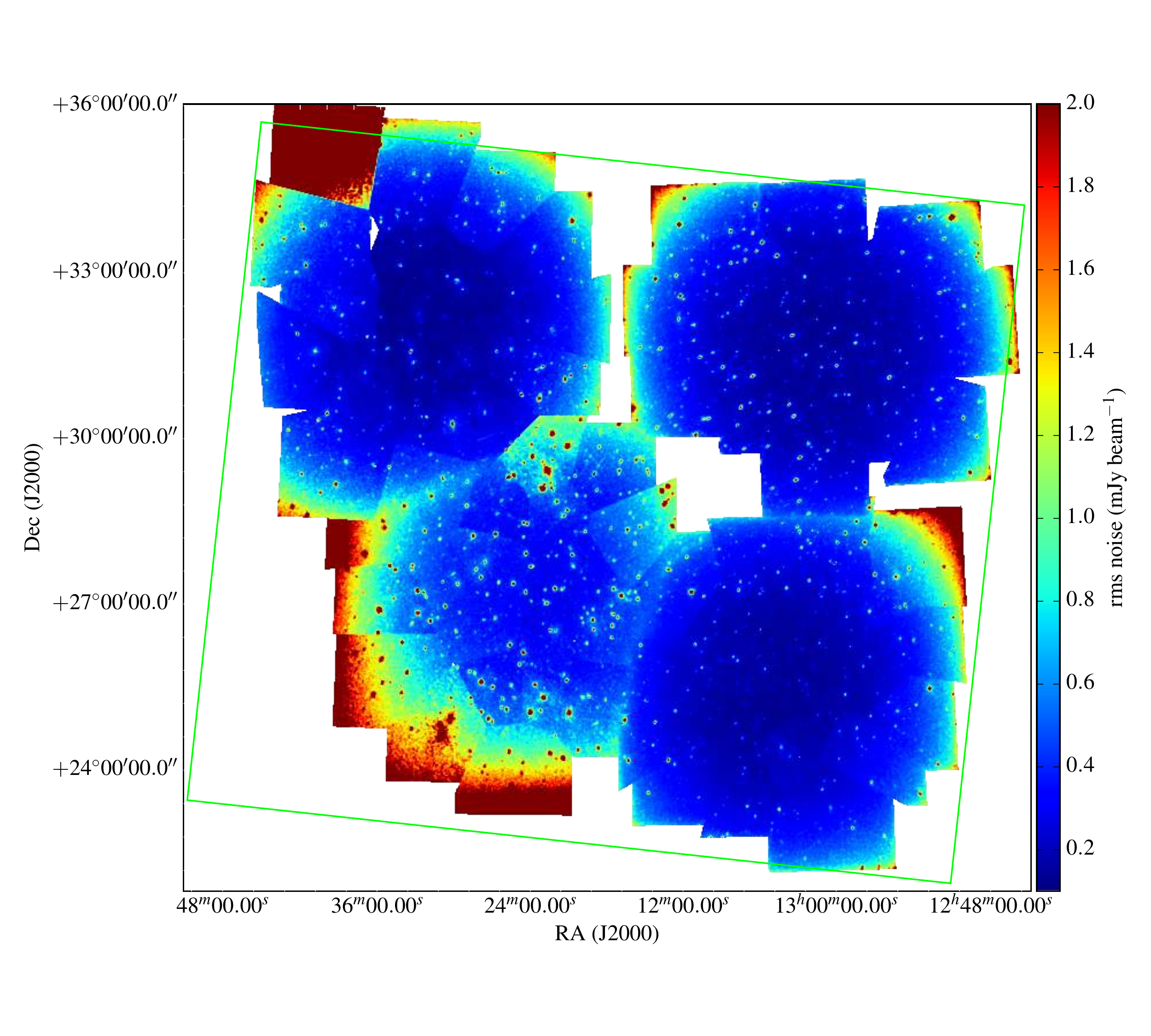}
\caption{Map showing the sky coverage and rms values of the four
  fields, constructed as described in the text. Colour levels run from
  100 $\mu$Jy to 2 mJy beam$^{-1}$. The green square shows the
  approximate boundary of the {\it Herschel} survey. The LOFAR survey
  is deeper (in rms terms) than FIRST, the previous most sensitive
  radio survey of this area, in the blue regions of the image. The
  many `point sources' in the image are the result of dynamic range
  limitations around bright objects, rather than the objects
  themselves: the pixel size in this image is 20 arcsec, significantly
  larger than the image resolution.}
\label{fig:rmsmap}
\end{figure*}

A map of the sky coverage of the images and the rms levels, generated
by resampling the full-resolution rms maps onto a grid with 20-arcsec
cell size, is shown in Fig. \ref{fig:rmsmap}. Where two fields
overlap, the best rms value is shown, for reasons explained in the
following section.

\subsection{Catalogue generation and completeness}

\label{sec:completeness}
The final source catalogue is made by combining the four per-field
catalogues. Ideally we would have combined the images of each field
and done source finding on a mosaiced image, but this proved
computationally intractable given the very large image cubes that
result from having six spectral windows. We therefore merged the
catalogues by identifying the areas of sky where there is overlap
between the fields and choosing those sources which are measured from
the region with the best rms values. This should ensure that there are
no duplicate sources in the final catalogue. The final master
catalogue contains 17,132 sources and is derived from images covering
a total of 142.7 square degrees of independently imaged sky, with
widely varying sensitivity as discussed above. Total HBA-band
(150-MHz) flux densities of catalogued sources detected using {\sc
  pybdsm} and a $5\sigma$ detection threshold range from a few hundred
$\mu$Jy to 20 Jy, with a median of 10 mJy.

For any systematic use of the catalogue it is necessary to investigate
its completeness. In the case of ideal, Gaussian noise and a catalogue
containing purely point sources this could simply be inferred from the
rms map, but neither of these things is true of the real catalogue. In
particular, the distribution of fitted deconvolved major axes in the
source catalogue shows a peak around 10 arcsec. This is probably the
result of several factors, including a certain fraction of genuinely
resolved sources, but we suspect that at least some of the apparent
broadening of these sources is imposed by the limitations of the
instrument and reduction and calibration procedure rather than being
physical. Part may be due to residual bandwidth or time-averaging
smearing in the individual facet images, though our lower angular
resolution (relative to the similar work of W16) helps to mitigate
these effects. We suspect that a significant fraction of the
broadening comes from residual phase errors in the facet-calibrated
images, particularly away from the calibration regions. This may be
compounded in our case by the effects of combining our multiple
spectral windows in the image plane -- no attempt was made to align
the images other than the self-calibration with an identical sky model
before facet calibration, and phase offsets between the spectral
windows will lead to blurring of the final image. Whatever the origin
of these effects, the fact that most sources are not pointlike in the
final catalogues needs to be taken into account in estimating the true
sensitivity of the data.

\begin{figure}
  \includegraphics[width=1.0\linewidth]{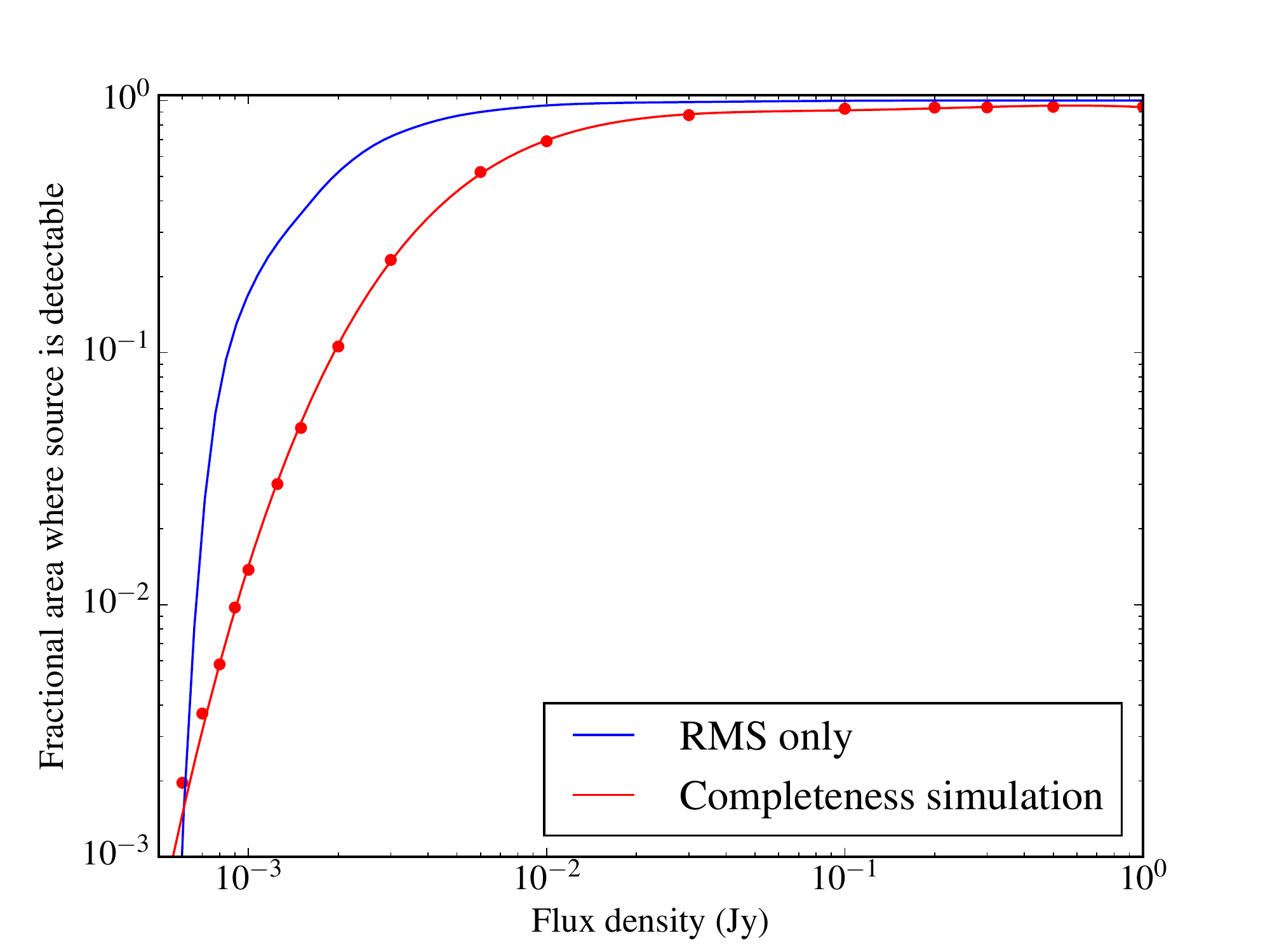}
\caption{Completeness function for the whole survey compared with the
  expectations from the rms map alone. The points are the results
  of simulations (Poisson errors, though present, are generally
  smaller than symbols and are not plotted) while the smooth curve
  shows the best-fitting 5th-order polynomial in log space used to
  approximate the completeness curve and interpolate to un-simulated
  flux densities.}
\label{fig:completeness}
\end{figure}

To assess this we therefore carried out completeness simulations in
the standard way in the image plane\footnote{In principle we should
  simulate the process all the way from the original observations,
  injecting sources in the $uv$ plane, corrupting them with simulated
  ionospheric and beam effects and repeating the facet calibration and
  imaging many times. However, although this would be a valuable
  exercise, it is computationally infeasible at present for the
  purposes of completeness simulation, and challenging even for a
  verification of the facet calibration process. Work being carried
  out along these lines in the Key Science Project will be described
  elsewhere.} [see, e.g., \cite{Heald+15} and W16] by adding in
simulated sources to the residual map for each field and recovering
them with {\sc pybdsm} with the same settings as used for the real
cataloguing. In our case, we assumed sources to be uniformly
distributed at random across the whole NGP area, and placed them on
the residual maps for the individual pointings based on the rms map
used for cataloguing. However, rather than placing point sources (i.e.
Gaussians with the parameters of the beam), we broadened the simulated
sources using a Gaussian blur where the broadening $\sigma$ was itself
drawn from an appropriate Gaussian distribution, chosen so as to
approximately reproduce in the extracted (output) catalogues the low
end of the observed distribution in deconvolved major and minor axes.
The use of the residual maps also naturally takes account of artefacts
around bright sources and other non-Gaussian features in the images,
such as any negative holes due to {\sc wsclean} aliasing effects.
We ran a number of simulations for each of a range of input source
flux densities, using between 10,000 and 30,000 simulated sources per run to
improve the statistics. We consider a source to be
matched if a source in the derived catalogue agrees with one in the
input catalogue to within 7 times the nominal error in RA and Dec and
20 times the nominal error in flux density. These criteria are deliberately
generous to reflect the fact that the errors on flux density and position from
off-source noise are generally underestimates. Noise peaks from the
residual map are removed from the catalogue before this comparison is
made to avoid false positives. It is also possible to recover false
detection rates in this way, but these are known to be very low (W16)
and so we do not discuss them further here.

The results are shown in Fig.\ \ref{fig:completeness}, where, for
comparison, the $5\sigma$ detection level for pure point sources based
on the rms map and the assumption of Gaussian noise is also shown. It
can be seen that the various effects we simulate have a strong effect
on completeness. The survey is complete, in the sense that a source of a
given flux density can be detected essentially anywhere, only above a
comparatively high flux density of $\sim 20$ mJy. At lower flux
densities, the completeness curve drops more steeply than the rms map
would imply. At 1 mJy, for example, the completeness curve implies a
probability of detection (for a source placed at random in the field)
ten times lower than would be inferred from the rms map. The curves
intersect again at very low flux densities ($\sim 0.5$ mJy), but we
suspect that the detection fraction here is artificially boosted by
Eddington bias (i.e. simulated sources placed on noise peaks in the
residual map are more likely to be recovered). The slight errors in
the completeness curve resulting from this are not problematic given
that there are so few sources with these flux densities in any case.
Also plotted in Fig.\ \ref{fig:completeness} is the best-fitting
5th-order polynomial in log space fitted to the results of the
simulations (taking account of the Poisson errors): this function
gives an adequate approximation to and interpolation of the
completeness curve, which we will make use of in later sections.

It is important to note that much of this incompleteness results from
the sparse sky coverage of the observations for this project, and the
poor quality of the Cycle 0 central field data. It is not
representative of the expectations for the Tier 1 (wide-area) LOFAR
sky survey: see W16 for a more representative completeness curve.

\subsection{Association, artefact rejection and optical identification}
\label{sec:optid}

The source catalogue was the starting point for our source association
and optical identification processes, which were carried out in
parallel. Optical identification was carried out using images and
catalogues from SDSS Data Release 12 \citep{Alam+15}, hereafter DR12.
Initially, we carried out a simple positional crossmatch for low-$z$
galaxies, selecting compact (deconvolved size $<10$ arcsec) LOFAR
sources whose position matched that of an optical source from the
MPA-JHU\footnote{The MPA-JHU catalogue is the Max Planck Institute for
  Astrophysics/Johns Hopkins University catalogue of bright SDSS Data
  Release 7 galaxies with spectroscopic redshifts: see
  \url{http://wwwmpa.mpa-garching.mpg.de/SDSS/DR7/}. This catalogue
  was used because the MPA-JHU catalogue forms the basis of the work
  on the radio/star-formation relation to be described by G\"urkan
  \etal} catalogue within 8 arcsec (chosen based on the distribution
of offsets). This identified 1,048 LOFAR sources, of which we would
expect around 30 to be chance coincidences given the number of MPA-JHU
sources in the survey area. We then visually inspected the LOFAR,
SDSS, FIRST and NVSS images for {\it all} the 16,084 remaining
sources, initially with a single author (one of GG, MJH or SCR)
inspecting each source. The person carrying out the visual inspection
was asked to associate individually detected LOFAR sources, i.e. to
say whether s/he believed that they were physically associated, to
identify any artefacts, and, for real sources, to specify any
plausible optical identification for the radio source. The NVSS images
were used only to confirm the reality of faint extended LOFAR sources,
which often show up well in the low-resolution NVSS data, but the
FIRST images had a more important role, as they turn out often to show
the flat-spectrum core of an extended LOFAR source making optical
identification far more robust. Identifications by one author were
cross-checked against those of another to ensure consistency and a
subset (consisting of a few hundred large, bright sources) of the
first pass of identifications were re-inspected visually by several
authors and some (a few per cent) corrected or rejected from the final
catalogue. The final outcomes of this process were (a) an associated,
artefact-free catalogue of 15,292 sources, all of which we believe to
be real physical objects, and (b) a catalogue of 6,227 objects with
plausible, single optical identifications with SDSS sources,
representing an identification fraction of just over 40 per cent.
(Note that around 50 sources with more than one equally plausible
optical ID are excluded from this catalogue; further observation would
be required to disambiguate these sources.) This identification
fraction is of interest because we can expect to achieve very similar
numbers in all parts of the Tier 1 LOFAR survey where SDSS provides
the optical catalogue. Forthcoming wide-area optical surveys such as
Pan-STARRS1 and, in the foreseeable future, LSST (for equatorial
fields), will improve on this optical ID rate.

Optical identification using shallow optical images can lead, and
historically has led, to misidentifications, where a plausible
foreground object is identified as the host instead of a true unseen
background source. This is particularly true when the LOFAR source is
large and no FIRST counterpart is seen. It is difficult to assess the
level of such misidentifications in our catalogue [likelihood-ratio
  based methods, such as those of \cite{Sutherland+Saunders92},
  require information about where plausible optical IDs could lie in a
  resolved radio source that is hard to put in quantitative form] but
as our resolution is relatively high, so that most sources are not
large in apparent angular size and do not have more than one plausible
optical ID, we expect it to be low. Sensitive high-frequency imaging
over the field, and/or deeper optical observations, would be needed to
make progress.

In what follows we refer to the raw, combined output from {\sc pybdsm}
as the `source catalogue', the product of the association process as
the `associated catalogue' and the reduced catalogue with SDSS optical
IDs as the `identified catalogue'. Sources in the associated or
identified catalogues that are composed of more than one source in the
source catalogue are referred to as `composite sources' (in total
2,938 sources from the original catalogue were associated to make
1,349 composite sources). The process of association renders the {\sc
  pybdsm}-derived peak flux densities meaningless (they are suspect in
any case because of the broadening effects discussed in the previous
subsection) and so in what follows unless otherwise stated the flux
density of a LOFAR source is its total flux density, derived either
directly from the source catalogue or by summing several associated
sources.

\section{Quality checks}
\label{sec:quality}

In this section we describe the tests carried out on the catalogues to
assess their suitability for further scientific analysis. From here
on, except where otherwise stated, we use only the associated and
identified catalogues.

\subsection{Flux scale tests: 7C crossmatch}

An initial check of the flux scale was carried out by crossmatching
the associated catalogue with the 7C catalogue \citep{Hales+07} over
the field. Unfortunately the NGP spans the southern boundary of 7C, so
we do not have complete coverage, though there is substantial overlap.
The crossmatching uses the same algorithm as that described by
\cite{Heald+15}, i.e. we use a simple maximum likelihood crossmatch
taking account of the formal positional errors in both catalogues and
using the correct (Rayleigh) distribution of the offsets, but not
taking into account any flux density information. Since 7C sources are
very sparse on the sky, any more complicated procedure is probably
unnecessary. Over the intersection of the 7C and LOFAR/NGP survey
areas, there are 735 7C sources, 694 of which (94 per cent) are
detected in the LOFAR images, with a mean positional offset of
$\delta_{RA} = 0.87 \pm 0.34$ arcsec and $\delta_{\rm Dec} = 0.28 \pm
0.32$ arcsec. The flux limit of 7C is a few hundred mJy, so we would
expect all 7C sources to be detected by LOFAR: in fact, the few
nominally unmatched 7C sources are either at the edges of one of the LOFAR
fields, where the sensitivity is very poor, or are actually close to a
LOFAR source but with discrepant co-ordinates, which could be
attributed to the very different resolutions of the surveys -- 7C has
a resolution of $70 \times 140$ arcsec at this declination. 7C is
complete above $\sim 0.4$ Jy at 150 MHz, and for sources above this
flux limit the mean ratio between 7C and LOFAR 150-MHz total flux
densities is $1.00 \pm 0.01$, showing excellent agreement between the
7C and LOFAR flux scales, though the scatter is larger than would be
expected from the nominal flux errors. We can conclude that there are
no serious global flux scale errors in the catalogue, at least in the
region covered by 7C (essentially the NE and NW fields).

\subsection{Flux scale tests: NVSS crossmatch}
\label{sec:nvss-flux}

\begin{figure}
  \includegraphics[width=1.0\linewidth]{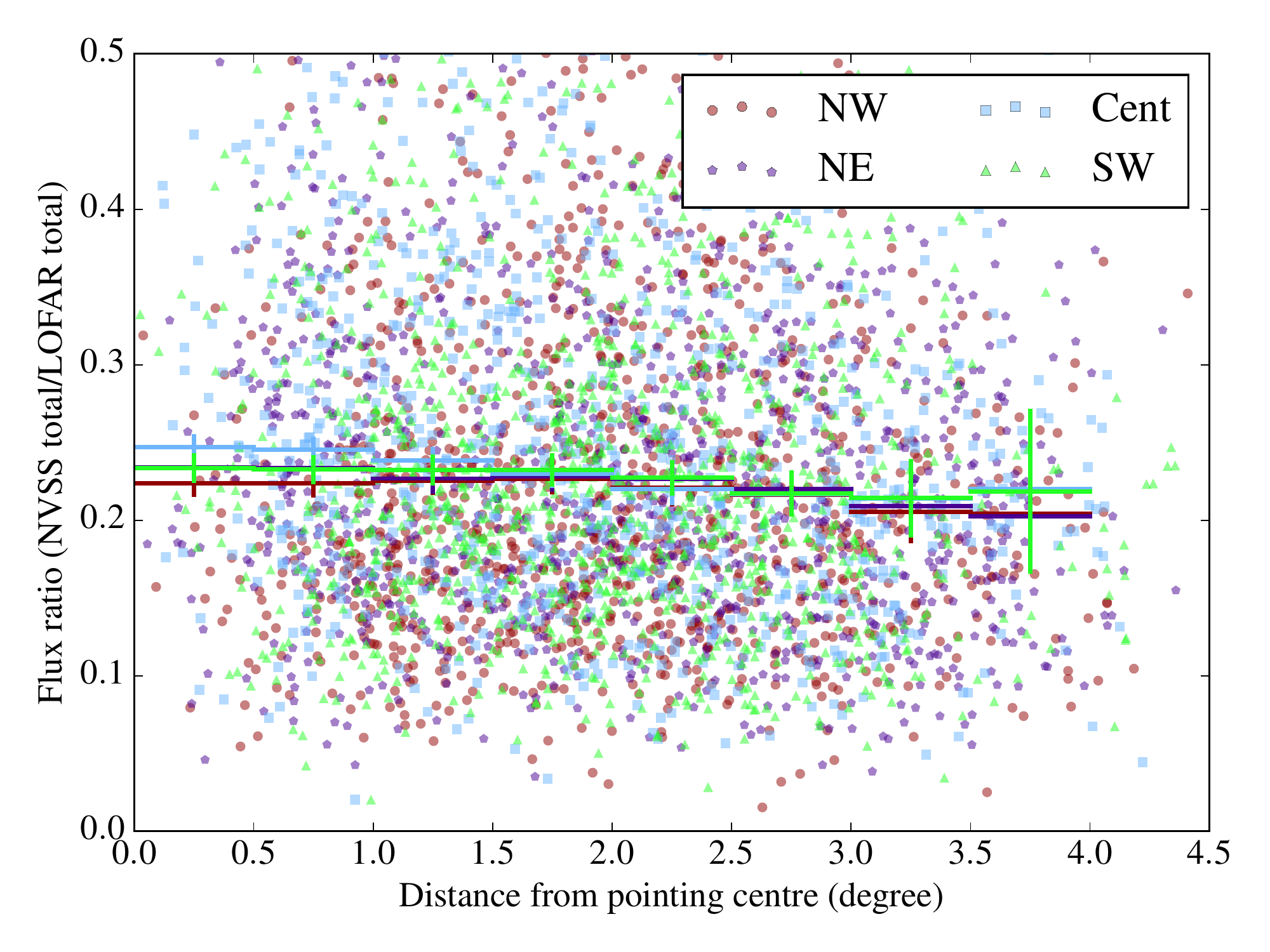}
  \caption{NVSS/LOFAR flux ratio as a function of radial distance from
  the pointing centre of each field. Points show individual matched
  sources, solid lines show the median in radial bins and its error
  on the approximation of Gaussian statistics.}
\label{fig:ratio-radius}
\end{figure}

The most suitable high-frequency survey for a direct comparison with
the LOFAR results is NVSS, which is sensitive to large-scale
structure, although its resolution is much lower than that of the
LOFAR images. To generate a suitable catalogue we extracted the NVSS
images from the image server and mosaiced them into a large image
covering the whole field. We then applied {\sc pybdsm} to this mosaic
with exactly the same settings as were used for the LOFAR catalogue;
this procedure allows us to measure accurate total flux densities for
extended sources, rather than inferring them from the peak flux
densities and Gaussian parametrization provided in the NVSS catalogue.
Filtering our {\sc pybdsm} catalogue to match the area coverage of the
LOFAR survey, we found 5,989 NVSS sources. These were then
crossmatched to the LOFAR data as for the 7C data, but adding a
Gaussian term to the likelihood crossmatching to favour sources where
the flux densities are consistent with the expected power law of
$\alpha \approx 0.7$ (i.e. a term proportional to $\exp(-(S_{\rm
  LOFAR} - (1400/151)^{0.7} S_{\rm NVSS})^2/\sigma^2)$: this helps to
reduce the incidence of spurious crossmatches) and also excluding
associations with a separation between NVSS and LOFAR positions of
greater than 1 arcmin. We obtained 4,629 matches: that is, as
expected, the vast majority of the NVSS sources have LOFAR
counterparts, with a mean positional offset of $0.4 \pm 0.1$ arcsec in
RA and $0.05 \pm 0.1$ in Dec. Counterparts are genuinely missing at
the edges of the LOFAR field, where the noise is high, but we have
verified by visual inspection that the comparatively large number of 
`unmatched' sources within the field are the result of disagreements
about source position (e.g. arising from structure resolved by LOFAR
but unresolved by NVSS) rather than from genuinely missing sources.
Similarly, most bright LOFAR sources have an NVSS counterpart. We
therefore do not regard the match rate of only 77 per cent as
problematic: visual inspection of the images could probably bring it
close to 100 per cent.

Our expectation is that NVSS should be uniformly calibrated (VLA flux
calibration uncertainties are 2-3 per cent, which should not introduce
much scatter into this comparison), so the flux ratios between NVSS
and the LOFAR catalogue allow an accurate check of the flux scale,
subject only to the possibility that the fields have genuinely
intrinsically different spectral index distributions, which could
happen, for example, if the SW field were affected by the presence of
the Coma cluster. For a further check of the flux scale and also its
dependence on radius we computed the median NVSS/LOFAR flux ratio for
all matched sources (median rather than mean to avoid effects from
strong outliers which might arise from misidentifications or extreme
intrinsic spectral indices) and also its dependence on distance from
the pointing centre for each facet in bins of 0.5 degree in radius. We
see (Fig.\ \ref{fig:ratio-radius}) that there are no significant flux
scale (or, equivalently, spectral index) offsets between fields. The
scatter is large, but much of this is imposed by the known dispersion
in spectral index (see below, Section \ref{sec:outofband}).

An encouraging result from the radial plot is that there is also no
significant systematic difference with radius, within the
uncertainties imposed by the scatter in the data. This suggests (a)
that the primary beam correction applied is adequate, and (b) that
bandwidth and time-averaging smearing at the edge of the field, beyond
2-3 degrees, do not seem to be having any detectable effect on the
LOFAR total flux densities. The same comparison was also carried out
using the peak flux densities of the LOFAR images and those of the
cross-matched FIRST sources (see below), which should be more
sensitive to smearing effects, again with no discernible radial
dependence of the ratios.

\subsection{Flux scale tests: TGSS crossmatch}
\label{sec:tgss}

\begin{figure}
\includegraphics[width=1.0\linewidth]{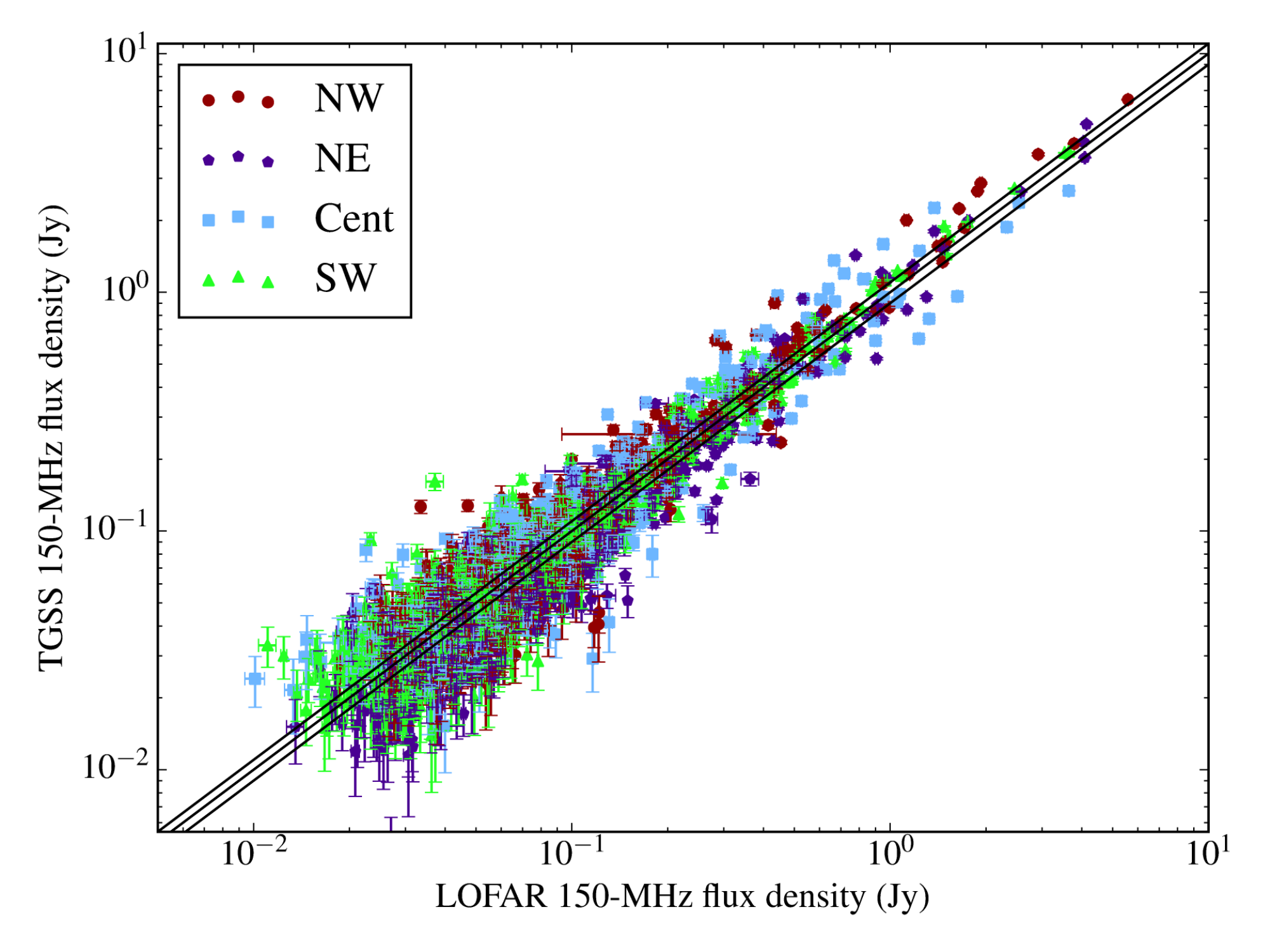}
\caption{TGSS against LOFAR flux densities, colour-coded by field.
  Error bars are plotted for all data points, but in many cases are
  smaller than the symbols. The central line shows the median ratio
  between the flux densities (1.08) and the dispersion that would be
  implied by 10 per cent calibration uncertainties on top of this. The
per-field flux scale offsets between LOFAR and TGSS data can be seen
as colour gradients across the plot.}
\label{fig:tgss-lofar}
\end{figure}

For comparison with a deeper survey than 7C at 151 MHz we make use of
the data from the TGSS survey made with the GMRT with a resolution of
$25 \times 25$ arcsec \citep{Intema+16}. As with the NVSS data, we
made a single large mosaic of the images, extracted flux densities
using {\sc pybdsm}, and then cross-matched positionally with the LOFAR
data. There are 2896 TGSS sources in the LOFAR field, of which almost
all (2449) can be cross-matched with LOFAR data. Surprisingly, given
the good agreement between LOFAR and NVSS flux scales (Section
\ref{sec:nvss-flux}), we see non-negligible per-field differences in
the mean LOFAR and TGSS flux densities. There is no overall flux scale offset
(as measured from median ratios of all matched sources), but the
median TGSS/LOFAR ratios for the individual fields vary between 0.86
and 1.10. GMRT flux calibration is itself not reliable to better than
about 10 per cent, and the overall medians will be dominated by the
sources close to the centre of each field, so it is perfectly possible
that much of this scatter comes from GMRT calibration uncertainties.
In addition, the GMRT's flux scale can be adversely affected by bright
sources in the field, and this is apparent, for example, in the flux
for the calibrator source 3C\,286, which is significantly offset in
the GMRT catalogue from the reference 150-MHz value of SH. We
therefore do not attempt to use the TGSS images to derive further
corrections to the per-field flux scale, but simply report the TGSS
comparison here for the benefit of future workers. Plotting the LOFAR
(corrected) total flux densities against TGSS flux densities (we
restrict the comparison to sources that should be unresolved to TGSS)
shows a good correlation, but, as with 7C, the scatter is larger than
would be expected from the nominal errors
(Fig.\ \ref{fig:tgss-lofar}), indicating some residual calibration
errors in either or both of the TGSS and LOFAR datasets. In the
absence of a detailed study of the TGSS flux calibration, we cannot
establish whether one or both of the datasets are responsible for
this.

\subsection{Positional accuracy tests: FIRST crossmatch}
\label{sec:first-crossmatch}

\begin{figure}
  \includegraphics[width=1.0\linewidth]{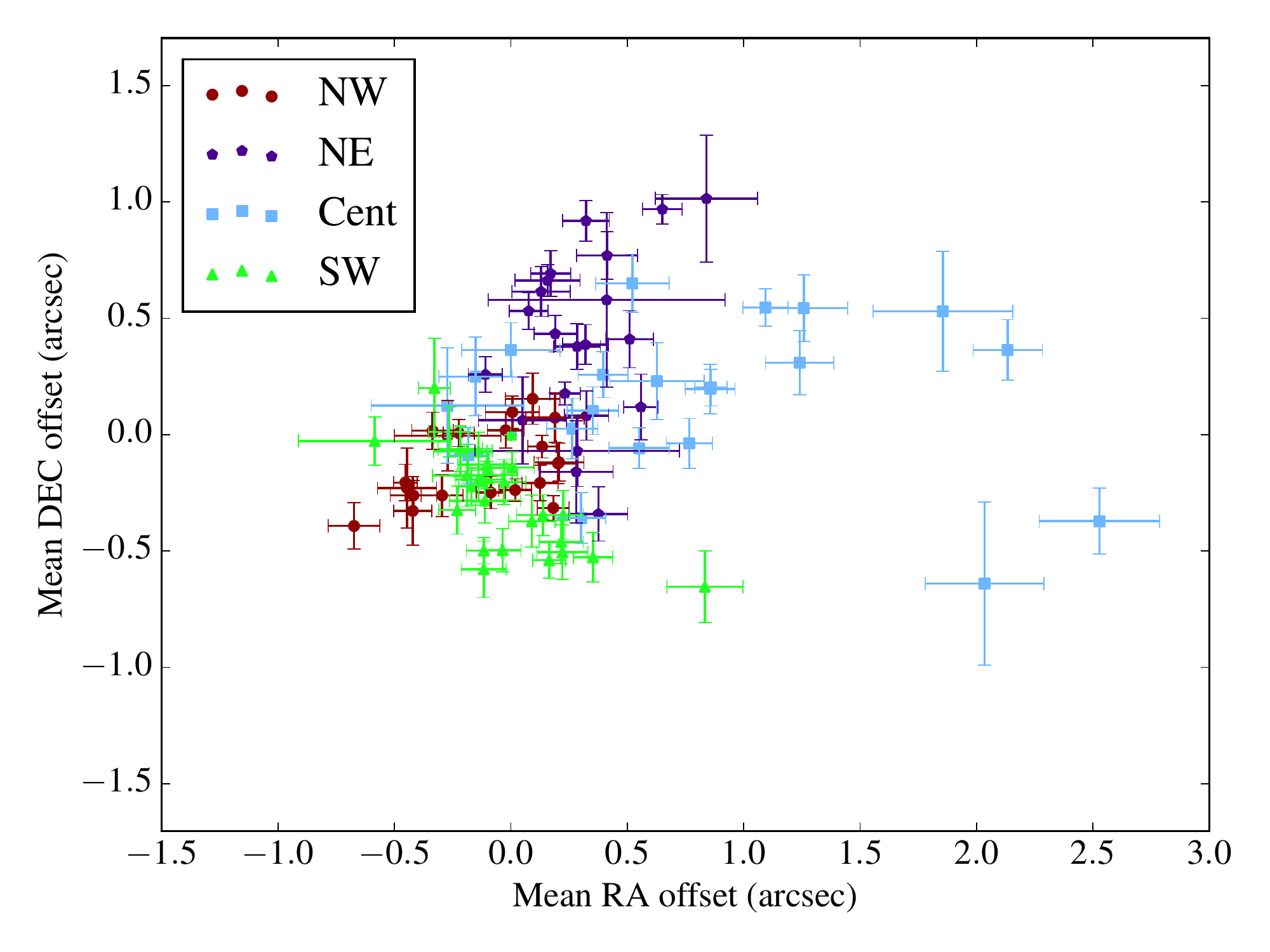}
\caption{FIRST/LOFAR offsets in the field. The mean offset for each
  facet is plotted individually. Error bars show the nominal errors on
  the mean offsets.}
\label{fig:offset-facet}
\end{figure}

The 7C crossmatch shows that there are no gross astrometric errors in
the catalogue, but to investigate positional accuracy in more detail
we need a larger sample with higher resolution. For this purpose we
cross-matched the source catalogue with the FIRST survey data in the
field. There are 9,856 FIRST sources in the survey area, after
filtering out sources with FIRST sidelobe probability (i.e.
probability of being an artefact) $>0.05$. We restricted
the crossmatching to compact LOFAR sources (fitted size less than 15
arcsec) with well-determined positions (nominal positional error less
than 5 arcsec). 3,319 LOFAR/FIRST matches were obtained by this
method, with a mean offset over the whole field of $\delta_{\rm RA} =
0.16 \pm 0.01$ arcsec and $\delta_{\rm DEC} = 0.01 \pm 0.01$ arcsec.
Using these matches, we can determine the mean LOFAR/FIRST offset
within each facet, shown in Fig.\ \ref{fig:offset-facet}. Some facets
have relatively few matches, so the results should be treated with
caution, but a couple of points are fairly clear. Firstly, the typical
offsets are small, a couple of arcsec at most: given that any offsets
are likely introduced by the phase self-calibration in the facet
calibration process, we would not expect them to be much larger than
the pixel size of 1.5 arcsec, as is observed. Secondly, fields in
which we had worse results with facet calibration also show larger
offsets; by far the largest offsets are seen in some facets of the
Central field, which, as discussed above, also has significantly
higher noise. This is consistent with the idea that the quality of the
initial direction-independent phase calibration has a strong effect on
the final facet calibration results: if the initial phase calibration
is poor, we expect offsets in the initial images for the first
(phase-only) facet self-calibration step, and we will never be able to
recover from these completely without an external reference source.

\subsection{In-band spectral index}
\label{sec:inband}

\begin{figure}
   \includegraphics[width=1.0\linewidth]{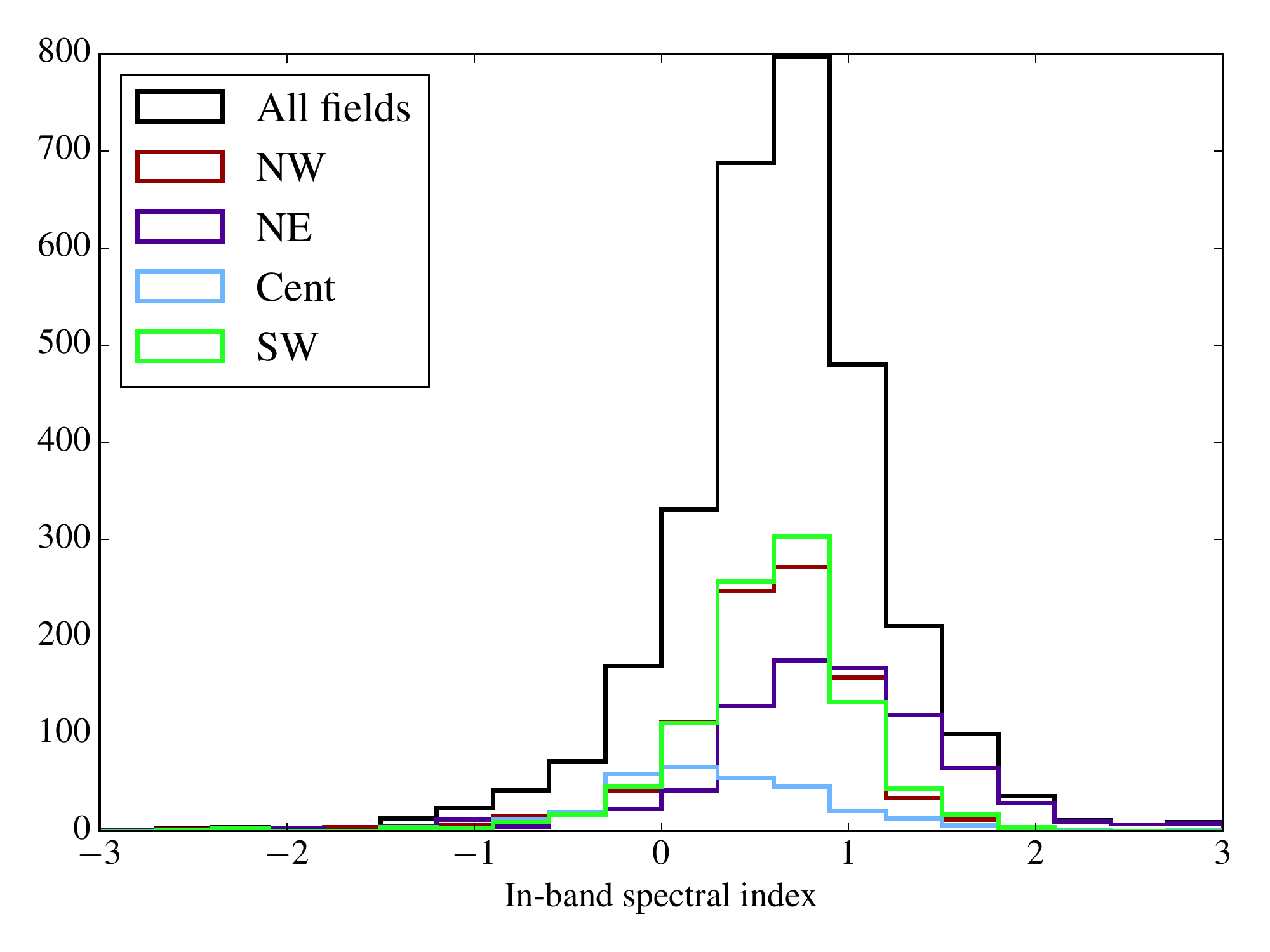}
\caption{In-band spectral index determined from LOFAR data. The overall histogram is shown together with the
histograms for each field.}
\label{fig:ibspix}
\end{figure}
 
We fitted power laws in frequency to the total flux densities for each source
in the associated catalogue. The in-band spectral index is a sensitive
test of the validity of the correction factors applied to the flux densities
in each field prior to combination, as even small calibration errors
will lead to large biases in in-band spectral index over the
relatively narrow HBA band alone. Many sources have poor $\chi^2$
values (suggesting that the errors in the catalogue are
underestimated) or large errors on the spectral index (estimated from
the fitting covariance matrix). The in-band spectral index
distribution for the overall associated catalogue and the four fields
is shown in Fig.\ \ref{fig:ibspix}, where we plot only sources with
nominal $1\sigma$ spectral index errors of $<0.2$ and exclude the
highest $\chi^2$ values ($\chi^2 > 80$). It can be seen that, although
the overall in-band spectral index distribution is reasonable and
peaked around the expected value (0.6--0.7), the catalogues for the
four fields have rather different distributions. The central field, in
particular, shows a peak at flat spectral index values which must be
the result of the generally poorer quality of the data in this field,
while the NE field has an excess of steep-spectrum sources.
By contrast, the normalization of the power-law fits at 150 MHz is
generally in good agreement with the broad-band total flux density we
measure. We conclude that in-band spectral indices cannot be reliably
compared between fields in this dataset, though sources with extreme
apparent in-band spectral index remain interesting topics for further
investigation. Reliable absolute in-band spectral index measurements
will require the LOFAR gain transfer problems to be solved by the use
of a correctly normalized beam model.

\begin{figure*}
   \includegraphics[width=0.48\linewidth]{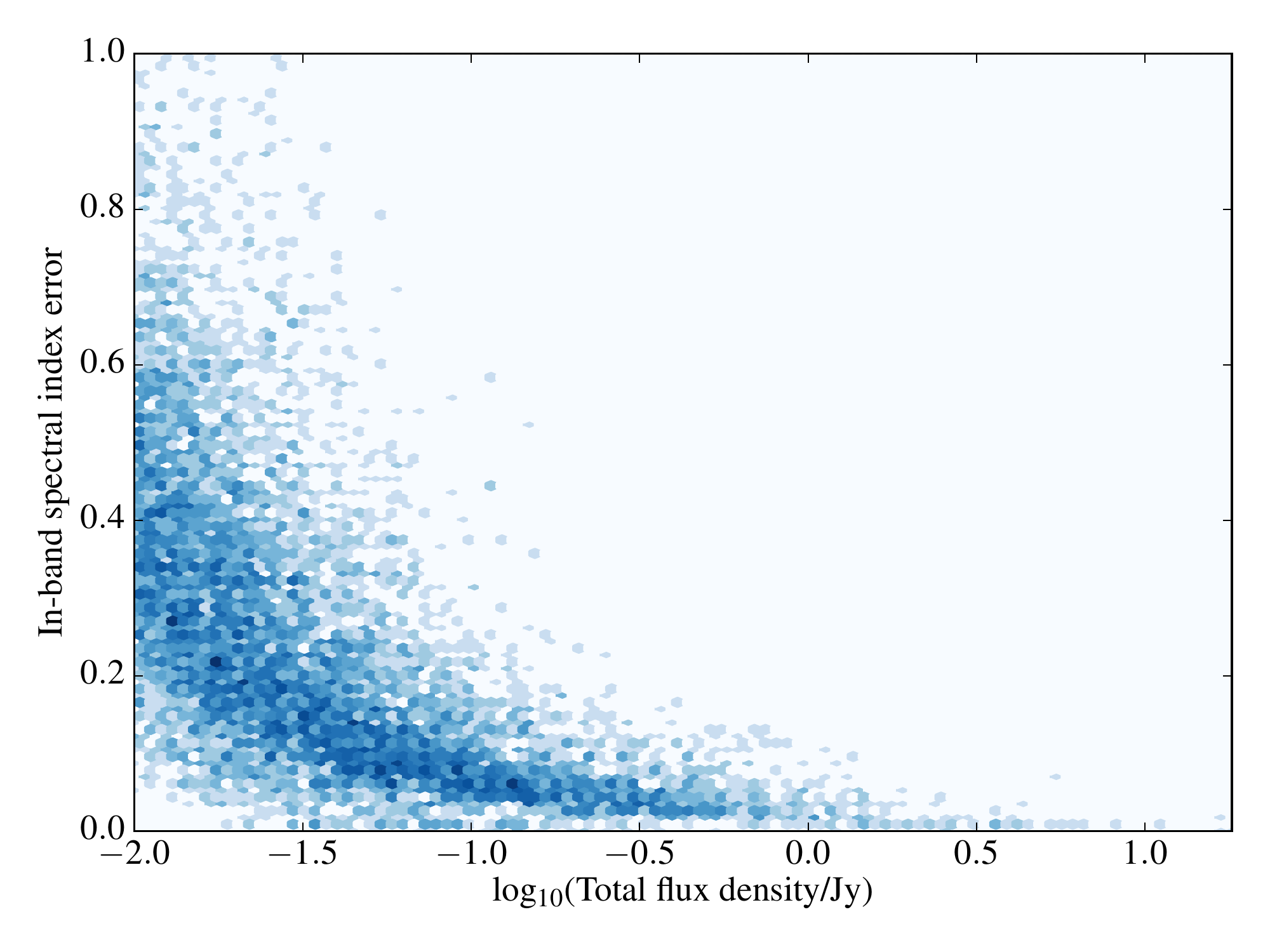}
   \includegraphics[width=0.48\linewidth]{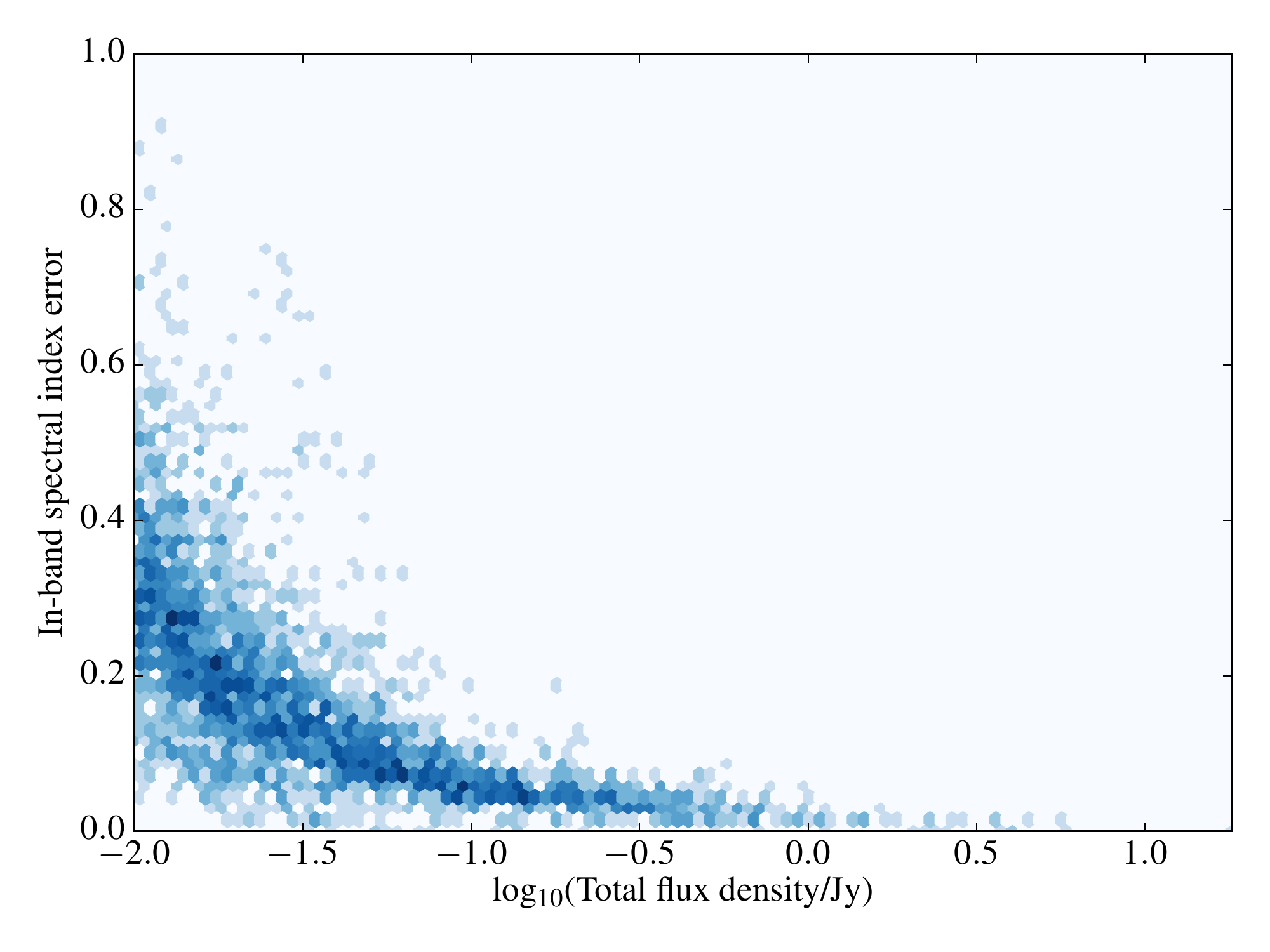}
\caption{Density plot of errors on the in-band spectral index as a
  function of total LOFAR flux density. Left: all sources and fields.
  Right: only the central 2 degrees of the NW, NE and SW fields are
  plotted.}
\label{fig:spixerr}
\end{figure*}

We can in addition comment on the errors on the in-band spectral index
to be expected from HBA data. Fig.\ \ref{fig:spixerr} shows the error
on in-band spectral index as a function of flux density for the
associated catalogue, both for the whole catalogue and for the inner 2
degrees of the three best fields, which should be more representative
of Tier 1 quality. It can be seen that errors are typically less than
$\pm 0.1$ only for bright sources, with flux densities $>100$ mJy,
even in the centres of the best fields. For almost all sources,
therefore, a much cleaner spectral index determination will be
obtained by comparing with NVSS, which will detect all but the
steepest-spectrum LOFAR sources with LOFAR flux densities above a few
tens of mJy. It will be possible to use in-band spectral index to
select sources which are extremely steep-spectrum (and so undetected
in NVSS) but this will only be reliable, even after LOFAR gain
calibration problems are solved, if they are also bright.

\subsection{Out-of-band spectral index}
\label{sec:outofband}

\begin{figure*}
  \includegraphics[width=0.33\linewidth]{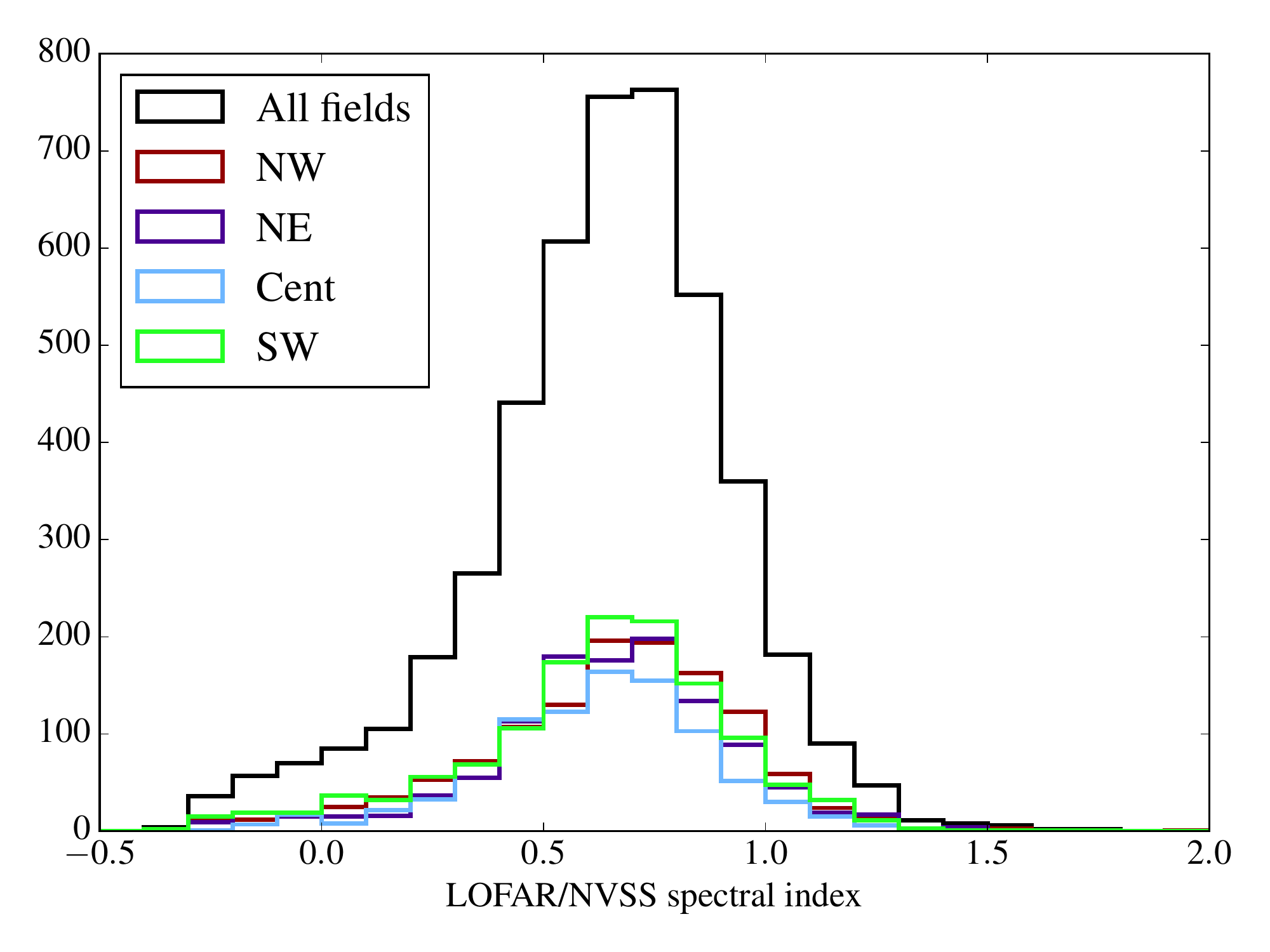}
  \includegraphics[width=0.33\linewidth]{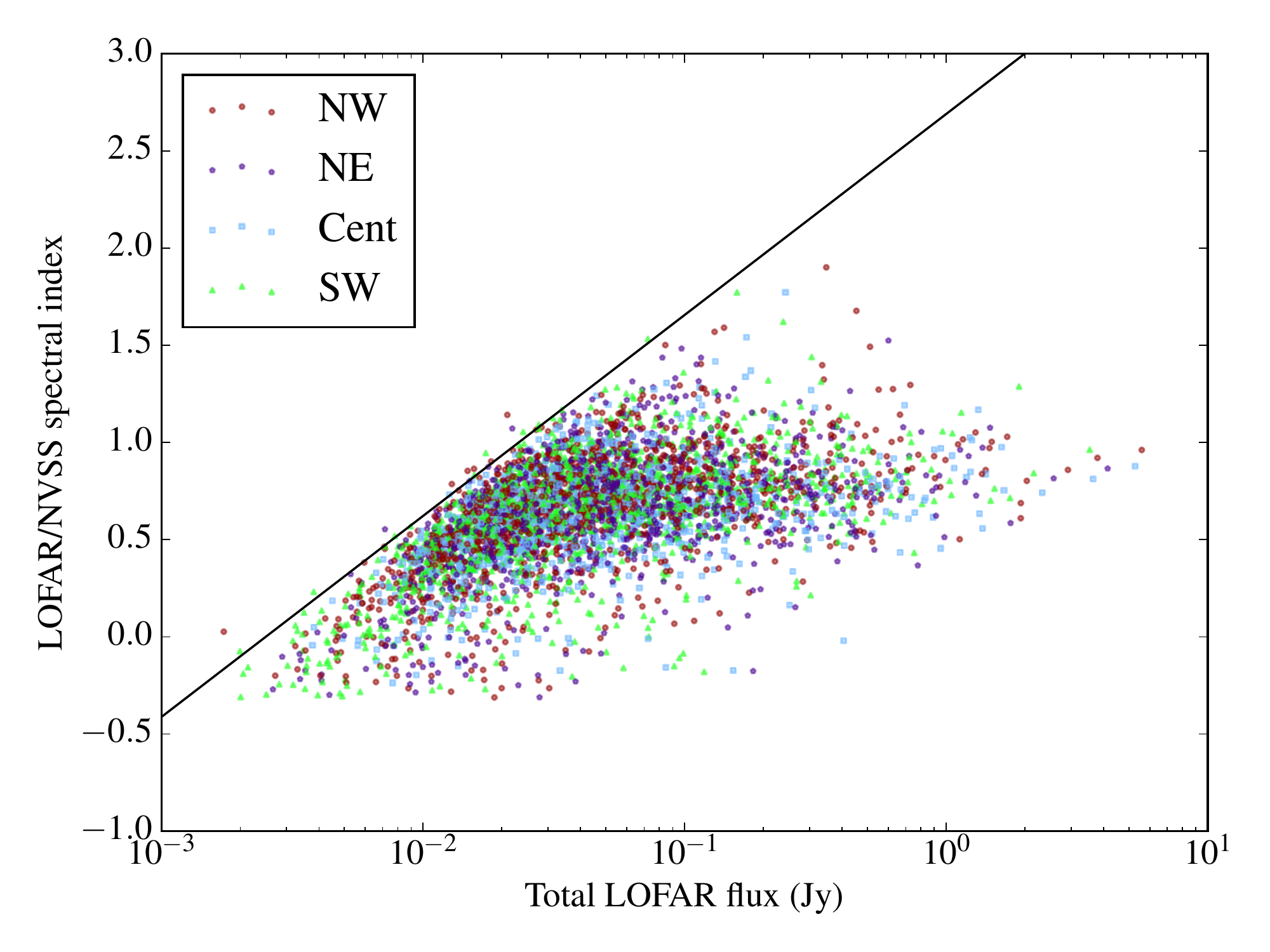}
  \includegraphics[width=0.33\linewidth]{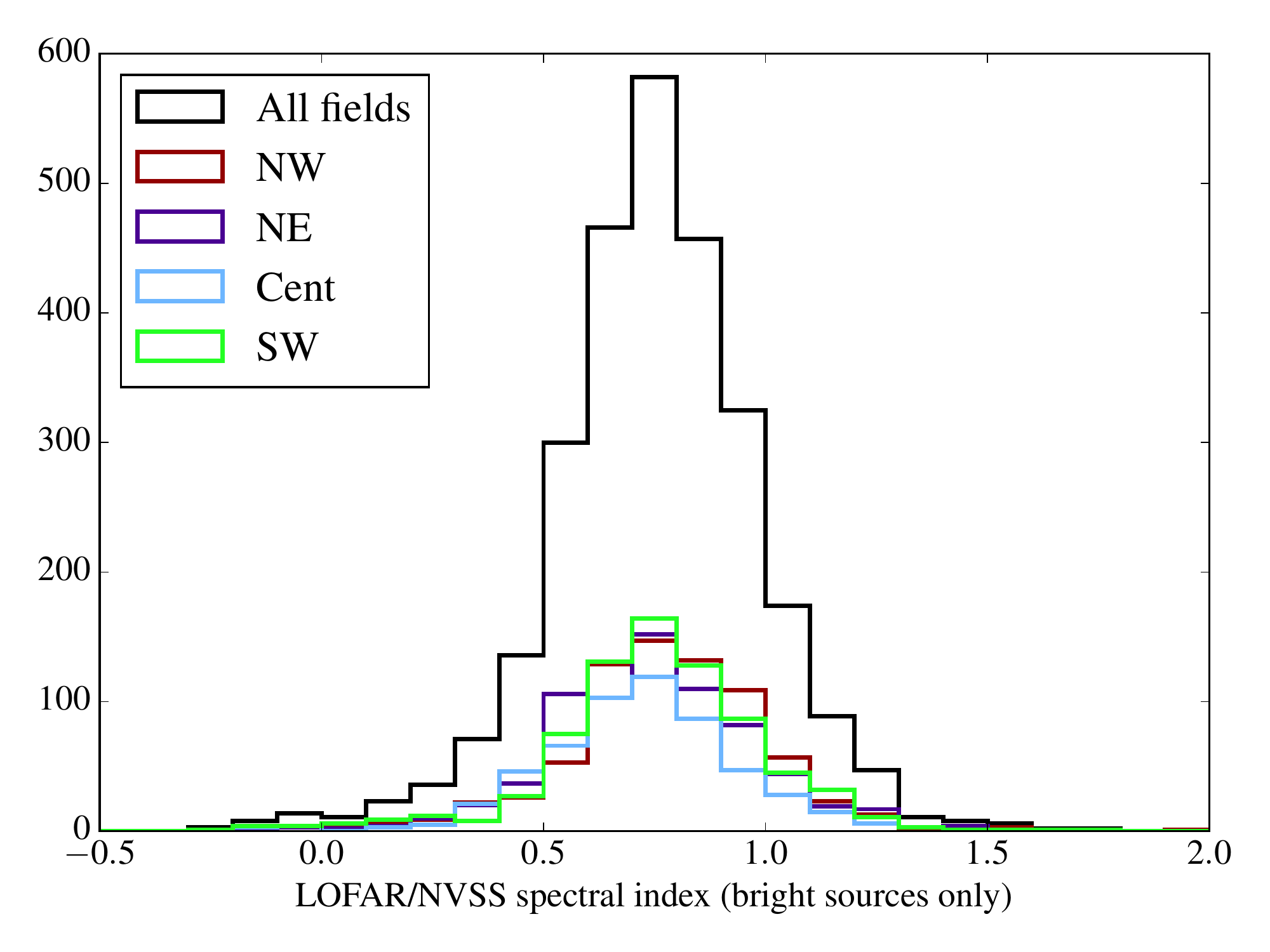}
  \caption{Left: histogram of LOFAR/NVSS spectral indices for all
  sources and for the individual fields. Centre: the relationship
  between spectral index and flux density: the solid line shows the region
  to the top left that cannot be populated by point sources given the NVSS
  sensitivity limit. Right: the histogram of spectral index for bright
  ($S_{150} > 30$ mJy) sources only.}
\label{fig:nvhist}
\end{figure*}

We use the NVSS/LOFAR crossmatch described above (Section
\ref{sec:nvss-flux}) to construct a distribution of spectral indices
between 150 MHz and 1.4 GHz (Fig.\ \ref{fig:nvhist}). The median
NVSS/LOFAR spectral index is 0.63, with almost no differences seen
between fields. It is important to note that the effective flux
density limit of $\sim 2.5$ mJy for point sources in the NVSS data
biases the global spectral index distribution to low (flat) values --
only flat-spectrum counterparts can be found to faint LOFAR sources
(Fig. \ref{fig:nvhist}). If we restrict ourselves to sources where
this bias is not significant, with LOFAR flux densities above $\sim
30$~mJy, the median spectral index becomes $0.755 \pm 0.005$ (errors
from bootstrap), in good agreement with other determinations of the
spectral index distribution around these frequencies \citep[e.g.,][and
  references therein]{Mauch+13}. Deeper 1.4-GHz data with comparable
$uv$ plane coverage to LOFAR's are required to investigate the
spectral index distribution of faint sources.

\begin{figure}
   \includegraphics[width=1.0\linewidth]{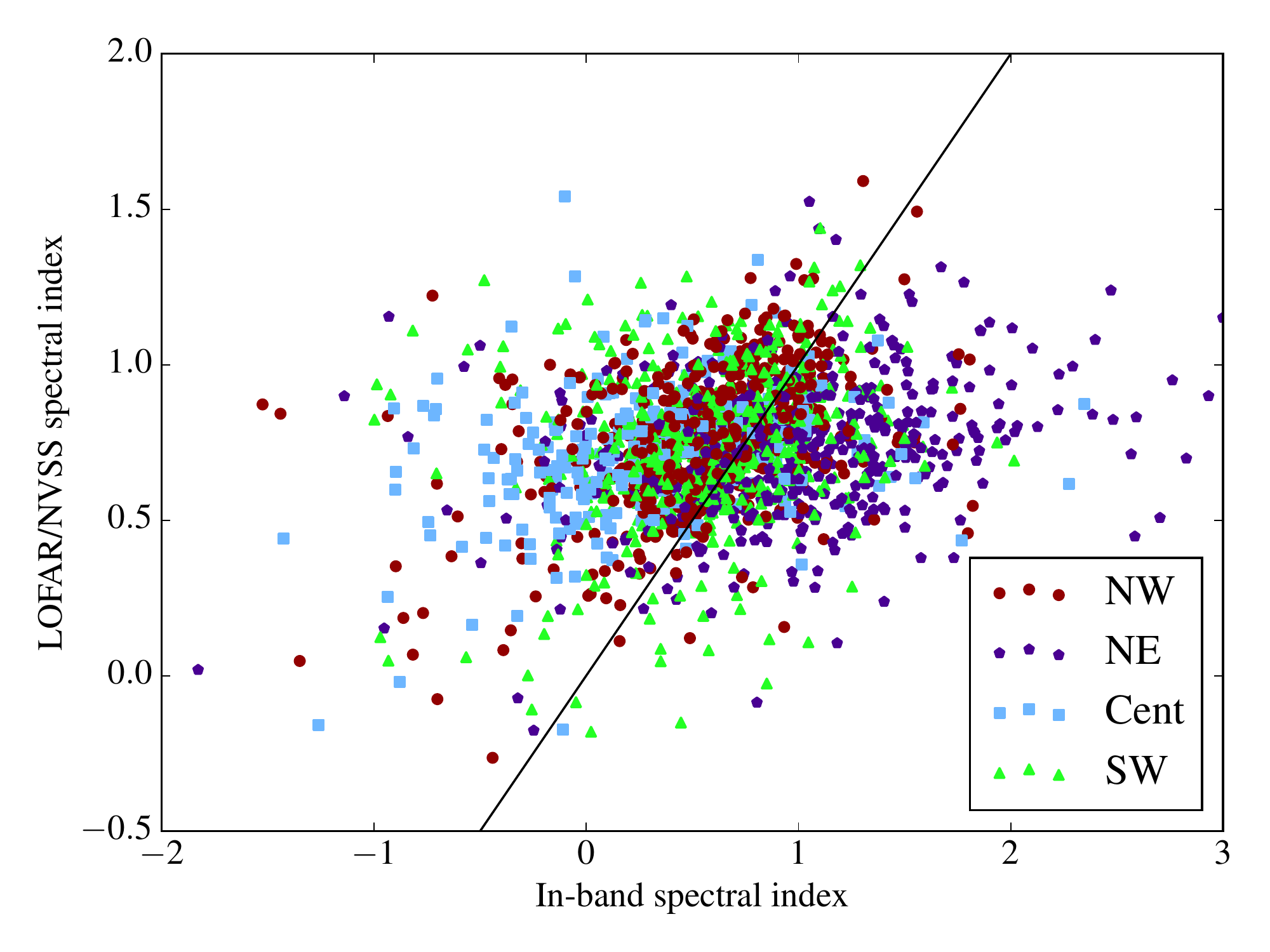}
\caption{The in-band and NVSS/LOFAR spectral indices compared. The solid
line shows equality between the two spectral indices. In general
increasing spectral steepness with frequency means that we
would expect points to lie above this line.}
\label{fig:spixcompare}
\end{figure}

With both the in-band and LOFAR/NVSS spectral indices in hand, we can
compare the two, and this comparison is shown in
Fig.\ \ref{fig:spixcompare}. Here we plot the $\sim 2,000$ sources
that have LOFAR flux density $>30$ mJy and also satisfy the
requirement that the nominal error on the in-band spectral index is $<0.1$
and the fit is acceptable. A general tendency for the in-band spectral
index to be flatter than the LOFAR/NVSS index is observed,
unsurprisingly, but many sources exhibit unrealistically steep (in the
NE field) or flat (in the Central field) in-band indices, and in
general the scatter in the plot is probably dominated by the known
per-field biases in in-band index. It is
possible to identify in this plot some individual sources
that plausibly have interestingly steep, inverted or curved spectra,
but the unreliability of the in-band index limits its use.

\subsection{The optical identifications}

As noted above, 6,227, or approximately 40 per cent, of the sources in
the associated catalogue have optical identifications with either
galaxies or point-like objects (presumably quasars) from the SDSS DR12
{\it photoobj} table. Of these, 1,934 have spectroscopic redshifts in
the {\it specobj} table and an additional 3,660 have photometric (but
not spectroscopic) redshifts, leaving 633 with no redshift information
(we discard objects with nominal errors $>0.3$ on the
photometric redshift). 263 objects are classed as pointlike in the
photometry catalogue based on the {\it prob\_psf} field, of which 89 have spectroscopic redshifts; the
pointlike objects with spectroscopic redshifts
are likely almost all quasars and we refer to them as quasars in what follows.

The highest spectroscopic redshift in the sample is for a quasar at
$z=5.2$, but no object that is not a quasar has a redshift much
greater than 1, as expected given the magnitude limits of SDSS; the
sharp cutoff in photometric redshifts at $z \approx 1$ is presumably a
consequence of the absence of $z>1$ objects from the training sets used in
SDSS photo-$z$ determination \citep{Beck+16}, but the locus of
magnitudes of radio-galaxy hosts with spectroscopic redshifts clearly
intercepts the SDSS $r$-band magnitude limit of 22.2 at this redshift
in any case. Detecting higher-redshift radio galaxies will require
deeper optical data. The spectroscopic coverage of the galaxies that
we do detect is excellent due to the presence of spectra from the
Baryon Oscillation Spectroscopic Survey (BOSS: \citealt{Dawson+13}) in
DR12, and as a result the number of objects with spectroscopic
redshifts is comparable to that in the FIRST/Galaxy and Mass Assembly
(GAMA: \citealt{Driver+09,Driver+11})-based sample of
\cite{Hardcastle+13}, although the distribution of redshifts is rather
different. The WEAVE-LOFAR
project\footnote{\url{http://star.herts.ac.uk/~dsmith/weavelofar.html}.}
\citep{Smith15} aims to obtain spectra and redshifts for essentially
all of the radio sources in the field.

\begin{figure*}
  \includegraphics[width=0.48\linewidth]{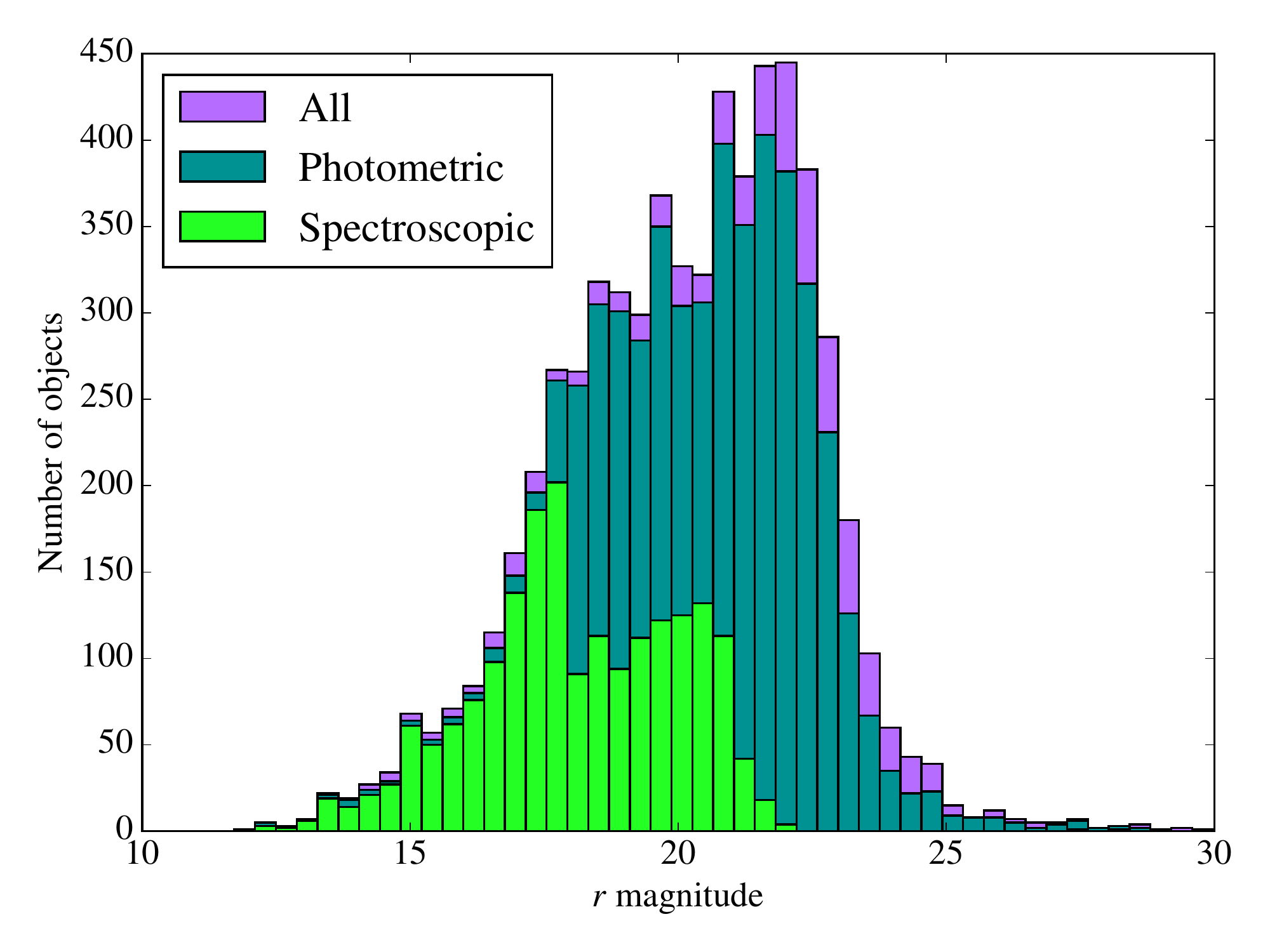}
  \includegraphics[width=0.48\linewidth]{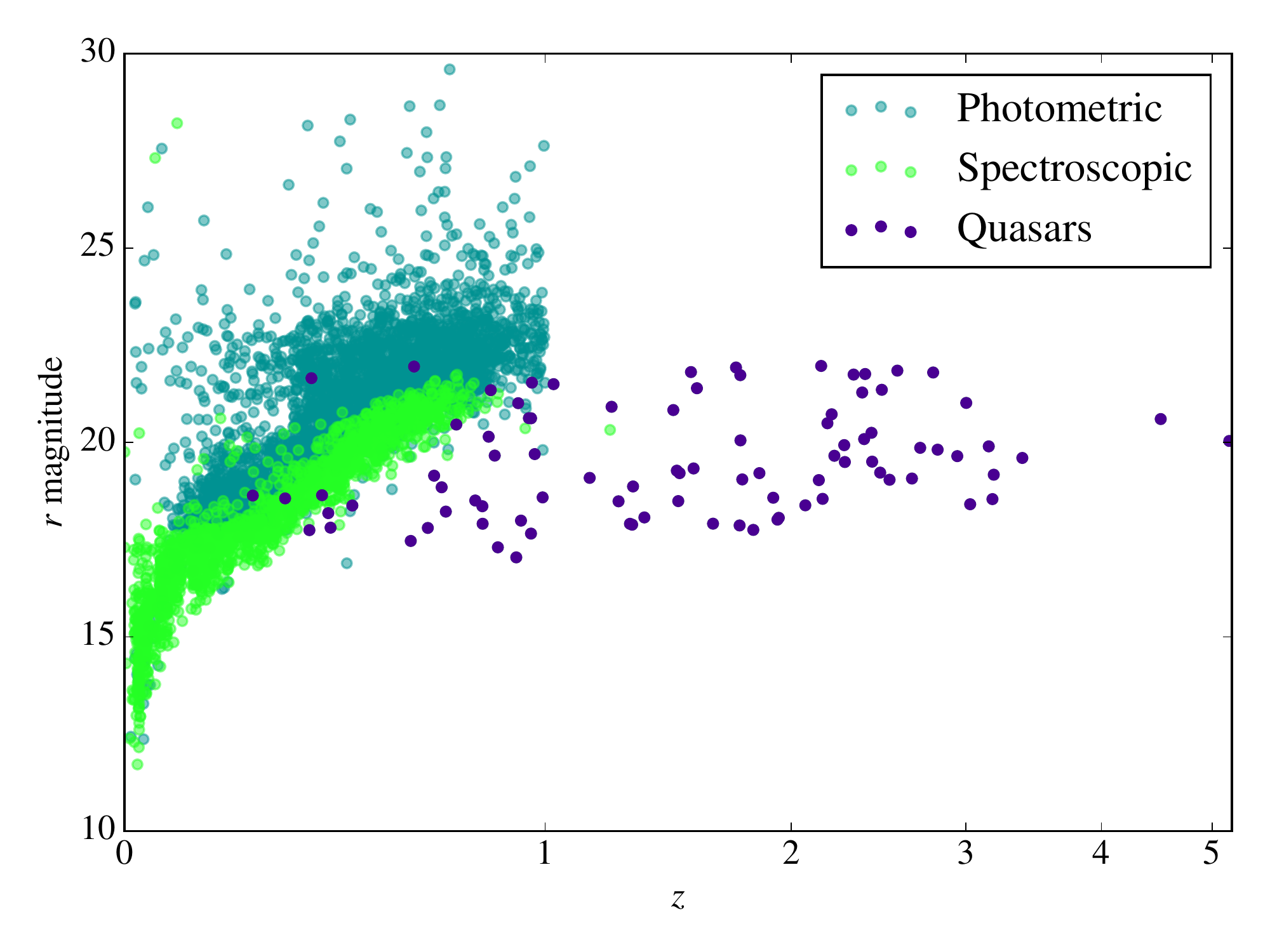}
  \caption{Left: the distribution of identified sources with spectroscopic,
    photometric, and no redshifts as a function of $r$-band magnitude.
    The second peak of spectroscopic redshifts at $r \approx 20$ is
    due to BOSS selection.
  Right: $r$-band magnitude as a function of redshift. Redshift is
  plotted on a $\log(1+z)$ scale but labelled with $z$ values.}
  \label{fig:magdist}
\end{figure*}

In Fig. \ref{fig:magdist} we plot the Petrosian $r$ magnitude from
SDSS for the optically identified sample, showing objects with
spectroscopic, photometric or no redshift. We see that the sample is
virtually spectroscopically complete at $r<17.7$ mag, and almost all
sources have a spectroscopic or photometric redshift at $r<19$ mag. A
clear lower limit in magnitude at a given redshift is seen, expected
since radio-loud AGN tend to be the most massive galaxies at any
redshift; the very few sources with an apparent magnitude too bright
for their redshift are likely to be due to erroneously high
photometric redshifts, but these are too small in number to
significantly affect our analyis. We also note a small population of
objects that are very faint in $r$, due presumably to SDSS photometric
errors in the $r$-band -- most of these objects have more reasonable
magnitudes in other SDSS bands.

\begin{figure*}
  \includegraphics[width=0.48\linewidth]{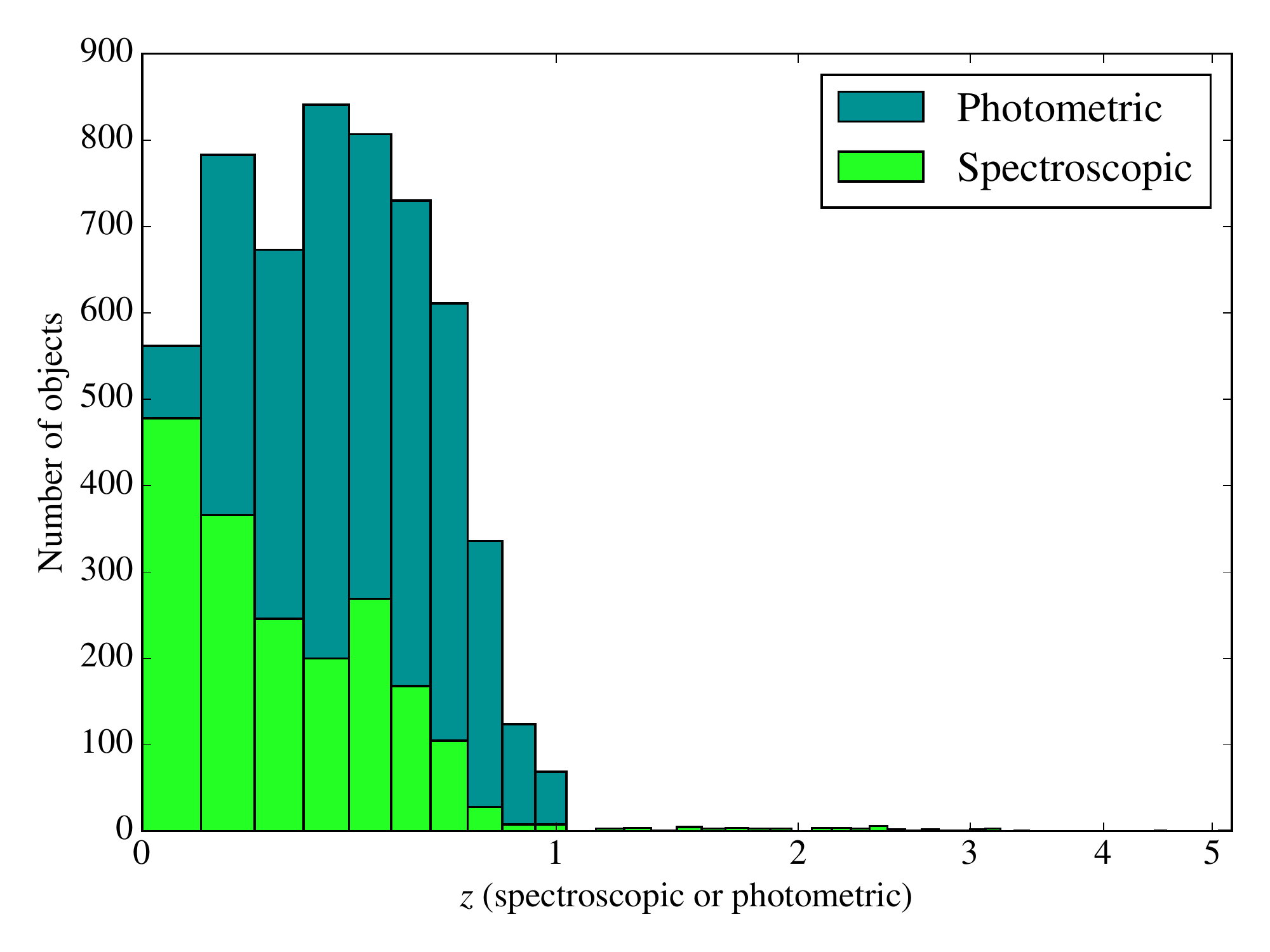}
  \includegraphics[width=0.48\linewidth]{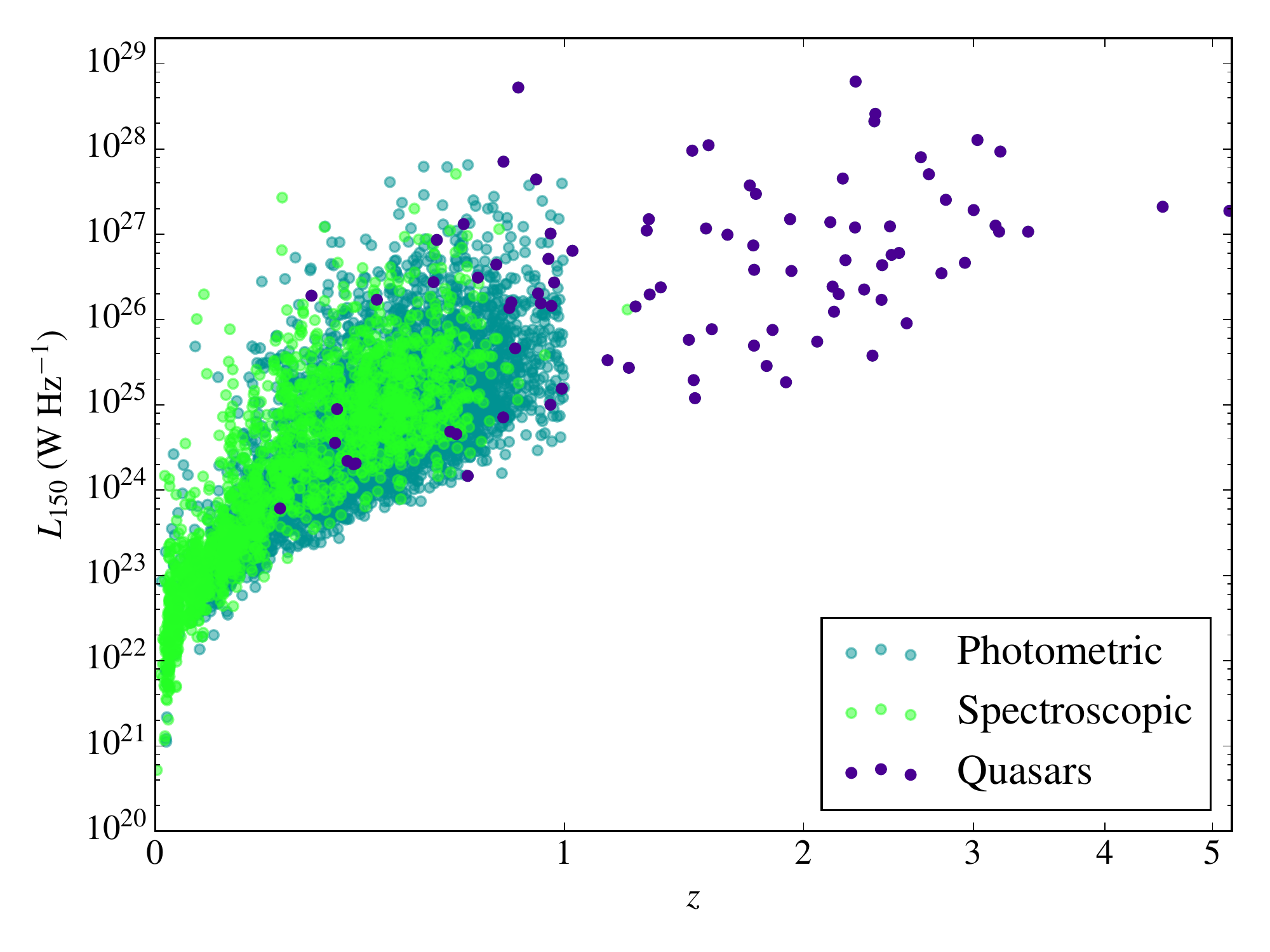}
  \caption{Left: redshift and right: radio luminosity distributions for the
    optically identified galaxy catalogue and the subsample with
    spectroscopic redshifts. Redshift is
  plotted on a $\log(1+z)$ scale but labelled with $z$ values.}
  \label{fig:zdist}
\end{figure*}
  
Fig. \ref{fig:zdist} shows the distribution of spectroscopic and
photometric redshifts in the galaxy sample, and the corresponding
radio luminosities (where we use a single spectral index of $\alpha =
0.7$ for K-correction). We see that the radio luminosities of the
optically identified sample span the range from $10^{21}$ W Hz$^{-1}$
(where we would expect star formation to be the dominant process)
through to well above $10^{26}$ W Hz$^{-1}$ (the nominal FRI/FRII
break luminosity at 150 MHz) even for the spectroscopic subsample. The
wide area and high sensitivity provided by LOFAR coupled with the
availability of spectroscopy for a large number of faint galaxies in
SDSS DR12 drives the wide range in radio luminosity that we observe.

\section{Initial science results}
\label{sec:science}
In this section we discuss some scientific conclusions that can
easily be drawn from the various catalogues that we have constructed.
Detailed analyses of all these topics will be presented in later papers.

\subsection{Source counts}
\label{sec:counts}

\begin{figure}
  \includegraphics[width=1.0\linewidth]{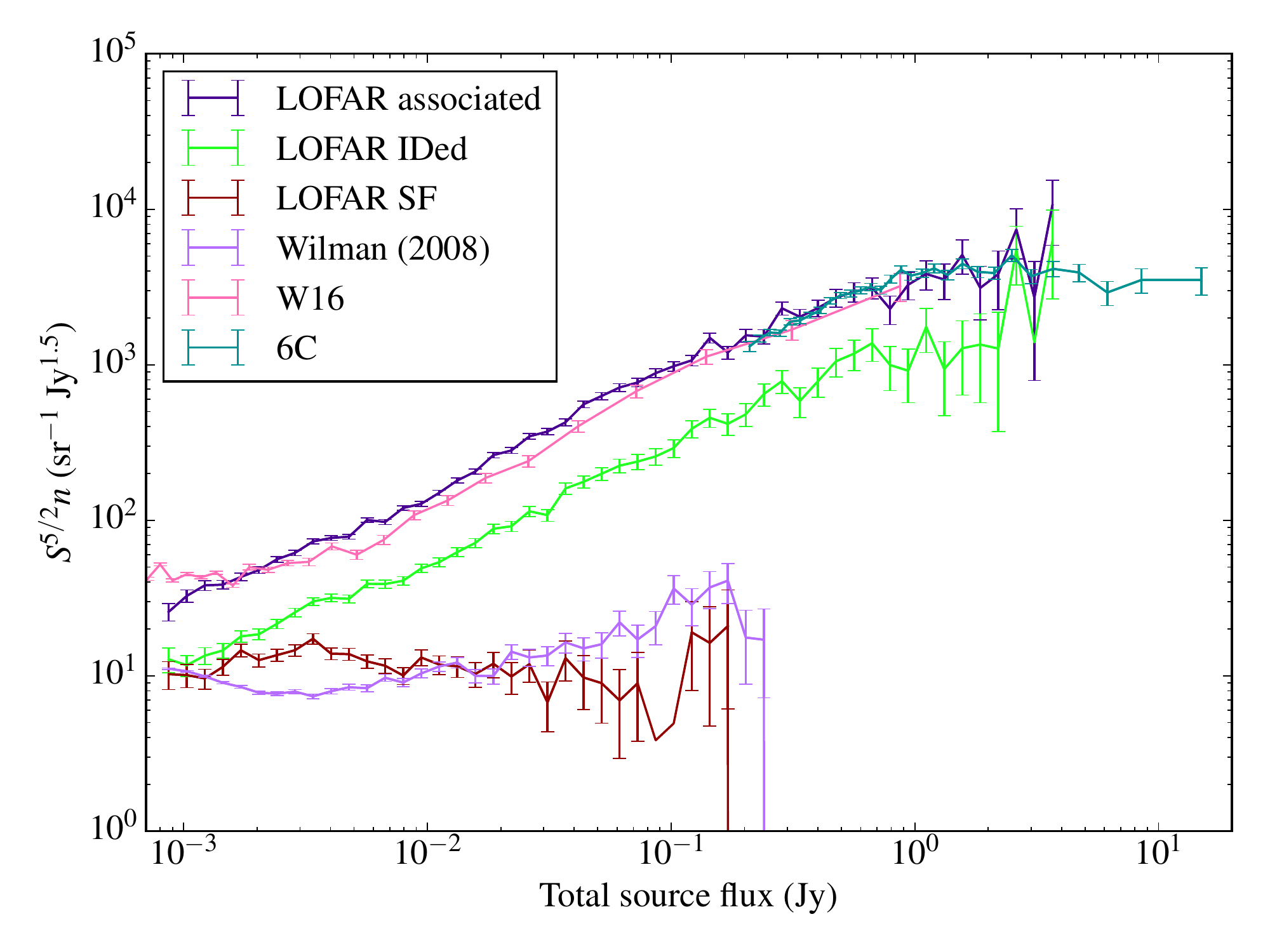}
  \caption{Euclidean-normalized differential source counts from the
    LOFAR associated catalogue after completeness correction.
    Overplotted are the 6C counts from \protect\cite{Hales+88},
    the LOFAR counts from W16, the
    counts for the identified catalogue, and the counts for objects
    classed as star-forming (see Section \ref{sec:sfg}) with the
    corresponding counts for SFGs from the simulations of
    \protect\cite{Wilman+08}. For clarity the very few points on the
    simulated SFG line with total flux density greater than 0.25 Jy are not
    plotted. Error bars are suppressed when there is only one count in
  the corresponding bin.}
  \label{fig:sourcecounts}
\end{figure}

The associated catalogue allows us to construct the standard
Euclidean-normalized differential source counts plot for the LOFAR
sample, and this is shown in Fig.\ \ref{fig:sourcecounts}. For
comparison at the bright end, we plot the 6C 151-MHz source counts of
\cite{Hales+88}. There is excellent agreement between the
normalization and slope of the 6C and LOFAR data where they overlap,
given the Poisson uncertainties on numbers of sources at the bright
end in the LOFAR data. Our source counts are corrected for
completeness (Section \ref{sec:completeness}) and of course take
account of physical associations between objects in the original
catalogue, but are not corrected for any other effects. W16, in their
similar but higher-resolution study, suggest that resolution bias,
i.e. the fact that resolved sources are less likely to be detected,
affects the counts significantly below a few mJy, where the SNR is
low, and this can be seen affecting the sub-mJy flux counts in the
comparison of their results with ours in Fig.\ \ref{fig:sourcecounts}; more detailed completeness simulations taking into account the
intrinsic distribution of source sizes would be necessary to have
confidence in the source counts at the very faint end of this plot.
Elsewhere our results are close to, but generally slightly above,
those of W16, which may be a result of our different approach to
completeness corrections.

\subsection{Cross-match with H-ATLAS}
\label{sec:herschel}
The H-ATLAS project produces maps and catalogues following the methods
described by \mbox{\cite{Pascale+11}} (SPIRE mapping), \cite{Ibar+10}
(PACS mapping) and \cite{Rigby+11} (cataloguing). An up-to-date
description of the process for the public data, shortly to be
released, will be provided by Valiante \etal\ (in prep.) and
descriptions of the NGP maps and catalogues will be provided by Smith
\etal\ (in prep.) and Maddox \etal\ (in prep.) respectively.

\begin{figure*}
  \includegraphics[width=0.48\linewidth]{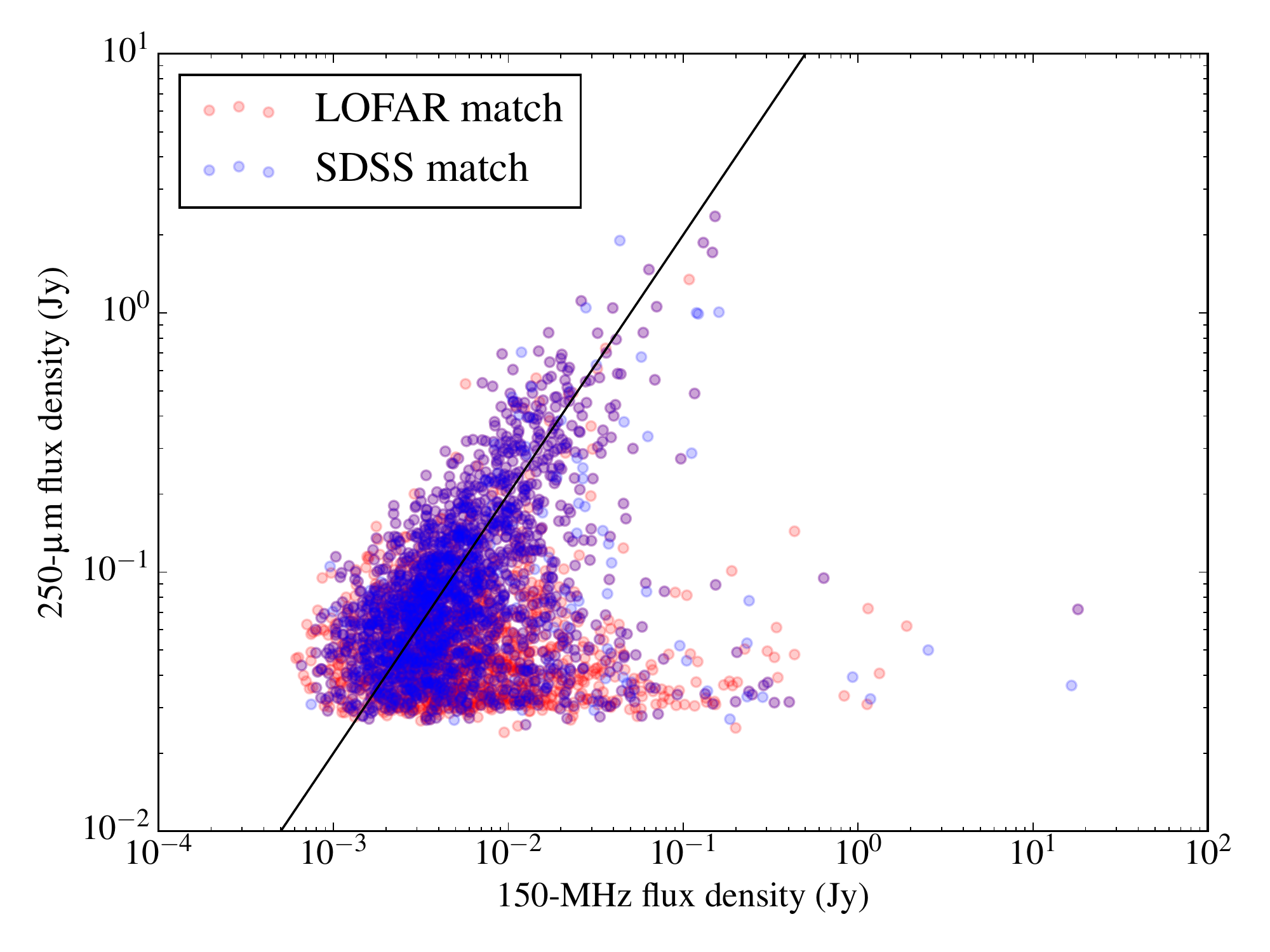}
  \includegraphics[width=0.48\linewidth]{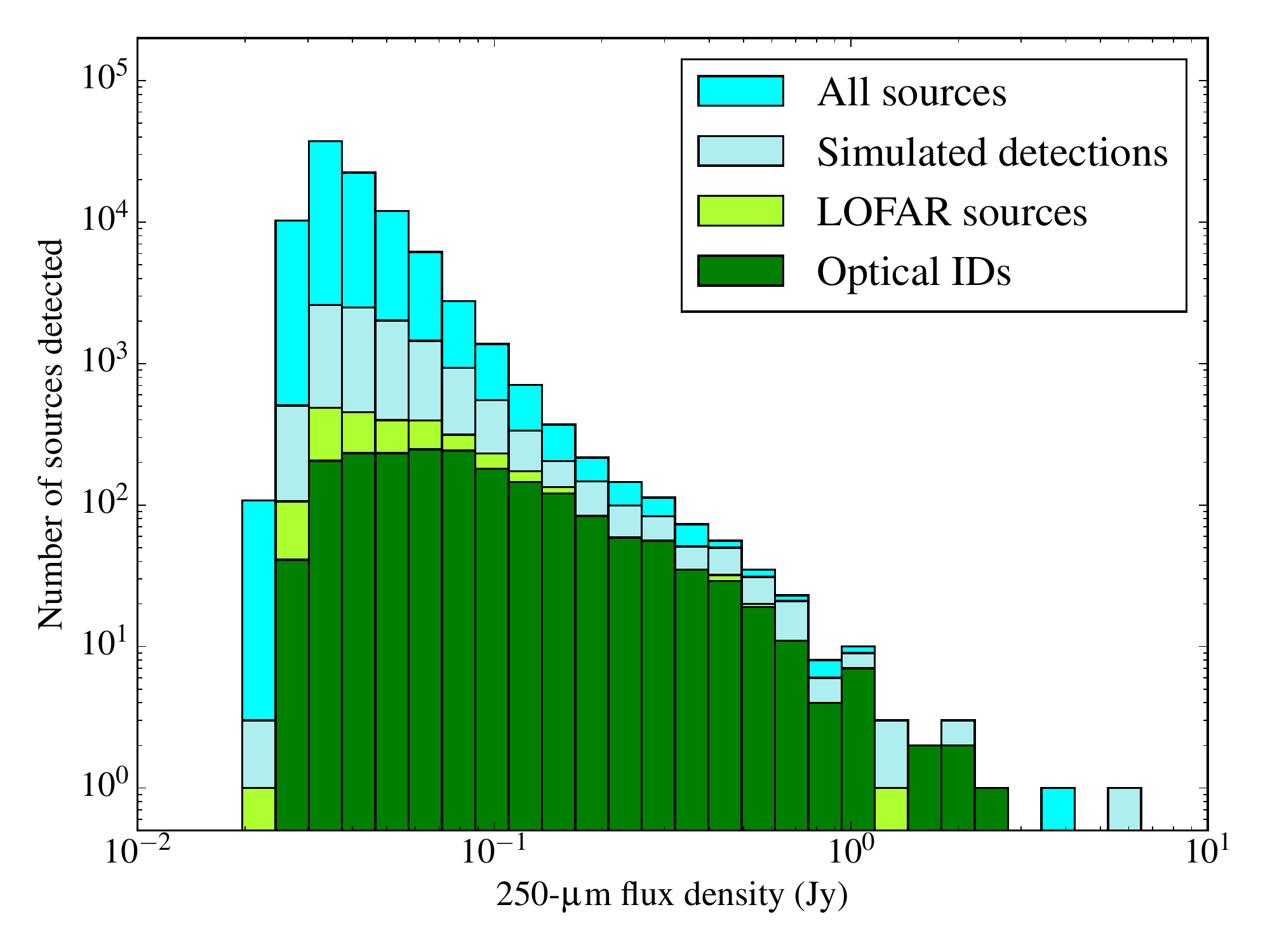}
\caption{Results of cross-matching {\it Herschel} and {\it LOFAR}
  sources. Left: LOFAR/H-ATLAS flux density plot showing positional
  crossmatches to the LOFAR positions (red) and positions of optical
  counterparts (blue). Right: detection histogram, showing the
  distribution of 250-$\mu$m flux densities in the whole area-matched
  sample, the simulated number of detections if $S_{\rm
    150\ MHz}$ were $0.05\, S_{\rm 250\ \mu m}$ for all sources, the distribution of flux densities of sources matched to
  LOFAR positions and the distribution for sources
  matched to SDSS sources. Note the logarithmic axes: at
  $S_{250} < 0.1$~Jy fewer than ten per cent of the {\it Herschel}
  sources are detected by LOFAR.}
\label{fig:herschel}
\end{figure*}

The currently available H-ATLAS catalogue of the NGP field contains
539,757 sources detected at approximately $2\sigma$ significance, of
which 443,500 overlap with the LOFAR images. For the purposes of flux
comparisons we restrict ourselves to sources with 250-$\mu$m
signal-to-noise (taking account of confusion noise) $> 4.0$, of which
there are 94,008 in the LOFAR field; this is a similar significance
level to the cut that will be applied in the forthcoming NGP data
release, and implies a typical 250-$\mu$m flux density limit of around
30 mJy. Clearly only a small fraction of these {\it Herschel} sources
are detected with LOFAR. We cross-matched on both LOFAR positions and
the positions of optical identifications, using the same
maximum-likelihood crossmatch as described above for radio catalogue
matches, with a maximum permitted offset of 8 arcsec. To do this we
take the error on {\it Herschel} positions to go as $1/SNR$,
normalizing to a positional error of 2.4 arcsec for a SNR of 4.5 based
on the results of Bourne \etal\ (in prep.) on the optical
crossmatching to the Phase 1 H-ATLAS data release. We find 2,994
matches to LOFAR positions and 1,957 matches to optical positions ---
the latter being more reliable as the optical positions are better
determined, but representing a smaller number of LOFAR objects as not
all have optical IDs. A flux-flux plot (Fig. \ref{fig:herschel}) shows
the expected two branches, one where there is a good correlation
between the radio and {\it Herschel} flux densities, and one where
there is none, representing respectively star-forming galaxies and
radio-loud AGN (some, but not all, of which will be detected in the
H-ATLAS images due to their star-formation activity). The flux-flux
relationship for the {\it detected} star-forming objects appears
approximately linear and could be represented by $S_{\rm 250\ \mu m}
\approx 20 S_{\rm 150\ MHz}$, as shown on Fig.\ \ref{fig:herschel};
such a relationship is consistent with the $z=0$ radio/far-infrared
(FIR) correlation observed at 1.4 GHz for sources detected in both
bands \citep{Jarvis+10,Smith+14} where the parameter $q_{250} =
\log_{10}(S_{\rm250\ \mu m}/S_{\rm 1.4\ GHz}) \approx 2.0$, assuming a
spectral index of 0.7 for these objects.

The LOFAR detection fraction (Fig.\ \ref{fig:herschel}) is low for all
{\it Herschel} flux densities after the brightest ones, but certainly
lower for the fainter objects, as would be expected given the
flux-flux relationship and the fact that the sensitivity of the LOFAR
images is not constant across the sky. It is interesting to ask
whether such an explanation {\it quantitatively} predicts the
detection fraction, which we can do if we assume that the flux-flux
relationship estimated above holds good for all H-ATLAS sources. We
can then use the LOFAR completeness curve to estimate which of the
H-ATLAS sources should have been detected in the LOFAR band. In fact
(Fig.\ \ref{fig:herschel}) we would expect to detect many more sources
(the simulations show this number to be around 12,000) than we
actually do if $S_{\rm 150\ MHz}$ were equal to $0.05\, S_{\rm
  250\ \mu m}$ for all {\it Herschel} sources. While the flux-flux
correlation we see in the data must be correct for the brightest
sources (we would be able to detect sources with, for example,
250-$\mu$m flux densities at the Jy level and mJy-level LOFAR flux
densities, but none exist) the true flux-flux relationship for the
bulk of {\it Herschel} sources needs to be at least a factor 2 below
the naive estimate derived from the correlation seen for the brightest
sources in order to come close to reproducing the actual detection
statistics. This is again consistent with the results of
\cite{Smith+14}, who showed that stacking radio luminosities including
sources not detected in the radio gave rise to $q_{250}$ values $\sim
2.5$. The implications here are important: even without analysing the
luminosity distribution, we can see that radio/FIR relations derived
from samples flux-limited in both radio and FIR are likely to be
strongly biased unless non-detections are taken into account, with
implications for the radio emission expected to be seen from
star-forming objects in the distant Universe. Here we do not speculate
whether this bias arising from the combined radio-FIR selection is due
to true radio deficiency in some star-forming galaxies or to other
effects such as the differing dust temperatures of objects selected at
250 $\mu$m \citep{Smith+14}. Later papers (G\"urkan et al in prep.;
Read et al in prep.) will discuss the relationship between radio
emission and star formation in more detail.

\subsection{AGN and star formation in the optically identified sample}
\label{sec:sfg}

\begin{figure}
  \includegraphics[width=\linewidth]{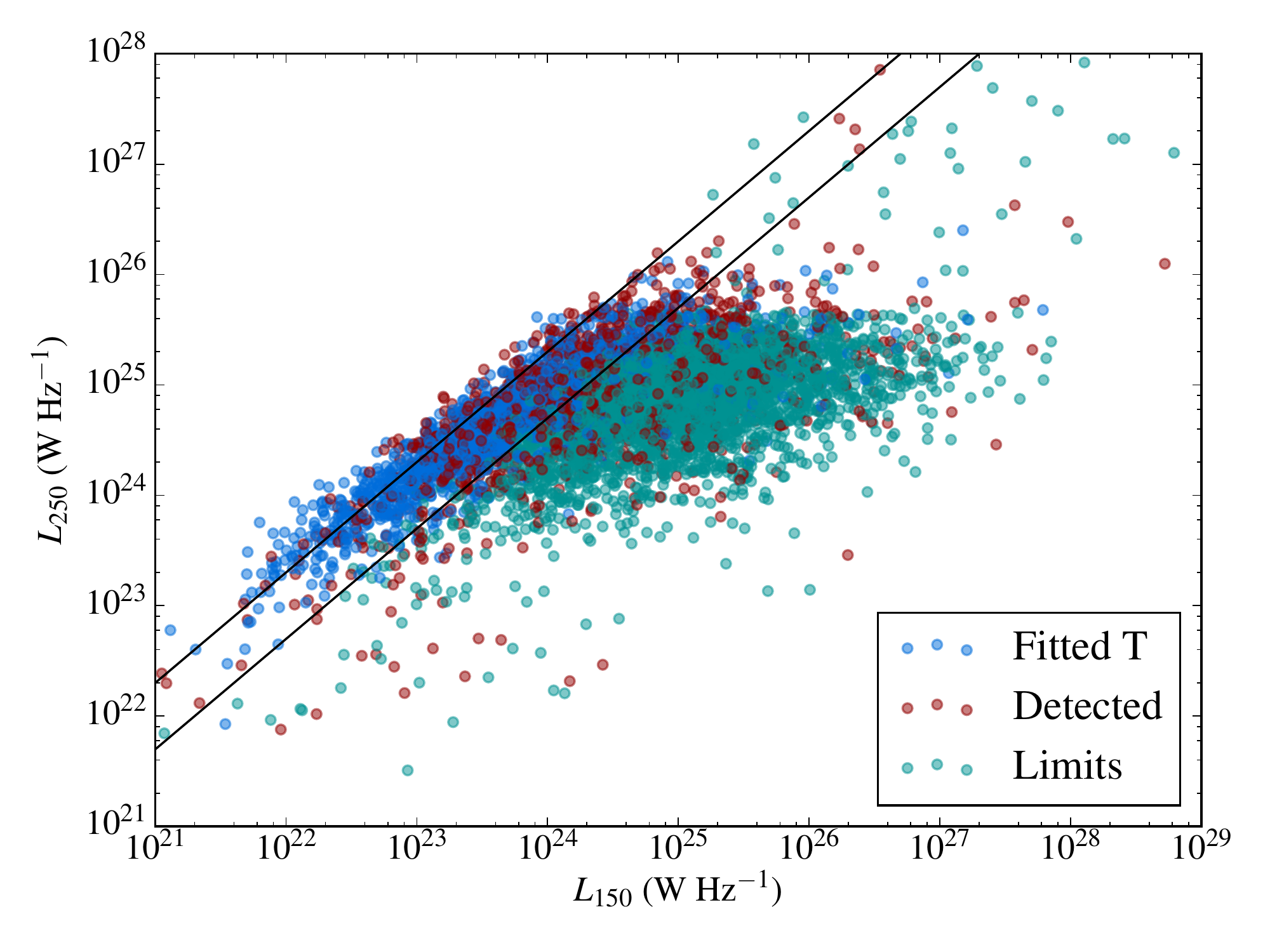}
  \caption{250-$\mu$m far-infrared luminosity, estimated as described
    in the text, as a function of LOFAR radio luminosity. Objects are
    colour coded depending on whether they have temperature
    measurements, are detected at the $2\sigma$ level at 250 $\mu$m but without a
    valid temperature measurement, or are $2\sigma$ upper limits at
    250 $\mu$m. The
    division into detected and non-detected sources at this level is
    for plotting purposes only and plays no part in the analysis.
    Solid lines show $q'_{250} = 20$ and $q'_{250} = 5$, where
    $q'_{250}$ is as defined in the text.}
  \label{fig:llirplt}
  \end{figure}

We made use of the {\it Herschel} data to separate AGN and star
formation in the optically identified sample. To do this we measured
{\it Herschel} flux densities from all five bands directly from the
H-ATLAS maps at the positions of all optical identifications with
redshifts in the manner described by \cite{Hardcastle+13}. We then
fitted modified black-body models with $\beta=1.8$ [the best-ftting
  value derived by \cite{Hardcastle+13} and \cite{Smith+13}] to all
objects with a $2\sigma$ detection in more than one {\it Herschel}
band, accepting fits with good $\chi^2$ and well-constrained
temperature, in the manner described by \cite{Hardcastle+13}. This
process gives us 1,434 dust temperatures and luminosities, with a mean
dust temperature of 24.5 K. For the remaining objects, we estimate the
250-$\mu$m IR luminosity, $L_{250}$, from the 250-$\mu$m flux density
alone, K-correcting using $\beta = 1.8$ and $T = 25$ K; we calculate a
luminosity in this way for all objects, including non-detections. The
temperature and $\beta$ parameters are only used here to provide a
K-correction at 250 $\mu$m, rather than to calculate an integrated
luminosity, and so the effects on the data should be very limited at
the low redshifts of the majority of objects in our sample. The
resulting radio-FIR luminosity plot is shown in
Fig.\ \ref{fig:llirplt}. A clear sequence of the radio-FIR correlation
can be seen, driven mostly by detected objects, as expected given the
results of the previous subsection; the correlation may be slightly
non-linear but at low luminosities/redshifts is broadly consistent
with a constant ratio of about a factor 20 between the two
luminosities. [It would not be surprising to see some non-linearity
  given the dependence of the radio-FIR correlation on dust
  temperature discussed by \cite{Smith+14}: once again, we defer
  detailed discussion of the radio/FIR correlation to G\"urkan et
  al.\ (in prep.).] Radio-loud AGN lie to the right of this
correlation, i.e. they have an excess in radio emission for a given
FIR luminosity. The scattering of points at high luminosities comes
from the high-$z$ quasar population, where the K-corrections almost
certainly break down to a large extent and where there may be some
contamination of the FIR from synchrotron emission.

\begin{figure*}
  \includegraphics[width=0.32\linewidth]{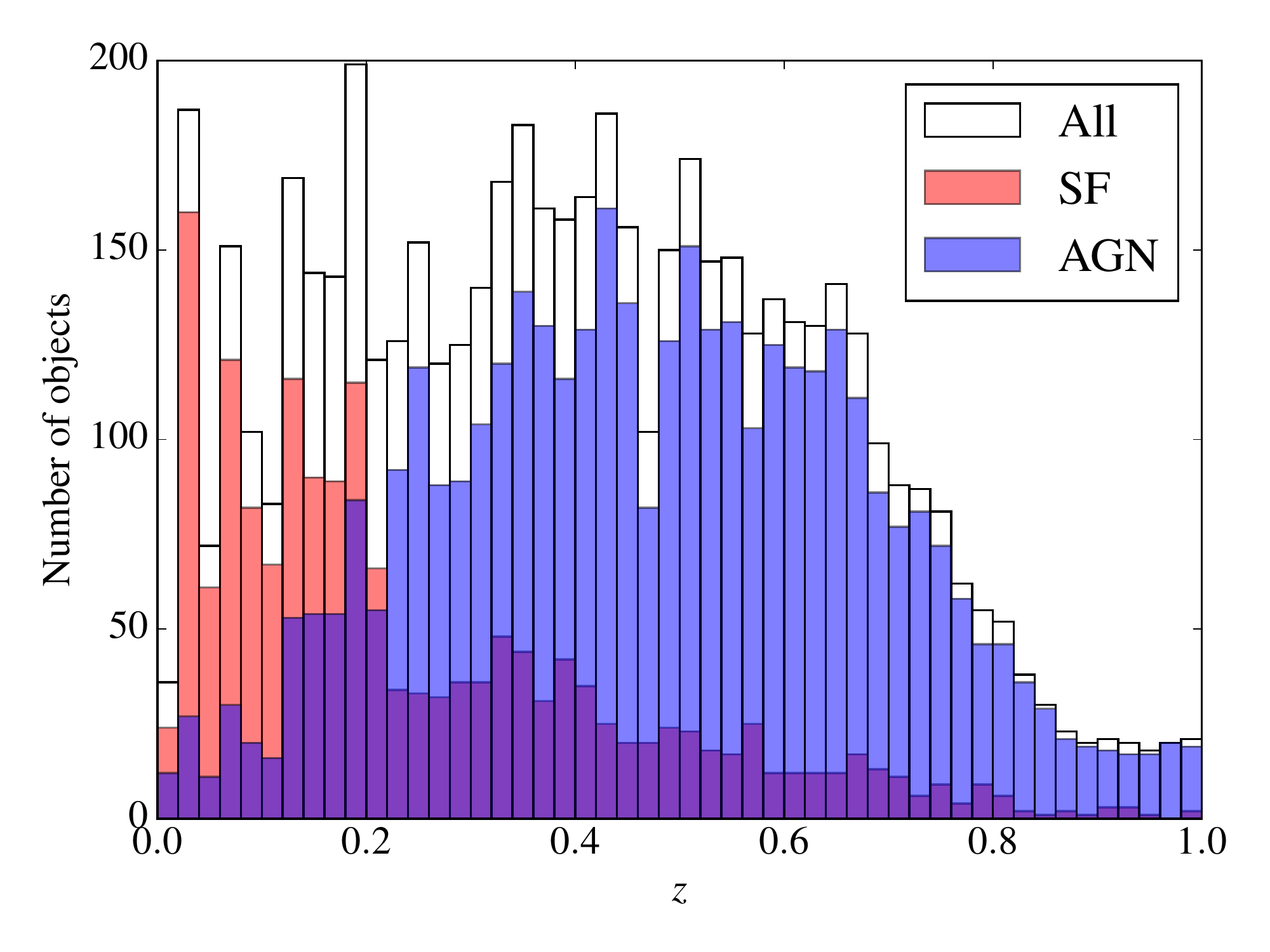}
  \includegraphics[width=0.32\linewidth]{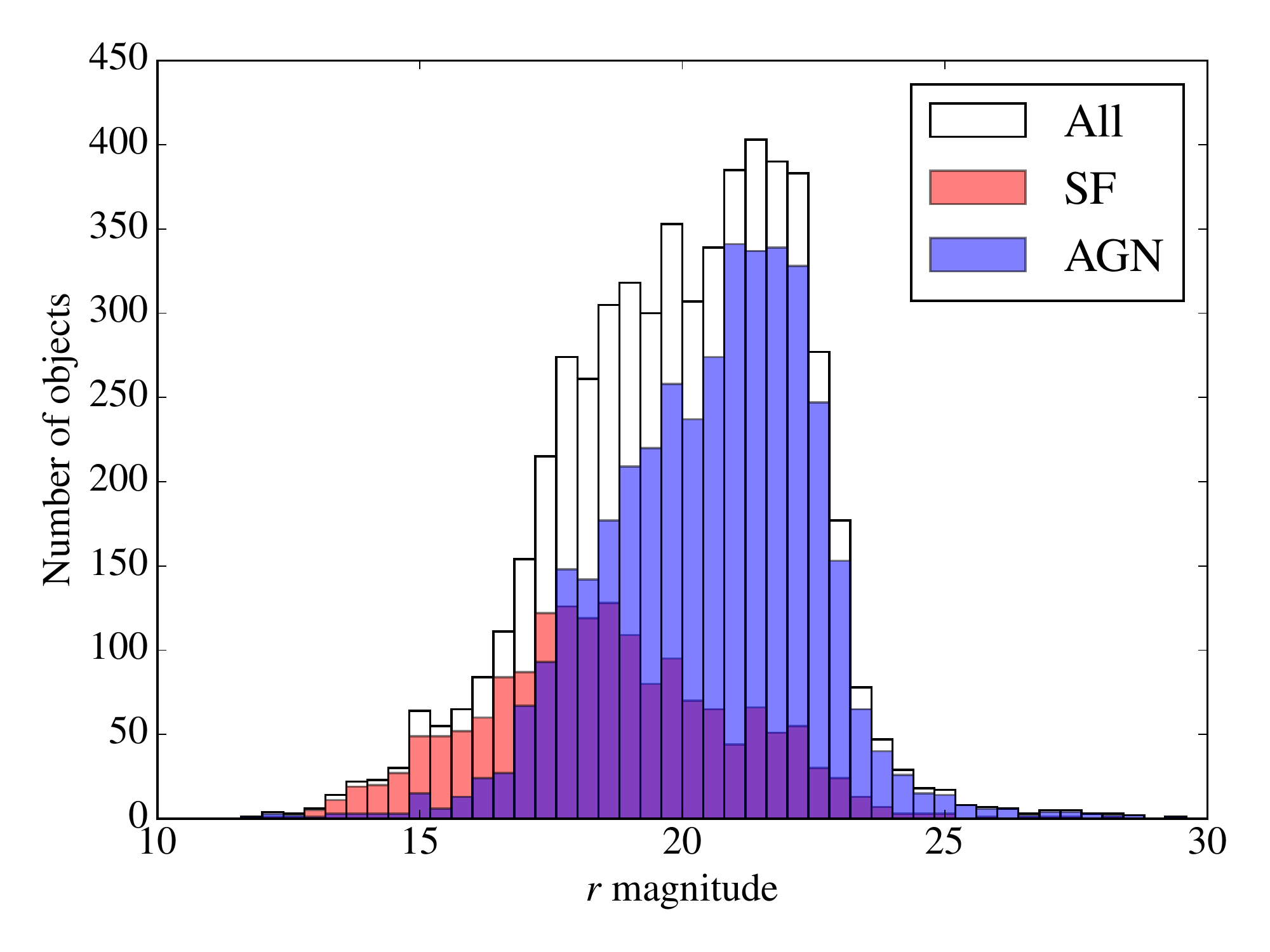}
  \includegraphics[width=0.32\linewidth]{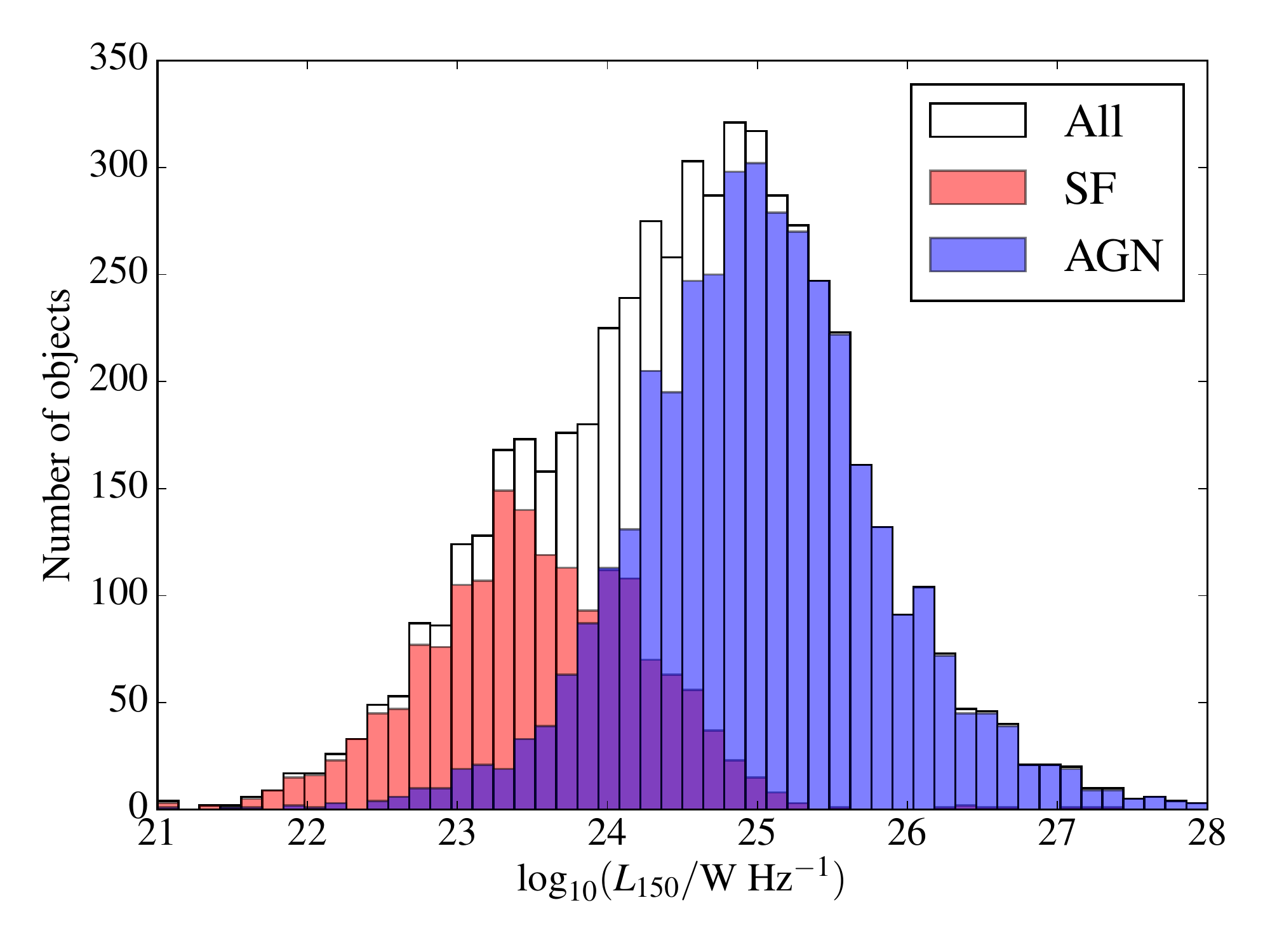}
  \caption{Distributions of key quantities for the whole sample and
    after AGN/star-formation separation as described in the text. In
    each histogram white shows the distribution of the parent sample,
    red star-forming objects, and blue AGN, with the overlapping
    regions of the two coloured histograms appearing in purple.}
  \label{fig:twopop}
\end{figure*}

To make a quantitative separation between the two classes of object we
define the quantity $q'_{250} = L_{250} / L_{150}$ -- we take the
ratio here rather than its log, as is more conventional, to allow for
the negative values of $L_{250}$ which may be assigned to {\it Herschel}
non-detections. We use the value of $L_{250}$ derived from temperature
fitting where available and from the 250-$\mu$m flux density otherwise. Then
we take a source to be an AGN if $q'_{250}<5$, and a star-forming
object otherwise (the division being indicated by a line on
Fig.\ \ref{fig:llirplt}). By this classification, 3,900 of the objects
with redshifts are AGN and the remaining 1,667 are star-forming
galaxies (SFGs). Consistent with expectation, these two populations
have very different distributions in redshift, galaxy magnitude and
150-MHz luminosity (Fig.\ \ref{fig:twopop}). The dividing line used
here is, of course, arbitrary, though it is chosen so as to isolate
the radio/FIR relation at low luminosities. We do not expect a clear
separation between the two classes in $q_{250}$ since radio-loud AGN
may occur in strongly star-forming galaxies. However, we checked the
classification by testing what fraction of sources in the two classes
are morphologically complex, using as a proxy for this multi-component
sources with a maximum component separation of $>20$ arcsec (to avoid
sources that are only moderately resolved by LOFAR). We find that of
the 275 such sources, all but 4 are in the AGN class, and of the four
extended objects classed as SFGs, 3 are genuinely extended very nearby
galaxies; only one is a clear double which should be classified as an
AGN, and that turns out to be one of the quasars that contaminate the
high-luminosity end of Fig.\ \ref{fig:llirplt}, 14 of which have
$q'_{250}$ above the SF threshold. These objects are easily excluded
from our SF catalogue and, apart from them, we do not appear to be
including in the SF class any significant number of double AGN,
suggesting, at least, that the SFG class is not strongly contaminated
by AGN. The fraction of morphologically complex sources increases
immediately below $q'_{250} = 5$, consistent with the idea that this
is a useful dividing line.

The source counts of objects classed as SFGs
(Fig.\ \ref{fig:sourcecounts}) show good agreement with the SKADS
model counts of \cite{Wilman+08} in both normalization at the lowest
flux densities and slope (i.e. flat when Euclidean-normalized): the
only difference is that we lack the sky area to find extremely bright
SFGs, and that we find slightly higher numbers of SFGs at flux
densities of 2--4 mJy. This suggests that we are correctly classifying
the vast majority of both SFGs and AGN. Misclassification of SFGs
would lead to inconsistencies in normalization; contamination of the
SFGs with AGN would lead to inconsistencies in slope. Residual
differences may be due to cosmic variance -- our sky area is
considerably smaller than that simulated by \cite{Wilman+08}. Our
results contrast with those of \cite{Simpson+12} and
\cite{Lindsay+14}, who both found a deficit of faint objects at low
$z$ at 1.4~GHz compared to the SKADS models; it is possible that this
is evidence that LOFAR's short baselines allow it to pick up a
population of low-z SF sources resolved out by high-resolution VLA
surveys. We note (Fig.\ \ref{fig:twopop}) that we continue to find
objects classed as SF up to the highest redshifts in our sample, and
up to radio luminosities of $10^{25}$ W Hz$^{-1}$; these objects must,
if correctly classified, be strongly star-forming galaxies with SFR of
hundreds of solar masses per year.

We conclude that {\it Herschel} data, where available, offer a
reliable and simple method of carrying out AGN/star-formation
separation in LOFAR data at Tier 1 depth.

\subsection{Luminosity functions}
\label{sec:lf}

We construct the low-redshift 150-MHz luminosity function from sources
with $r<19$ mag, excluding quasars. Below this limit, 1809 of our 1917
candidate identifications (95 per cent) have redshifts (1190
spectroscopic) and so we are able to construct a luminosity function
without much normalization uncertainty. 1017 of the 1809 are classed
as SFGs by the $q'_{250}$ criterion, the rest are AGN. We drop at this
point 2 AGN with photometric redshifts that are clear outliers on the
$r$-$z$ plot (Fig.\ \ref{fig:magdist}) leaving a sample of 1017 SFGs
and 790 AGN with a maximum redshift just over $0.4$. This large
redshift range means that we may be somewhat affected by cosmological
evolution; the median SFG redshift is 0.12 and for AGN it is 0.24. We
return to this point below. We expect that there are very few
unidentified radio sources which should in fact be identified with
$r<19$ galaxies, setting aside small gaps in the DR12 photometric
catalogues around bright stars and the like, so that a luminosity
function with these constraints should be representative of the true
source population.

In order to calculate the luminosity function we must deal with the
effects of K-correction in the optical. We first of all calculated
absolute $g$ and $r$ magnitudes for our targets using the methods of
\cite{Chilingarian+10}, correcting for an average Galactic reddening
using {\sc astroquery} to query the IRSA Dust Extinction
Service\footnote{\url{http://irsa.ipac.caltech.edu/applications/DUST/index.html}}
and retrieve dust extinction calculated according to
\cite{Schlafly+Finkbeiner11}. A colour-magnitude diagram constructed
for the $r<19$ objects shows a good separation into red sequence and
blue cloud largely dominated by AGN and star-forming objects
respectively, as expected.

\begin{figure*}
  \includegraphics[width=0.48\linewidth]{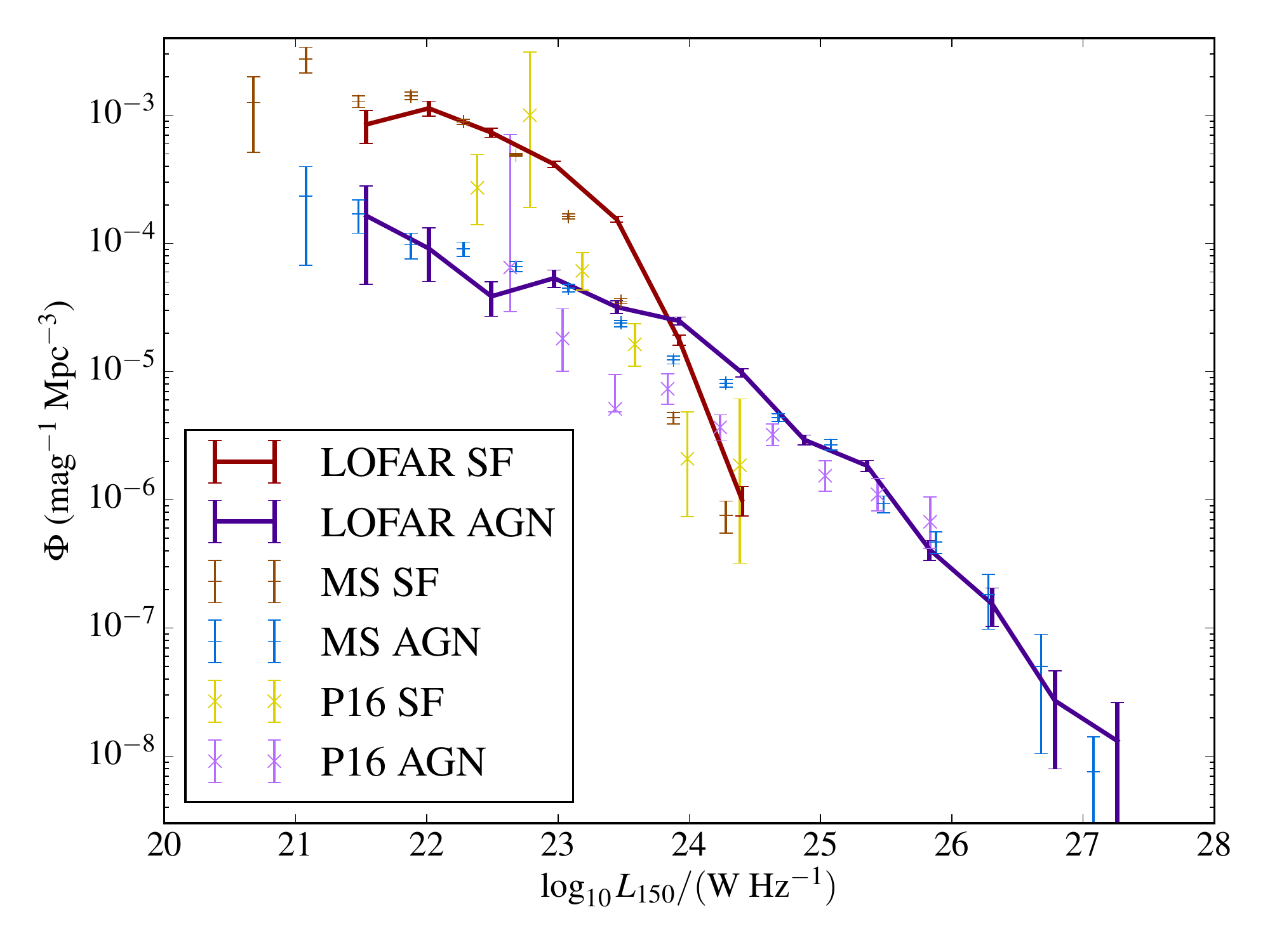}
  \includegraphics[width=0.48\linewidth]{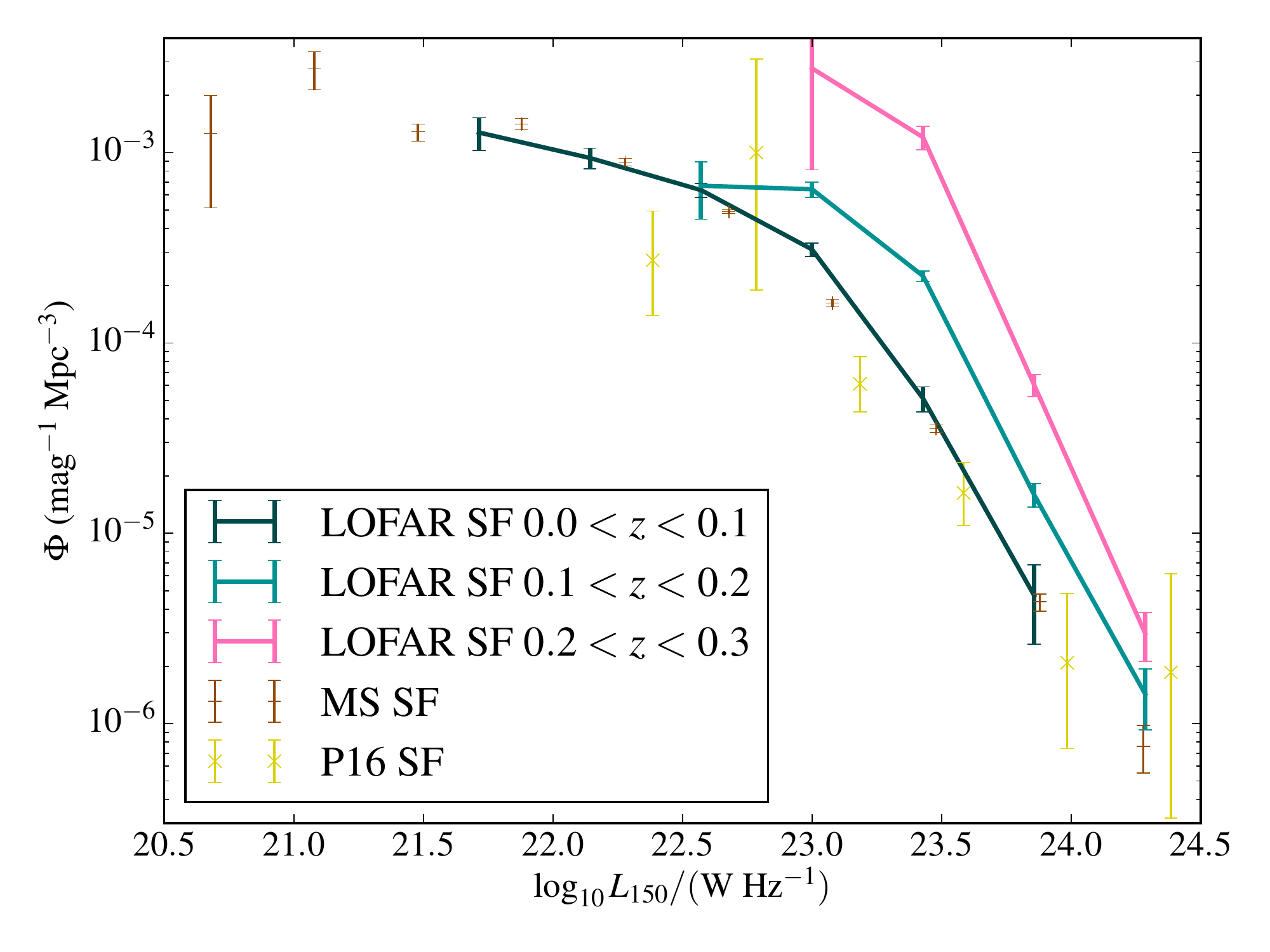}
  \caption{Left: the 150-MHz luminosity function in the H-ATLAS field
    dividing the sources into SFG and AGN as described in the text.
    Solid lines show the LOFAR-derived luminosity function: light
    points with error bars
    show the 1.4-GHz luminosity function from
    \protect\cite{Mauch+Sadler07} and the 325-MHz luminosity function
    of \protect\cite{Prescott+16}, scaled assuming a constant
    $\alpha =0.7$. Error bars are Poissonian from number counts in
    bins only. Right: the same plot, but for the SFG only, dividing
    into three redshift bins.}
  \label{fig:lf}
  \end{figure*}

We then computed the luminosity function in the standard way, i.e. by
binning $1/V_{\rm max}$, where $V_{\rm max}$ is the maximum volume out
to which a source can be seen given the radio and optical limits. We
calculate $V_{\rm max}$ separately for optical and radio and adopt the
smaller of the two. For the radio, $V_{\rm max}$ is calculated as
$\int d_{\rm max} \mathrm{d}A$, where we use the completeness function
described in Section \ref{sec:completeness} to compute the area over
which sources can be seen out to a given depth. For the optical, the
survey is assumed to be uniform, but we invert the approximations of
\cite{Chilingarian+10} to derive K-corrections as a function of
redshift and intrinsic ($M_g - M_r$) colour.

The results for the whole sample are shown in Fig.\ \ref{fig:lf} (left
panel). For AGN, we see overall good agreement with a scaled ($\alpha
= 0.7$) version of the luminosity functions of \cite{Mauch+Sadler07}
at 1.4 GHz or \cite{Prescott+16} at 325 MHz, implying little variation
in the spectral index as a function of radio luminosity. However, the
luminosity function for SFG clearly has an excess with respect to the
literature catalogues at higher luminosities. We attribute this to
redshift evolution of the SFG population. This can clearly be seen in
the right-hand panel of Fig.\ \ref{fig:lf} where we plot the SFG
luminosity function only, broken down into redshift bins. A strong
positive luminosity function evolution with redshift is evident, and
the lowest-redshift luminosity function is now in excellent agreement
with that of \cite{Mauch+Sadler07}. By contrast, we have verified that
the AGN luminosity function shows no significant variation out to
$z=0.4$, the limit of our magnitude-limited sample, presumably because
the AGN in our sample are largely low-excitation radio galaxies which
are expected to show only weak cosmological evolution
\citep{Best+14,Pracy+16}. A future paper (Williams \etal\ 2016b) will
discuss the luminosity function evolution for radio-loud AGN, both
low-excitation and high-excitation using the deeper optical data in
the Bo\"otes field.

The strong radio luminosity function evolution we see for SFGs is
striking. We naturally expect some evolution given the known overall
evolution of the star-formation density of the universe
\citep[e.g.,][]{Madau+96,Hopkins+Beacom06}. In wide-area surveys hints
of positive evolution have been seen for some time
\citep{Machalski+Godlowski00,Condon+02}. However, most work in this
area has focussed on deep fields, and this has shown that the
luminosity function \citep[e.g.][]{Haarsma+00,Smolcic+09,McAlpine+13},
the specific star-formation rate estimated from the radio
\citep[e.g.][]{Karim+11,Zwart+14} and the total radio-estimated
star-formation rate density \citep[e.g][]{Seymour+08} all evolve
positively in the redshift range $z=0$ -- 1. What is unusual about our
sample, other than the fact that it is calculated at 150 MHz, is that
it has the area to see this evolution directly at low redshift,
coupled with the ability of the H-ATLAS data to allow AGN/SF
separation over such a wide area. Although the error bars are large,
Fig.\ \ref{fig:lf} implies that pure luminosity evolution has the form
$\sim (1+z)^5$, which is steeper than the $\sim (1+z)^{2.5}$ found in
most earlier work on the radio luminosity, suggesting either some
difference in our selection or a real change in the redshift
dependence at low radio luminosity and $z$. The corresponding positive
evolution at low $z$ in the far-IR is relatively well known
\citep{Dye+10,Magnelli+13,Gruppioni+13} and seems to imply a similarly
strong evolution with redshift \citep{Dye+10}, but, unlike the far-IR
where dust mass evolution may also be implicated \citep{Dunne+11}, the
radio data -- if contaminating AGN can be removed -- provide an
unambiguous tracer of star formation evolution comparable to the
ultraviolet or H$\alpha$. More optical identifications and
spectroscopic redshifts for objects in the NGP field, and LOFAR
observations of the equatorial H-ATLAS fields, will enable us to
investigate this evolution of the low-frequency luminosity function to
higher redshift in future, and to compare to the results at 1.4 GHz
and to the evolution of other star formation tracers \citep[see,
  e.g.,][]{Mancuso+15}.
\vspace{10pt}
\subsection{The power/linear-size diagram and the incidence of giant sources}
\label{sec:pdd}

\begin{figure}
  \includegraphics[width=1.0\linewidth]{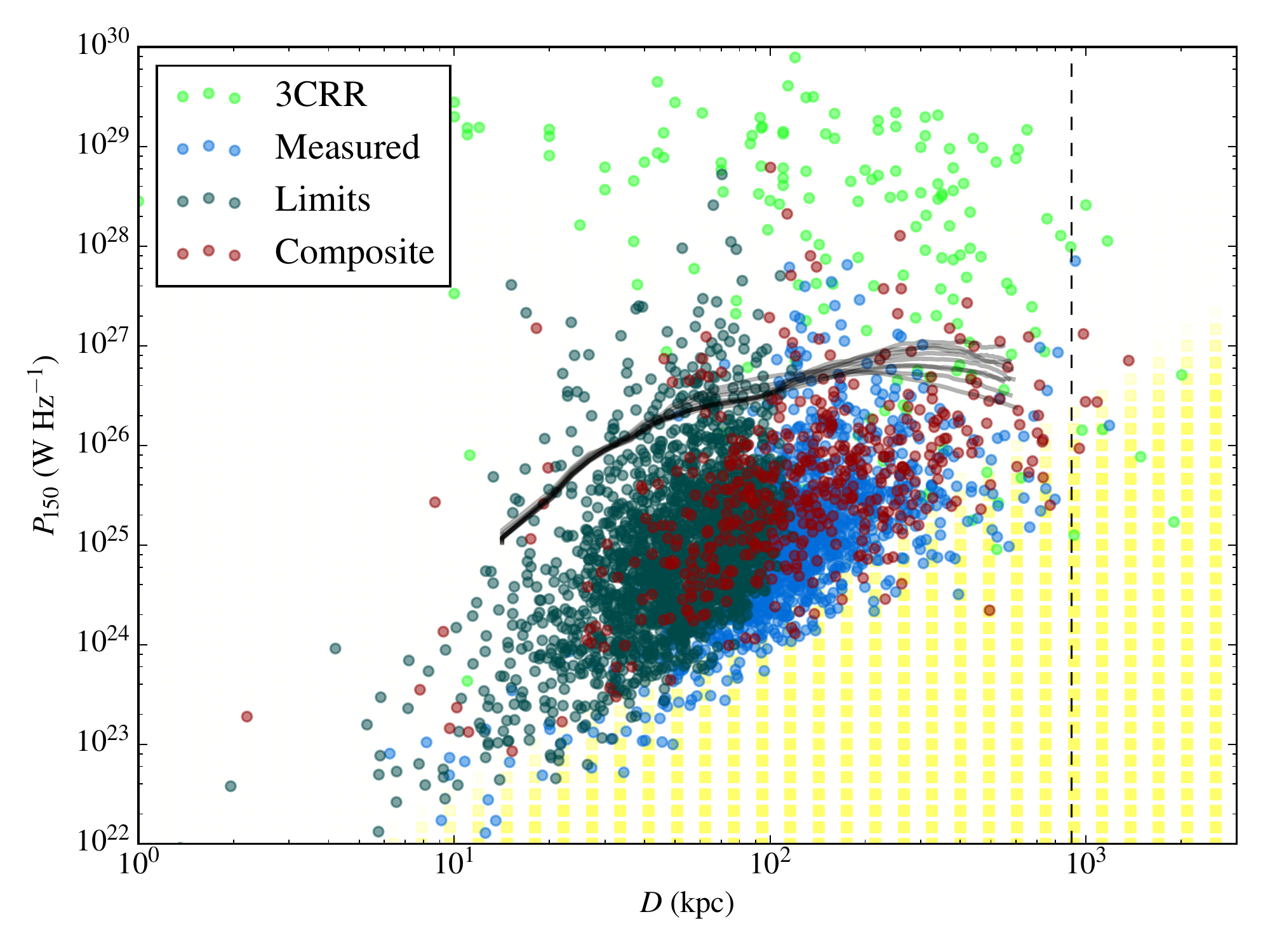}
  \caption{The power/linear-size diagram for AGN in our survey and for
    3CRR objects. For the LOFAR AGN, we plot separately sources with
    angular sizes smaller than 15 arcsec as measured by {\sc pybdsm},
    where the physical sizes should probably be considered upper
    limits; sources with larger sizes, probably at least somewhat
    meaningfully measured by {\sc pybdsm}: and composite sources, whose
    largest component separation is used as a proxy for their size.
    3CRR sizes are all measured from high-resolution radio maps. Boxes
    indicate the region where some or all LOFAR sources in the range
    $z=0$ to $z=0.8$ with uniform surface brightness would drop below
    the detection threshold for our images. The vertical dashed line
    indicates our giant-radio-galaxy selection criterion (see the text
    for details). The grey curves represent tracks in the diagram for
    a source with a jet power of $10^{38}$ W, in various environments,
    derived from the modelling of \protect\cite{Hardcastle+Krause14}.}
  \label{fig:pdd}
  \end{figure}

The radio power-linear size plot or $P$-$D$ diagram for radio-loud
AGN, introduced by \cite{Baldwin82}, is an important diagnostic of
radio galaxy evolution. A new-born radio source will start at $P=0$,
$D=0$ and (barring strong interactions with the external medium) is
expected to have ${\rm d}D/{\rm d}t > 0$ throughout its active
lifetime, as the expansion of the source is driven by the ram pressure
at the head of the jets: thus linear size is an important proxy of
age, though the relationship between the two is determined by the
source environment. The radio power is a function of the energy
density in electrons and magnetic field in the lobes and of their
volume, and so has a more complex relationship with source age,
particularly when the effects of radiative losses are taken into
account. Theoretical or numerical models of radio galaxy evolution
(e.g., \citealt{Kaiser+97}; \citealt{Blundell+99};
\citealt{Manolakou+Kirk02}; \citealt{Hardcastle+Krause13}; \citealt{Turner+Shabala15})
predict tracks in the $P$-$D$ diagram for individual sources,
depending on source environment and jet power; survey observations
provide an instantaneous snapshot of the positions of many sources on
their $P$-$D$ tracks. Observations of large samples can constrain
models directly if they contain sources that are extreme with respect
to the predicted tracks, e.g. very large or very powerful objects;
more importantly, we may hope in future to use observed
(multifrequency) radio power and size in large samples to infer
properties such as jet power, age and environment on a per-source
basis from the theoretical predictions. Inferring these properties for
the large samples of radio-loud AGN expected to be generated by
next-generation radio surveys, including those with LOFAR, will provide crucial
input into our understanding of the `feedback' processes believed to
control the evolution of the most massive galaxies, as discussed in
Section \ref{sec:intro}.

To construct a $P$-$D$ diagram clearly estimates of both $P$ and $D$
are necessary, i.e. in observational terms we need measurements of
radio flux density, largest angular size and redshift. We have
estimates of the angular sizes of our sources from the cataloguing
process, but these need to be treated with caution for several
reasons. As discussed in Section \ref{sec:completeness}, the
deconvolved major axes estimated by {\sc pybdsm} for objects fitted
with a single Gaussian tend to be overestimates, as small residual
phase errors or offsets between the different spectral windows will
make a source that is really unresolved to LOFAR appear marginally
resolved in the images. For this reason, we consider all deconvolved
sizes $<15$ arcsec to be unreliable, where the threshold is chosen
based on the distribution of apparent source sizes and on visual
inspection of the images. For sources where {\sc pybdsm} associates
more than one component, the size estimates are probably slightly more
reliable, but these are a minority. For sources associated by us, we
record the largest angular separation between the {\sc pybdsm}
positions of any pair of components, but this is only a crude estimate
of the true angular size, being most reliable for edge-brightened
FRII-type sources. All of these automatically generated sizes would
benefit from verification by human inspection. Nevertheless they
provide an interesting starting point for consideration of the sample
power/linear-size plot. This is shown for the sources classed as AGN
in Fig.\ \ref{fig:pdd}. We overlay on this plot the equivalent values
for the 3CRR sample\footnote{Data from
  \url{http://3crr.extragalactic.info/}.} \citep{Laing+83}, which,
with its flux density limit of 10.9 Jy at 178 MHz, represents the most
luminous radio AGN in the Universe at any particular redshift; it can
be seen that there is significant overlap between the two,
unsurprising since a number of 3CRR sources are present in our survey,
but that, also as expected, the LOFAR survey picks up many more
low-luminosity AGN. We also overlay, for illustrative purposes only,
the theoretical tracks for a source with a jet power of $10^{38}$ W in
various different environments for sizes between 10 and 600 kpc,
taking account of radiative losses, derived from the MHD simulations
of \cite{Hardcastle+Krause14}, which, if accurate, indicate that the
most luminous large sources seen in the LOFAR survey have jet powers
around the value simulated in that work; however, some of the
luminous, compact sources we see in the LOFAR surveys with $D \la 20$
kpc, $P \approx 10^{27}$ W Hz$^{-1}$ may well be young sources with
significantly higher jet power that will eventually evolve, if their
jets remain active, to 3CRR-like luminosities of $10^{28}$ --
$10^{29}$ W Hz$^{-1}$. Thus we see the potential of the LOFAR data to
allow us to construct a true jet kinetic luminosity function over
several orders of magnitude in jet power, something we expect to
return to in future papers.

Also plotted in Fig.\ \ref{fig:pdd} are the expected regions where
sources cannot be detected, given the surface brightness limits at
full resolution and sensitivity, and on the assumption of uniform
source surface brightness, considering the redshift range 0--0.8 in
which most of our sources lie. We see that we are capable of detecting
(and do detect) sources with $\sim 100$-kpc sizes down to almost the
lowest radio luminosities at which we detect AGN, but we expect to be
significantly biased against low-luminosity large sources because of
our surface brightness limitations. At high radio luminosities, we
would expect to be able to detect all but the most extreme giant
radio galaxies, where we adopt the standard definition in which the
projected linear size of a giant is $>1$ Mpc. Giant radio galaxies are
of particular interest because they represent one of the extremes in
$P$-$D$ space: they must be particularly long-lived sources and their
very existence places constraints on models of e.g. the possible
active lifetime of jets. For this reason a number of searches for
giant sources in existing low-frequency surveys have been carried out
\citep[e.g.][]{Cotter+96,Lara+01,Machalski+01,Schoenmakers+01}. Such
searches have generally used large {\it angular} size as a proxy for
giant status, and then followed up optically and/or with spectroscopy
to identify physically large sources, thus favouring low-redshift
giants. For example, \cite{Lara+01} find a sky density of 1 giant per
$\sim 300$ square degrees, with an initial selection criterion of an
angular size exceeding 4 arcmin in NVSS.

Our optically identified sample and the fact that we are capable of
detecting powerful giants allows a direct approach to the problem. As
Fig.\ \ref{fig:pdd} shows, we do detect 7 sources with sizes $\sim 1$
Mpc: in counting these we use a selection at 900 kpc to allow for the
fact that the angular size values from component association are
generally slightly underestimated, since they are the separations
between the centres of the associated components, not their edges. Of
these 7, all but one seem likely to be bona fide powerful giants (see
Appendix \ref{app:tmbg}), implying a density of such $P \ga 3\times
10^{26}$ W Hz$^{-1}$ sources on the sky of at least 1 per 20 square
degrees. These are, of course, only the optically identified giants,
and we would expect to be biased against optical IDs of luminous
sources, which will tend to be at high redshift, as well as of large
sources, which are inherently difficult to identify. Our smallest (in
angular size) giant is 140 arcsec in length, and there are a further
10 composite sources with sizes $>2$~arcmin in the associated
catalogue, many of which may be high-$z$ giants. Although the numbers
are small, these are substantially higher sky densities than were
found by \cite{Lara+01}, and suggest that the Tier 1 surveys will be a
fruitful hunting ground for giant sources.

The surface brightness limitations in these full-resolution
observations suggest that it would be useful to re-image the
facet-calibrated data at low resolution (20--30 arcsec) to allow a
search for low-surface brightness sources: \cite{Saripalli+12} have
found a high detection rate of relatively low-luminosity large sources in a
small sky area with good surface brightness sensitivity. As noted above,
measurements of the numbers of giants as a function of radio
luminosity and redshift provide important constraints on models of
radio source evolution, and we plan to revisit the implications of the
population of large sources in the LOFAR surveys in a future paper.

\subsection{Remnant AGN}
\label{sec:remnant}
One of the key uncertainties in AGN evolution models is what happens
when the jets are switched off. At this point, about half the energy
that has ever been transported up the jets remains in the lobes, at
least for powerful double objects \citep{Hardcastle+Krause13} and so
the question of whether, and where, that energy is transferred to the
external medium is one of great interest. However, the detection of
sources in the post-switch-off phase, so-called remnant or relic AGN,
has been surprisingly difficult. There are some well-known objects
that appear to have no current AGN activity, for example B2 0924+30
\citep{Cordey87} or 3C\,319 \citep{Hardcastle+97}, showing no
flat-spectrum arcsec-scale core (the self-absorbed base of a currently
active jet) and, where data are available, no AGN activity at any
other waveband. But such objects are rare \citep{Giovannini+88},
making up no more than 7 per cent of the low-frequency classical
double (FRII) population selected from 3CRR at $z<1.0$, for example
\citep{Mullin+08}, though this fraction may be environment-dependent
\citep{Murgia+11}. In fact true remnants, where AGN activity has
completely ceased, seem to be somewhat rarer than double-double or
restarting radio galaxies \citep{Schoenmakers+00,Saripalli+12}
despite the fact that double-doubles should be a fairly short-lived
phenomenon as the newly active lobes will merge into the pre-existing
plasma \citep{Konar+13,Konar+Hardcastle13}, implying a very rapid
fading process for remnants (cf.\ \citealt{Kaiser+Cotter02}). To date,
however, statistical information on the remnant population has mainly
come from studies of bright flux-limited samples like 3CRR, and as
radio galaxies are expected to fade significantly as they age due to
the effects of adiabatic expansion and radiative losses, it is clear
that such samples may be biased against remnant sources.

It has long been suggested that remnant or relic AGN\footnote{Relic
  AGN should not be confused with the `radio relics' found in clusters
  of galaxies, whose origin is for the most part not directly in AGN
  activity; we use the term `remnant', as adopted by
  \cite{Brienza+16a}, to avoid this confusion.} would have steep
spectra \citep[e.g.][]{Parma+07,Murgia+11} and so would be detectable
in greater numbers in sensitive surveys of the low-frequency sky.
LOFAR should be extremely sensitive to remnant AGN, which are also
expected to be physically large. \cite{Brienza+16b} set out possible
methods for identifying remnant AGN in LOFAR fields, which include
spectral selection (i.e., looking for steep-spectrum sources),
morphological selection (looking for sources with little or no compact
structure) and what might be termed core selection (looking for
sources with no identifiable radio core). Each of these methods has
its advantages and disadvantages. \cite{Brienza+16b} show that
morphological selection appears to be more efficient in the Lockman
Hole field than spectral selection, for the criteria they use, and one
morphologically selected remnant in another LOFAR field has been
followed up in detail, confirming its remnant nature
\citep{Brienza+16a}. However, it is possible that remnants with
recently switched off AGN would be missed by both the morphological
and spectral selection methods; by contrast, all genuine remnant
sources, whatever their age, would be expected to have no nuclear jet
and so no arcsec-scale, flat-spectrum core. As an initial test for the
efficiency of remnant surveys using this core selection criterion at
Tier 1 depth and resolution, we have selected from the identified
catalogue all sources which are (1) bright (total flux density $>80$
mJy at 150 MHz), ensuring that the sample is flux-complete across the
survey region and that there is a good chance of seeing a core at high
frequency, (2) well resolved ($>40$ arcsec), and (3) classed as AGN on
the radio/FIR relation, for an initial visual search for remnants.

Of 127 such objects (after removing a few objects where it is doubtful
that they are truly extended), we can see no evidence for a currently
active core in the FIRST images in 38, a potential remnant fraction of
30 per cent. Examples of candidate remnant sources, together with some
comparable sources where a FIRST core is seen, are shown in Appendix
\ref{app:remnant}. We do not include in our remnant count any source where
FIRST emission is coincident with the optical ID, even if there is no
clear evidence of a point source; so the true remnant fraction could
be slightly higher than we quote above. Sources without a core in
FIRST are also less likely to be optically identified, hence again
biasing our estimate low. On the other hand, we do not exclude sources
on the basis of showing apparent compact hotspots in the FIRST images.
Even if truly compact, something we cannot really assess on the basis
of the FIRST images, such features may persist for more than a
light-travel time along the lobes after the jet turns off, and so do
not imply that the jet is still active. The fact that the fraction we
measure is higher than for 3CRR sources is consistent with the idea
that remnants might be more detectable in more sensitive surveys,
although it is clear that remnants do not dominate the LOFAR sky at
these flux density levels.

The main limitation on this conclusion is the fact that the FIRST
images are not particularly sensitive to cores. If we define the core
prominence as the ratio of core flux density at 1.4 GHz [the exact
  frequency is unimportant since radio galaxy core spectra are flat up
  to high frequencies: \citep[e.g.][]{Hardcastle+Looney08,Whittam+13}]
to total flux density at 150 MHz, then FIRST's $3\sigma$ upper limit
on core prominence for the faintest objects corresponds to $0.4/80 = 5
\times 10^{-3}$, while we know that the {\it median} core prominence
for 3CRR objects with detected cores is $\sim 3 \times 10^{-4}$
\citep{Mullin+08}. 3CRR objects are selected to be the brightest
low-frequency sources on the sky and would be expected to have
systematically low prominences, so this is an unfair comparison, but
clearly it is possible that even moderately faint radio cores are
escaping detection in our calculation of the remnant fraction above.
We therefore regard the remnant fraction we have derived above as an
upper limit. Even so, such a limit is interesting, as it requires the
typical fading timescale for remnants at 150 MHz to be at most $\sim
30$ per cent of the active time. Sensitive radio followup of remnant
candidates will be necessary to constrain the remnant fraction
further.

\begin{figure}
  \includegraphics[width=1.0\linewidth]{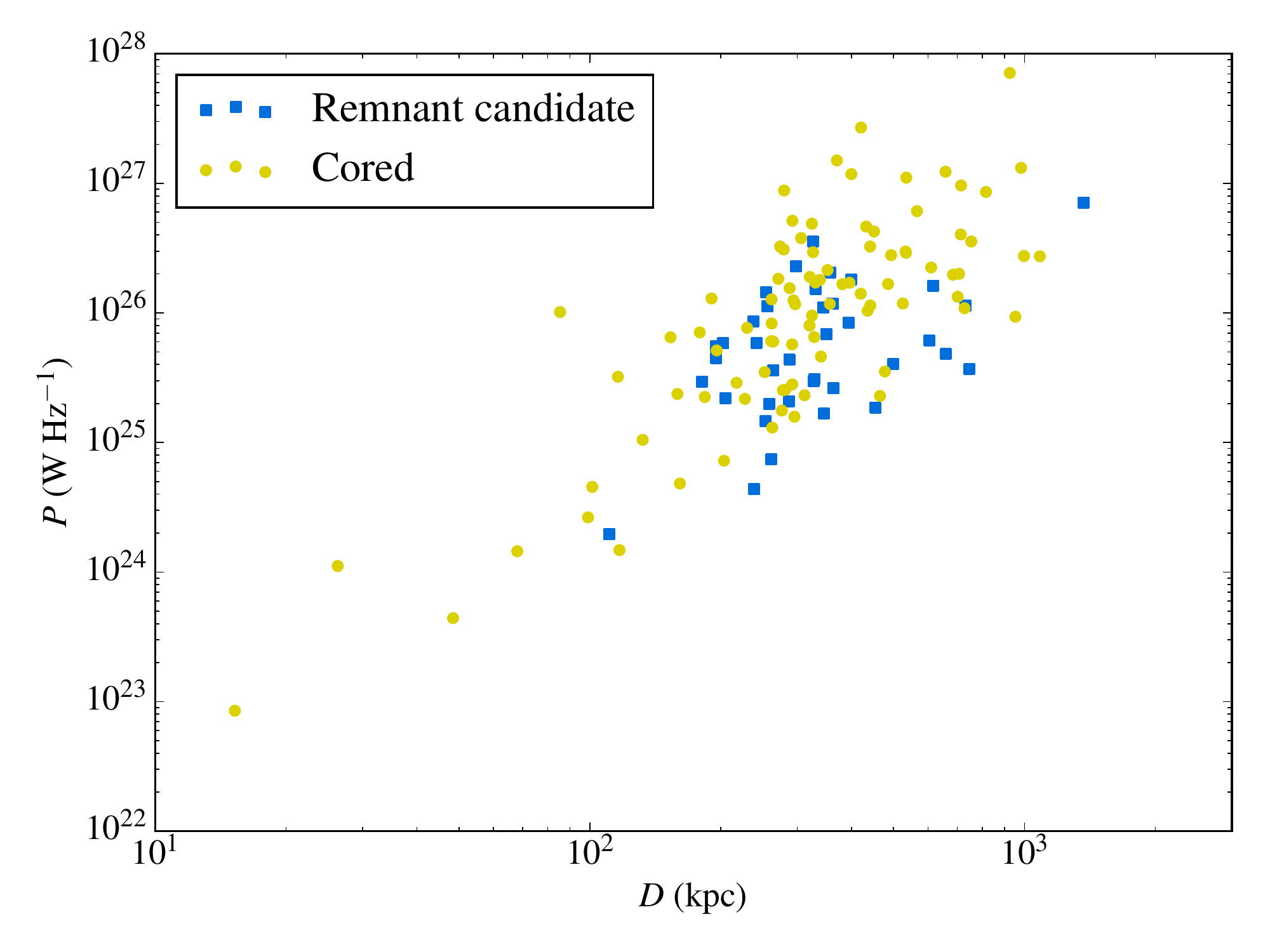}
  \caption{The power/linear size diagram for the bright, resolved
    subsample discussed in the text.}
  \label{fig:pdd-80}
  \end{figure}

A power/linear-size diagram for the 80-mJy, 40-arcsec subsample
(Fig.\ \ref{fig:pdd-80}) shows that the remnant candidates do not
occupy any particularly special position with respect to the cored
sources, but there is a slight tendency for them to have lower radio
luminosity for a given size, in the sense that all remnant candidates
with sizes $>300$ kpc, and many of those below that size, lie at the
very lowest end of the radio luminosity distribution. This would
support the idea that remnants fade rapidly even at low frequencies
once the jets switch off. There is no apparent difference between the
in-band spectral indices of the two samples, but as noted above, these
are very unreliable, and some remnants by our definition would be
expected to have flat radio spectra anyway. We defer an investigation
of the LOFAR/NVSS spectral index for the bright identified sample to a
future paper, as this will require individual measurements from LOFAR
and NVSS maps for all sources.

\section{Summary and conclusions}

We have presented details of the observations, data reduction and
quality assessment for a survey consisting of four LOFAR HBA pointings
in the H-ATLAS NGP area. Although this survey does not have uniform
noise, it is otherwise expected to be reasonably representative in
data quality of the much larger `Tier 1' survey currently being
carried out by the LOFAR Surveys Key Science project. Key points from
the discussion of the data reduction and quality assessment are as
follows:
\begin{itemize}
  \item We have been able to image over 140 square degrees at 150 MHz
    at a resolution of better than 10 arcsec and with rms noise in the
    broad-band 150-MHz maps
    ranging between 100 $\mu$Jy beam$^{-1}$ and $\sim 2$ mJy
    beam$^{-1}$, thus covering almost all of the NGP field as surveyed
    by H-ATLAS. The fact that this can be done in a total of 34 hours'
    observing illustrates the capabilities of LOFAR for wide-area,
    deep surveys.
  \item The LOFAR flux scale remains problematic. The method we have
    developed to calibrate flux densities in the field -- by
    cross-matching a large number of objects to existing low-frequency
    catalogues (Section \ref{sec:fluxcal}) -- works well so long as there
    are enough objects in existing catalogues to make a statistical
    comparison possible, but this is not the case for the whole sky.
  \item Even after correcting as well as possible for the flux scale
    issues, HBA in-band spectral indices are unreliable because of the
    limited frequency range spanned by the data, to the extent that
    these are unlikely to be useful for all but the brightest sources
    (Section \ref{sec:inband}).
    However, reasonable HBA/NVSS spectral index distributions are
    obtained for compact sources (Section \ref{sec:outofband}).
  \item Per-facet positional offsets introduced by the
    self-calibration in facet calibration are small but can be
    significant in poor-quality data. Perhaps more significant is the
    effect that we take to be the residual blurring of sources by
    inadequate phase calibration, leading to a loss of peak flux density and
    incompleteness at low flux densities (Section
    \ref{sec:completeness}). This may be exacerbated for us by the
    image-plane combination of multiple spectral windows.
\end{itemize}

For data reduction of the Tier 1 surveys it would probably be
preferable to carry out our flux scaling method {\it before} facet
calibration, which would then allow the facet calibration to be run
using all bands simultaneously as described by vW16 while still having
correct flux densities. In-band spectral indices, if desired, could be derived
by re-imaging after facet calibration. It is not yet clear whether the
facet-based approach of vW16 is fundamentally limited in terms of the
calibration quality that can be achieved away from the calibration
point, i.e. whether some sort of phase screen interpolation or an
algorithm that can fit to much smaller facets will be required.
Nevertheless, the technique represents a significant advance towards
the exploitation of LOFAR's capabilities and in this paper we have
demonstrated that applying it to large areas is technically and
computationally feasible.

We have also constructed an optically identified catalogue using SDSS
galaxy catalogues and spectroscopy, and have used it to investigate
the science that can be done with a combination of LOFAR, SDSS and
{\it Herschel} data. Some key points from this analysis are as
follows:
\begin{itemize}
  \item We achieve a roughly 40 per cent optical identification rate
    for the LOFAR catalogue using SDSS together with FIRST to assist
    with identifications (Section \ref{sec:optid}). This is the result
    of a labour-intensive process in which the vast majority of
    sources were visually inspected in several bands [a similar
    approach for the Bo\"otes field will be described by Williams
    \etal\ (2016b)]. Clearly if this
    process is to be scaled up to the many hundreds of LOFAR pointings
    already in hand in Tier 1, it will require significant automation.
    On the other hand, the catalogues would benefit from further
    visual inspection, for example to provide quality checks on large
    sources found by {\sc pybdsm} and to measure flux density and
    source size more accurately. Generating high-quality catalogues
    from Tier 1 data will remain labour-intensive even if some of the
    optical identification process can be streamlined.
  \item It is important to note that there are many resolved sources
    without optical IDs, and, as discussed in Section
    \ref{sec:remnant}, many of these may not have radio cores in
    FIRST. This may represent a challenge for spectroscopic followup
    projects such as WEAVE-LOFAR \citep{Smith15} which rely on
    accurate positions for their targets.
  \item Comparing with the H-ATLAS images, we recover the well-known
    radio/FIR relation in flux/flux and luminosity/luminosity plots
    using the 250-$\mu$m {\it Herschel} data
    (Sections \ref{sec:herschel}, \ref{sec:sfg}) but see evidence that
    there is a population of radio-faint SFGs with radio flux densities well
    below their expected values on this correlation, consistent with
    earlier empirical work \citep{Smith+14}. On the other hand, we are
    able to use the fact that sources with larger radio luminosities
    than would be expected from their FIR emission must be radio-loud
    AGN to perform an efficient SFG/AGN separation in the optically
    identified catalogue. Radio source count analysis not only shows
    consistency with existing 150-MHz determinations where available
    (Section \ref{sec:counts}) but, for the SFGs, shows good agreement
    (Section \ref{sec:sfg}) with the widely used models of
    \cite{Wilman+08}.
  \item We present a 150-MHz radio luminosity function derived from
    $\sim 2,000$ objects with $r<19$ mag, which shows good agreement
    with the expectations from higher frequencies (Section
    \ref{sec:lf}). Strong luminosity function evolution with redshift
    is seen for the SFGs.
  \item The power/linear-size diagram for the overall sample (Section
    \ref{sec:pdd}) shows that we are still insensitive to very large
    sources at low radio luminosities, something which may need to be
    addressed in Tier 1 data processing by an additional imaging step
    at low resolution. However, we measure a sky density of genuine
    optically identified powerful giant radio galaxies ($L_{\rm 150}
    \ga 10^{26}$ W Hz$^{-1}$, $D \ga 1$ Mpc) which is
    high compared to some estimates in the literature, thanks to our
    good optical identification rate out to relatively high redshifts.
  \item We carry out an initial search for candidate remnant sources,
    where the jets have switched off, in a bright, resolved subsample
    (Section \ref{sec:remnant}). Up to 30 per cent of the sample
    sources show no FIRST core, which might imply a lifetime in the
    remnant phase comparable to that in the active phase. However,
    many of these remnant candidates may have radio cores below the
    FIRST detection limit: sensitive high-frequency observations will
    be necessary to refine the upper limit on remnants in LOFAR
    samples.
\end{itemize}

Subsequent papers will address many of these points in more detail.

\section*{Acknowledgments}

We thank Shea Brown for providing the WSRT image of the Coma cluster,
V. N. Pandey for providing the model of 3C\,196, and
Michal Michalowski and Eduardo Ibar for helpful comments on the paper.
We thank the anonymous referee and scientific editor, whose
  comments improved the paper's presentation.

MJH and WLW acknowledge support from the UK Science and Technology
Facilities Council [ST/M001008/1]. GG thanks the University of
Hertfordshire for a research studentship. PNB is grateful for support
from the UK STFC via grant ST/M001229/1. RM gratefully acknowledges
support from the European Research Council under the European Union's
Seventh Framework Programme (FP/2007-2013) ERC Advanced Grant
RADIOLIFE-320745. TWS acknowledges support from the ERC Advanced
Investigator programme NewClusters 321271. LD and SJM acknowledge
support from ERC Advanced and Consolidator grants Cosmic ISM and
Cosmic Dust. GJW gratefully acknowledges
support from the Leverhulme Trust.

This research has made use of the University of Hertfordshire
high-performance computing facility
(\url{http://stri-cluster.herts.ac.uk/}) and the LOFAR-UK computing
facility located at the University of Hertfordshire and supported by
STFC [ST/P000096/1]. This research
made use of {\sc Astropy}, a community-developed core Python package
for astronomy \citep{AstropyCollaboration13} hosted at
\url{http://www.astropy.org/}, of {\sc APLpy}, an open-source
astronomical plotting package for Python hosted at
\url{http://aplpy.github.com/}, and of {\sc topcat} and {\sc stilts} \citep{Taylor05}.

LOFAR, the Low Frequency Array designed and constructed by ASTRON, has
facilities in several countries, that are owned by various parties
(each with their own funding sources), and that are collectively
operated by the International LOFAR Telescope (ILT) foundation under a
joint scientific policy.

The {\it Herschel}-ATLAS is a project with {\it Herschel}, which is an ESA space
observatory with science instruments provided by European-led
Principal Investigator consortia and with important participation from
NASA. The H-ATLAS website is \url{http://www.h-atlas.org/}.

Funding for SDSS-III has been provided by the Alfred P. Sloan
Foundation, the Participating Institutions, the National Science
Foundation, and the U.S. Department of Energy Office of Science. The
SDSS-III web site is \url{http://www.sdss3.org/}.

SDSS-III is managed by the Astrophysical Research Consortium for the
Participating Institutions of the SDSS-III Collaboration including the
University of Arizona, the Brazilian Participation Group, Brookhaven
National Laboratory, Carnegie Mellon University, University of
Florida, the French Participation Group, the German Participation
Group, Harvard University, the Instituto de Astrofisica de Canarias,
the Michigan State/Notre Dame/JINA Participation Group, Johns Hopkins
University, Lawrence Berkeley National Laboratory, Max Planck
Institute for Astrophysics, Max Planck Institute for Extraterrestrial
Physics, New Mexico State University, New York University, Ohio State
University, Pennsylvania State University, University of Portsmouth,
Princeton University, the Spanish Participation Group, University of
Tokyo, University of Utah, Vanderbilt University, University of
Virginia, University of Washington, and Yale University.

The National Radio Astronomy Observatory (NRAO)
is a facility of the National Science Foundation operated under
cooperative agreement by Associated Universities, Inc.

\bibliographystyle{mnras}
\renewcommand{\refname}{REFERENCES}
\bibliography{../bib/mjh,../bib/cards}

\begin{thebibliography}{}
\makeatletter
\relax
\def\mn@urlcharsother{\let\do\@makeother \do\$\do\&\do\#\do\^\do\_\do\%\do\~}
\def\mn@doi{\begingroup\mn@urlcharsother \@ifnextchar [ {\mn@doi@}
  {\mn@doi@[]}}
\def\mn@doi@[#1]#2{\def\@tempa{#1}\ifx\@tempa\@empty \href
  {http://dx.doi.org/#2} {doi:#2}\else \href {http://dx.doi.org/#2} {#1}\fi
  \endgroup}
\def\mn@eprint#1#2{\mn@eprint@#1:#2::\@nil}
\def\mn@eprint@arXiv#1{\href {http://arxiv.org/abs/#1} {{\tt arXiv:#1}}}
\def\mn@eprint@dblp#1{\href {http://dblp.uni-trier.de/rec/bibtex/#1.xml}
  {dblp:#1}}
\def\mn@eprint@#1:#2:#3:#4\@nil{\def\@tempa {#1}\def\@tempb {#2}\def\@tempc
  {#3}\ifx \@tempc \@empty \let \@tempc \@tempb \let \@tempb \@tempa \fi \ifx
  \@tempb \@empty \def\@tempb {arXiv}\fi \@ifundefined
  {mn@eprint@\@tempb}{\@tempb:\@tempc}{\expandafter \expandafter \csname
  mn@eprint@\@tempb\endcsname \expandafter{\@tempc}}}

\bibitem[\protect\citeauthoryear{{Alam} et~al.,}{{Alam} et~al.}{2015}]{Alam+15}
{Alam} S.,  et~al., 2015, \mn@doi [\apjs] {10.1088/0067-0049/219/1/12}, \href
  {http://adsabs.harvard.edu/abs/2015ApJS..219...12A} {219, 12}

\bibitem[\protect\citeauthoryear{{Astropy Collaboration} et~al.,}{{Astropy
  Collaboration} et~al.}{2013}]{AstropyCollaboration13}
{Astropy Collaboration} et~al., 2013, \mn@doi [\aap]
  {10.1051/0004-6361/201322068}, \href
  {http://adsabs.harvard.edu/abs/2013A%26A...558A..33A} {558, A33}

\bibitem[\protect\citeauthoryear{{Baldwin}}{{Baldwin}}{1982}]{Baldwin82}
{Baldwin} J.~E.,  1982, in {Heeschen} D.~S.,  {Wade} C.~M.,  eds,  IAU
  Symposium Vol. 97, Extragalactic Radio Sources. pp 21--24

\bibitem[\protect\citeauthoryear{{Beck}, {Dobos}, {Budav{\'a}ri}, {Szalay}  \&
  {Csabai}}{{Beck} et~al.}{2016}]{Beck+16}
{Beck} R.,  {Dobos} L.,  {Budav{\'a}ri} T.,  {Szalay} A.~S.,   {Csabai} I.,
  2016, preprint, \href {http://adsabs.harvard.edu/abs/2016arXiv160309708B} {}
  (\mn@eprint {arXiv} {1603.09708})

\bibitem[\protect\citeauthoryear{{Becker}, {White}  \& {Helfand}}{{Becker}
  et~al.}{1995}]{Becker+95}
{Becker} R.~H.,  {White} R.~L.,   {Helfand} D.~J.,  1995, \apj, 450, 559

\bibitem[\protect\citeauthoryear{{Best}, {Kauffmann}, {Heckman}, {Brinchmann},
  {Charlot}, {Ivezi\'c}  \& {White}}{{Best} et~al.}{2005}]{Best+05}
{Best} P.~N.,  {Kauffmann} G.,  {Heckman} T.~M.,  {Brinchmann} J.,  {Charlot}
  S.,  {Ivezi\'c} Z.,   {White} S.~D.~M.,  2005, \mnras, 362, 25

\bibitem[\protect\citeauthoryear{{Best}, {Ker}, {Simpson}, {Rigby}  \&
  {Sabater}}{{Best} et~al.}{2014}]{Best+14}
{Best} P.~N.,  {Ker} L.~M.,  {Simpson} C.,  {Rigby} E.~E.,   {Sabater} J.,
  2014, \mn@doi [\mnras] {10.1093/mnras/stu1776}, \href
  {http://adsabs.harvard.edu/abs/2014MNRAS.445..955B} {445, 955}

\bibitem[\protect\citeauthoryear{{Blundell}, {Rawlings}  \&
  {Willott}}{{Blundell} et~al.}{1999}]{Blundell+99}
{Blundell} K.~M.,  {Rawlings} S.,   {Willott} C.~J.,  1999, \mn@doi [\aj]
  {10.1086/300721}, \href {http://adsabs.harvard.edu/abs/1999AJ....117..677B}
  {117, 677}

\bibitem[\protect\citeauthoryear{{Bonfield} et~al.,}{{Bonfield}
  et~al.}{2011}]{Bonfield+11}
{Bonfield} D.~G.,  et~al., 2011, \mn@doi [\mnras]
  {10.1111/j.1365-2966.2011.18826.x}, \href
  {http://adsabs.harvard.edu/abs/2011MNRAS.416...13B} {416, 13}

\bibitem[\protect\citeauthoryear{{Brienza}, {Mahony}, {Morganti}, {Prandoni}
  \& {Godfrey}}{{Brienza} et~al.}{2016a}]{Brienza+16b}
{Brienza} M.,  {Mahony} E.,  {Morganti} R.,  {Prandoni} I.,   {Godfrey} L.,
  2016a, preprint, \href {http://adsabs.harvard.edu/abs/2016arXiv160301837B} {}
  (\mn@eprint {arXiv} {1603.01837})

\bibitem[\protect\citeauthoryear{{Brienza} et~al.,}{{Brienza}
  et~al.}{2016b}]{Brienza+16a}
{Brienza} M.,  et~al., 2016b, \mn@doi [\aap] {10.1051/0004-6361/201526754},
  \href {http://adsabs.harvard.edu/abs/2016A%26A...585A..29B} {585, A29}

\bibitem[\protect\citeauthoryear{{Brown} \& {Rudnick}}{{Brown} \&
  {Rudnick}}{2011}]{Brown+Rudnick11}
{Brown} S.,  {Rudnick} L.,  2011, \mn@doi [\mnras]
  {10.1111/j.1365-2966.2010.17738.x}, \href
  {http://adsabs.harvard.edu/abs/2011MNRAS.412....2B} {412, 2}

\bibitem[\protect\citeauthoryear{{Chilingarian}, {Melchior}  \&
  {Zolotukhin}}{{Chilingarian} et~al.}{2010}]{Chilingarian+10}
{Chilingarian} I.~V.,  {Melchior} A.-L.,   {Zolotukhin} I.~Y.,  2010, \mn@doi
  [\mnras] {10.1111/j.1365-2966.2010.16506.x}, \href
  {http://adsabs.harvard.edu/abs/2010MNRAS.405.1409C} {405, 1409}

\bibitem[\protect\citeauthoryear{{Colla} et~al.,}{{Colla}
  et~al.}{1973}]{Colla+73}
{Colla} G.,  et~al., 1973, \aaps, \href
  {http://adsabs.harvard.edu/abs/1973A%26AS...11..291C} {11, 291}

\bibitem[\protect\citeauthoryear{{Condon}, {Cotton}, {Greisen}, {Yin},
  {Perley}, {Taylor}  \& {Broderick}}{{Condon} et~al.}{1998}]{Condon+98}
{Condon} J.~J.,  {Cotton} W.~D.,  {Greisen} E.~W.,  {Yin} Q.~F.,  {Perley}
  R.~A.,  {Taylor} G.~B.,   {Broderick} J.~J.,  1998, \aj, 115, 1693

\bibitem[\protect\citeauthoryear{{Condon}, {Cotton}  \& {Broderick}}{{Condon}
  et~al.}{2002}]{Condon+02}
{Condon} J.~J.,  {Cotton} W.~D.,   {Broderick} J.~J.,  2002, \mn@doi [\aj]
  {10.1086/341650}, \href {http://adsabs.harvard.edu/abs/2002AJ....124..675C}
  {124, 675}

\bibitem[\protect\citeauthoryear{{Cordey}}{{Cordey}}{1987}]{Cordey87}
{Cordey} R.~A.,  1987, \mnras, 227, 695

\bibitem[\protect\citeauthoryear{{Cotter}, {Rawlings}  \& {Saunders}}{{Cotter}
  et~al.}{1996}]{Cotter+96}
{Cotter} G.,  {Rawlings} S.,   {Saunders} R.,  1996, \mn@doi [\mnras]
  {10.1093/mnras/281.3.1081}, \href
  {http://adsabs.harvard.edu/abs/1996MNRAS.281.1081C} {281, 1081}

\bibitem[\protect\citeauthoryear{{Croton} et~al.,}{{Croton}
  et~al.}{2006}]{Croton+06}
{Croton} D.,  et~al., 2006, \mnras, 365, 111

\bibitem[\protect\citeauthoryear{{Cruz} et~al.,}{{Cruz} et~al.}{2006}]{Cruz+06}
{Cruz} M.~J.,  et~al., 2006, \mn@doi [\mnras]
  {10.1111/j.1365-2966.2006.11101.x}, \href
  {http://adsabs.harvard.edu/abs/2006MNRAS.373.1531C} {373, 1531}

\bibitem[\protect\citeauthoryear{{Dawson} et~al.,}{{Dawson}
  et~al.}{2013}]{Dawson+13}
{Dawson} K.~S.,  et~al., 2013, \mn@doi [\aj] {10.1088/0004-6256/145/1/10},
  \href {http://adsabs.harvard.edu/abs/2013AJ....145...10D} {145, 10}

\bibitem[\protect\citeauthoryear{{Driver} et~al.,}{{Driver}
  et~al.}{2009}]{Driver+09}
{Driver} S.~P.,  et~al., 2009, \mn@doi [Astronomy and Geophysics]
  {10.1111/j.1468-4004.2009.50512.x}, \href
  {http://adsabs.harvard.edu/abs/2009A%26G....50e..12D} {50, 050000}

\bibitem[\protect\citeauthoryear{{Driver} et~al.,}{{Driver}
  et~al.}{2011}]{Driver+11}
{Driver} S.~P.,  et~al., 2011, \mn@doi [\mnras]
  {10.1111/j.1365-2966.2010.18188.x}, \href
  {http://adsabs.harvard.edu/abs/2011MNRAS.413..971D} {413, 971}

\bibitem[\protect\citeauthoryear{{Dunne} et~al.,}{{Dunne}
  et~al.}{2011}]{Dunne+11}
{Dunne} L.,  et~al., 2011, \mn@doi [\mnras] {10.1111/j.1365-2966.2011.19363.x},
  \href {http://adsabs.harvard.edu/abs/2011MNRAS.417.1510D} {417, 1510}

\bibitem[\protect\citeauthoryear{{Dye} et~al.,}{{Dye} et~al.}{2010}]{Dye+10}
{Dye} S.,  et~al., 2010, \mn@doi [\aap] {10.1051/0004-6361/201014614}, \href
  {http://adsabs.harvard.edu/abs/2010A%26A...518L..10D} {518, L10}

\bibitem[\protect\citeauthoryear{{Eales}}{{Eales}}{1985}]{Eales85}
{Eales} S.~A.,  1985, \mnras, 213, 899

\bibitem[\protect\citeauthoryear{{Eales} et~al.,}{{Eales}
  et~al.}{2010}]{Eales+10}
{Eales} S.,  et~al., 2010, \mn@doi [\pasp] {10.1086/653086}, \href
  {http://adsabs.harvard.edu/abs/2010PASP..122..499E} {122, 499}

\bibitem[\protect\citeauthoryear{{Eisenstein} et~al.,}{{Eisenstein}
  et~al.}{2011}]{Eisenstein+11}
{Eisenstein} D.~J.,  et~al., 2011, \mn@doi [\aj] {10.1088/0004-6256/142/3/72},
  \href {http://adsabs.harvard.edu/abs/2011AJ....142...72E} {142, 72}

\bibitem[\protect\citeauthoryear{{Foreman-Mackey}, {Hogg}, {Lang}  \&
  {Goodman}}{{Foreman-Mackey} et~al.}{2013}]{Foreman-Mackey+13}
{Foreman-Mackey} D.,  {Hogg} D.~W.,  {Lang} D.,   {Goodman} J.,  2013, \mn@doi
  [\pasp] {10.1086/670067}, \href
  {http://adsabs.harvard.edu/abs/2013PASP..125..306F} {125, 306}

\bibitem[\protect\citeauthoryear{{Giovannini}, {Feretti}, {Gregorini}  \&
  {Parma}}{{Giovannini} et~al.}{1988}]{Giovannini+88}
{Giovannini} G.,  {Feretti} L.,  {Gregorini} L.,   {Parma} P.,  1988, \aap,
  199, 73

\bibitem[\protect\citeauthoryear{{Griffin} et~al.,}{{Griffin}
  et~al.}{2010}]{Griffin+10}
{Griffin} M.~J.,  et~al., 2010, \mn@doi [\aap] {10.1051/0004-6361/201014519},
  \href {http://adsabs.harvard.edu/abs/2010A%26A...518L...3G} {518, L3}

\bibitem[\protect\citeauthoryear{{Gruppioni} et~al.,}{{Gruppioni}
  et~al.}{2013}]{Gruppioni+13}
{Gruppioni} C.,  et~al., 2013, \mn@doi [\mnras] {10.1093/mnras/stt308}, \href
  {http://adsabs.harvard.edu/abs/2013MNRAS.432...23G} {432, 23}

\bibitem[\protect\citeauthoryear{{G{\"u}rkan} et~al.,}{{G{\"u}rkan}
  et~al.}{2015}]{Gurkan+15}
{G{\"u}rkan} G.,  et~al., 2015, \mn@doi [\mnras] {10.1093/mnras/stv1502}, \href
  {http://adsabs.harvard.edu/abs/2015MNRAS.452.3776G} {452, 3776}

\bibitem[\protect\citeauthoryear{{Haarsma}, {Partridge}, {Windhorst}  \&
  {Richards}}{{Haarsma} et~al.}{2000}]{Haarsma+00}
{Haarsma} D.~B.,  {Partridge} R.~B.,  {Windhorst} R.~A.,   {Richards} E.~A.,
  2000, \mn@doi [\apj] {10.1086/317225}, \href
  {http://adsabs.harvard.edu/abs/2000ApJ...544..641H} {544, 641}

\bibitem[\protect\citeauthoryear{{Hales}, {Baldwin}  \& {Warner}}{{Hales}
  et~al.}{1988}]{Hales+88}
{Hales} S.~E.~G.,  {Baldwin} J.~E.,   {Warner} P.~J.,  1988, \mnras, \href
  {http://adsabs.harvard.edu/abs/1988MNRAS.234..919H} {234, 919}

\bibitem[\protect\citeauthoryear{{Hales}, {Riley}, {Waldram}, {Warner}  \&
  {Baldwin}}{{Hales} et~al.}{2007}]{Hales+07}
{Hales} S.~E.~G.,  {Riley} J.~M.,  {Waldram} E.~M.,  {Warner} P.~J.,
  {Baldwin} J.~E.,  2007, \mn@doi [\mnras] {10.1111/j.1365-2966.2007.12392.x},
  \href {http://adsabs.harvard.edu/abs/2007MNRAS.382.1639H} {382, 1639}

\bibitem[\protect\citeauthoryear{{Hardcastle} \& {Krause}}{{Hardcastle} \&
  {Krause}}{2013}]{Hardcastle+Krause13}
{Hardcastle} M.~J.,  {Krause} M.~G.~H.,  2013, \mn@doi [\mnras]
  {10.1093/mnras/sts564}, \href
  {http://adsabs.harvard.edu/abs/2013MNRAS.430..174H} {430, 174}

\bibitem[\protect\citeauthoryear{{Hardcastle} \& {Krause}}{{Hardcastle} \&
  {Krause}}{2014}]{Hardcastle+Krause14}
{Hardcastle} M.~J.,  {Krause} M.~G.~H.,  2014, \mn@doi [\mnras]
  {10.1093/mnras/stu1229}, \href
  {http://adsabs.harvard.edu/abs/2014arXiv1406.5300H} {443, 1482}

\bibitem[\protect\citeauthoryear{{Hardcastle} \& {Looney}}{{Hardcastle} \&
  {Looney}}{2008}]{Hardcastle+Looney08}
{Hardcastle} M.~J.,  {Looney} L.~W.,  2008, \mnras, 388, 176

\bibitem[\protect\citeauthoryear{{Hardcastle}, {Alexander}, {Pooley}  \&
  {Riley}}{{Hardcastle} et~al.}{1997}]{Hardcastle+97}
{Hardcastle} M.~J.,  {Alexander} P.,  {Pooley} G.~G.,   {Riley} J.~M.,  1997,
  \mnras, 288, 859

\bibitem[\protect\citeauthoryear{{Hardcastle}, {Croston}  \&
  {Kraft}}{{Hardcastle} et~al.}{2007}]{Hardcastle+07}
{Hardcastle} M.~J.,  {Croston} J.~H.,   {Kraft} R.~P.,  2007, \apj, 669, 893

\bibitem[\protect\citeauthoryear{{Hardcastle}, {Evans}  \&
  {Croston}}{{Hardcastle} et~al.}{2009}]{Hardcastle+09}
{Hardcastle} M.~J.,  {Evans} D.~A.,   {Croston} J.~H.,  2009, \mnras, 396, 1929

\bibitem[\protect\citeauthoryear{{Hardcastle} et~al.,}{{Hardcastle}
  et~al.}{2010}]{Hardcastle+10b}
{Hardcastle} M.~J.,  et~al., 2010, \mn@doi [\mnras]
  {10.1111/j.1365-2966.2010.17791.x}, \href
  {http://adsabs.harvard.edu/abs/2010MNRAS.409..122H} {409, 122}

\bibitem[\protect\citeauthoryear{{Hardcastle} et~al.,}{{Hardcastle}
  et~al.}{2013}]{Hardcastle+13}
{Hardcastle} M.~J.,  et~al., 2013, \mn@doi [\mnras] {10.1093/mnras/sts510},
  \href {http://adsabs.harvard.edu/abs/2013MNRAS.429.2407H} {429, 2407}

\bibitem[\protect\citeauthoryear{{Heald} et~al.,}{{Heald}
  et~al.}{2015}]{Heald+15}
{Heald} G.~H.,  et~al., 2015, \mn@doi [\aap] {10.1051/0004-6361/201425210},
  \href {http://adsabs.harvard.edu/abs/2015A%26A...582A.123H} {582, A123}

\bibitem[\protect\citeauthoryear{{Helou}, {Soifer}  \&
  {Rowan-Robinson}}{{Helou} et~al.}{1985}]{Helou+85}
{Helou} G.,  {Soifer} B.~T.,   {Rowan-Robinson} M.,  1985, \apj, 298, L7

\bibitem[\protect\citeauthoryear{{Hopkins} \& {Beacom}}{{Hopkins} \&
  {Beacom}}{2006}]{Hopkins+Beacom06}
{Hopkins} A.~M.,  {Beacom} J.~F.,  2006, \mn@doi [\apj] {10.1086/506610}, \href
  {http://adsabs.harvard.edu/abs/2006ApJ...651..142H} {651, 142}

\bibitem[\protect\citeauthoryear{{Ibar} et~al.,}{{Ibar} et~al.}{2008}]{Ibar+08}
{Ibar} E.,  et~al., 2008, \mn@doi [\mnras] {10.1111/j.1365-2966.2008.13077.x},
  \href {http://adsabs.harvard.edu/abs/2008MNRAS.386..953I} {386, 953}

\bibitem[\protect\citeauthoryear{{Ibar}, {Ivison}, {Biggs}, {Lal}, {Best}  \&
  {Green}}{{Ibar} et~al.}{2009}]{Ibar+09}
{Ibar} E.,  {Ivison} R.~J.,  {Biggs} A.~D.,  {Lal} D.~V.,  {Best} P.~N.,
  {Green} D.~A.,  2009, \mn@doi [\mnras] {10.1111/j.1365-2966.2009.14866.x},
  \href {http://adsabs.harvard.edu/abs/2009MNRAS.397..281I} {397, 281}

\bibitem[\protect\citeauthoryear{{Ibar} et~al.,}{{Ibar} et~al.}{2010}]{Ibar+10}
{Ibar} E.,  et~al., 2010, \mn@doi [\mnras] {10.1111/j.1365-2966.2010.17620.x},
  \href {http://adsabs.harvard.edu/abs/2010MNRAS.409...38I} {409, 38}

\bibitem[\protect\citeauthoryear{{Intema}, {Jagannathan}, {Mooley}  \&
  {Frail}}{{Intema} et~al.}{2016}]{Intema+16}
{Intema} H.~T.,  {Jagannathan} P.,  {Mooley} K.~P.,   {Frail} D.~A.,  2016,
  preprint, \href {http://adsabs.harvard.edu/abs/2016arXiv160304368I} {}
  (\mn@eprint {arXiv} {1603.04368})

\bibitem[\protect\citeauthoryear{{Ivison} et~al.,}{{Ivison}
  et~al.}{2010a}]{Ivison+10}
{Ivison} R.~J.,  et~al., 2010a, \mn@doi [\mnras]
  {10.1111/j.1365-2966.2009.15918.x}, \href
  {http://adsabs.harvard.edu/abs/2010MNRAS.402..245I} {402, 245}

\bibitem[\protect\citeauthoryear{{Ivison} et~al.,}{{Ivison}
  et~al.}{2010b}]{Ivison+10b}
{Ivison} R.~J.,  et~al., 2010b, \mn@doi [\aap] {10.1051/0004-6361/201014552},
  \href {http://adsabs.harvard.edu/abs/2010A%26A...518L..31I} {518, L31}

\bibitem[\protect\citeauthoryear{{Jarvis} et~al.,}{{Jarvis}
  et~al.}{2010}]{Jarvis+10}
{Jarvis} M.~J.,  et~al., 2010, \mn@doi [\mnras]
  {10.1111/j.1365-2966.2010.17772.x}, \href
  {http://adsabs.harvard.edu/abs/2010MNRAS.409...92J} {409, 92}

\bibitem[\protect\citeauthoryear{{Kaiser} \& {Cotter}}{{Kaiser} \&
  {Cotter}}{2002}]{Kaiser+Cotter02}
{Kaiser} C.~R.,  {Cotter} G.,  2002, \mn@doi [\mnras]
  {10.1046/j.1365-8711.2002.05799.x}, \href
  {http://adsabs.harvard.edu/abs/2002MNRAS.336..649K} {336, 649}

\bibitem[\protect\citeauthoryear{{Kaiser}, {Dennett-Thorpe}  \&
  {Alexander}}{{Kaiser} et~al.}{1997}]{Kaiser+97}
{Kaiser} C.~R.,  {Dennett-Thorpe} J.,   {Alexander} P.,  1997, \mnras, 292, 723

\bibitem[\protect\citeauthoryear{{Kalfountzou} et~al.,}{{Kalfountzou}
  et~al.}{2014}]{Kalfountzou+14}
{Kalfountzou} E.,  et~al., 2014, \mn@doi [\mnras] {10.1093/mnras/stu782}, \href
  {http://adsabs.harvard.edu/abs/2014MNRAS.442.1181K} {442, 1181}

\bibitem[\protect\citeauthoryear{{Karim} et~al.,}{{Karim}
  et~al.}{2011}]{Karim+11}
{Karim} A.,  et~al., 2011, \mn@doi [\apj] {10.1088/0004-637X/730/2/61}, \href
  {http://adsabs.harvard.edu/abs/2011ApJ...730...61K} {730, 61}

\bibitem[\protect\citeauthoryear{{Konar} \& {Hardcastle}}{{Konar} \&
  {Hardcastle}}{2013}]{Konar+Hardcastle13}
{Konar} C.,  {Hardcastle} M.~J.,  2013, \mn@doi [\mnras]
  {10.1093/mnras/stt1676}, \href
  {http://adsabs.harvard.edu/abs/2013MNRAS.436.1595K} {436, 1595}

\bibitem[\protect\citeauthoryear{{Konar}, {Hardcastle}, {Jamrozy}  \&
  {Croston}}{{Konar} et~al.}{2013}]{Konar+13}
{Konar} C.,  {Hardcastle} M.~J.,  {Jamrozy} M.,   {Croston} J.~H.,  2013,
  \mn@doi [\mnras] {10.1093/mnras/stt040}, \href
  {http://adsabs.harvard.edu/abs/2013MNRAS.430.2137K} {430, 2137}

\bibitem[\protect\citeauthoryear{{Lacki}, {Thompson}  \& {Quataert}}{{Lacki}
  et~al.}{2010}]{Lacki+10}
{Lacki} B.~C.,  {Thompson} T.~A.,   {Quataert} E.,  2010, \mn@doi [\apj]
  {10.1088/0004-637X/717/1/1}, \href
  {http://adsabs.harvard.edu/abs/2010ApJ...717....1L} {717, 1}

\bibitem[\protect\citeauthoryear{{Laing}, {Riley}  \& {Longair}}{{Laing}
  et~al.}{1983}]{Laing+83}
{Laing} R.~A.,  {Riley} J.~M.,   {Longair} M.~S.,  1983, \mnras, 204, 151

\bibitem[\protect\citeauthoryear{{Lane}, {Cotton}, {van Velzen}, {Clarke},
  {Kassim}, {Helmboldt}, {Lazio}  \& {Cohen}}{{Lane} et~al.}{2014}]{Lane+14}
{Lane} W.~M.,  {Cotton} W.~D.,  {van Velzen} S.,  {Clarke} T.~E.,  {Kassim}
  N.~E.,  {Helmboldt} J.~F.,  {Lazio} T.~J.~W.,   {Cohen} A.~S.,  2014, \mn@doi
  [\mnras] {10.1093/mnras/stu256}, \href
  {http://adsabs.harvard.edu/abs/2014MNRAS.440..327L} {440, 327}

\bibitem[\protect\citeauthoryear{{Lara}, {Cotton}, {Feretti}, {Giovannini},
  {Marcaide}, {M{\'a}rquez}  \& {Venturi}}{{Lara} et~al.}{2001}]{Lara+01}
{Lara} L.,  {Cotton} W.~D.,  {Feretti} L.,  {Giovannini} G.,  {Marcaide} J.~M.,
   {M{\'a}rquez} I.,   {Venturi} T.,  2001, \mn@doi [\aap]
  {10.1051/0004-6361:20010254}, \href
  {http://adsabs.harvard.edu/abs/2001A%26A...370..409L} {370, 409}

\bibitem[\protect\citeauthoryear{{Leahy}}{{Leahy}}{1993}]{Leahy93}
{Leahy} J.~P.,  1993, in {R\"oser H.-J., Meisenheimer K.} ed., {Jets in
  Extragalactic Radio Sources}. {Springer-Verlag, Heidelberg}, p.~1

\bibitem[\protect\citeauthoryear{{Lindsay} et~al.,}{{Lindsay}
  et~al.}{2014}]{Lindsay+14}
{Lindsay} S.~N.,  et~al., 2014, \mn@doi [\mnras] {10.1093/mnras/stu354}, \href
  {http://adsabs.harvard.edu/abs/2014MNRAS.440.1527L} {440, 1527}

\bibitem[\protect\citeauthoryear{{Machalski} \& {Godlowski}}{{Machalski} \&
  {Godlowski}}{2000}]{Machalski+Godlowski00}
{Machalski} J.,  {Godlowski} W.,  2000, \aap, 360, 463

\bibitem[\protect\citeauthoryear{{Machalski}, {Jamrozy}  \& {Zola}}{{Machalski}
  et~al.}{2001}]{Machalski+01}
{Machalski} J.,  {Jamrozy} M.,   {Zola} S.,  2001, \mn@doi [\aap]
  {10.1051/0004-6361:20010352}, \href
  {http://adsabs.harvard.edu/abs/2001A%26A...371..445M} {371, 445}

\bibitem[\protect\citeauthoryear{{Madau}, {Ferguson}, {Dickinson},
  {Giavalisco}, {Steidel}  \& {Fruchter}}{{Madau} et~al.}{1996}]{Madau+96}
{Madau} P.,  {Ferguson} H.~C.,  {Dickinson} M.~E.,  {Giavalisco} M.,  {Steidel}
  C.~C.,   {Fruchter} A.,  1996, \mnras, 283, 1388

\bibitem[\protect\citeauthoryear{{Magnelli} et~al.,}{{Magnelli}
  et~al.}{2013}]{Magnelli+13}
{Magnelli} B.,  et~al., 2013, \mn@doi [\aap] {10.1051/0004-6361/201321371},
  \href {http://adsabs.harvard.edu/abs/2013A%26A...553A.132M} {553, A132}

\bibitem[\protect\citeauthoryear{{Mancuso} et~al.,}{{Mancuso}
  et~al.}{2015}]{Mancuso+15}
{Mancuso} C.,  et~al., 2015, \mn@doi [\apj] {10.1088/0004-637X/810/1/72}, \href
  {http://adsabs.harvard.edu/abs/2015ApJ...810...72M} {810, 72}

\bibitem[\protect\citeauthoryear{{Manolakou} \& {Kirk}}{{Manolakou} \&
  {Kirk}}{2002}]{Manolakou+Kirk02}
{Manolakou} K.,  {Kirk} J.~G.,  2002, \mn@doi [\aap]
  {10.1051/0004-6361:20020780}, \href
  {http://adsabs.harvard.edu/abs/2002A%26A...391..127M} {391, 127}

\bibitem[\protect\citeauthoryear{{Mauch} \& {Sadler}}{{Mauch} \&
  {Sadler}}{2007}]{Mauch+Sadler07}
{Mauch} T.,  {Sadler} E.,  2007, \mnras, 375, 931

\bibitem[\protect\citeauthoryear{{Mauch}, {Kl{\"o}ckner}, {Rawlings}, {Jarvis},
  {Hardcastle}, {Obreschkow}, {Saikia}  \& {Thompson}}{{Mauch}
  et~al.}{2013}]{Mauch+13}
{Mauch} T.,  {Kl{\"o}ckner} H.-R.,  {Rawlings} S.,  {Jarvis} M.,  {Hardcastle}
  M.~J.,  {Obreschkow} D.,  {Saikia} D.~J.,   {Thompson} M.~A.,  2013, \mn@doi
  [\mnras] {10.1093/mnras/stt1323}, \href
  {http://adsabs.harvard.edu/abs/2013MNRAS.435..650M} {435, 650}

\bibitem[\protect\citeauthoryear{{McAlpine}, {Jarvis}  \&
  {Bonfield}}{{McAlpine} et~al.}{2013}]{McAlpine+13}
{McAlpine} K.,  {Jarvis} M.~J.,   {Bonfield} D.~G.,  2013, \mn@doi [\mnras]
  {10.1093/mnras/stt1638}, \href
  {http://adsabs.harvard.edu/abs/2013MNRAS.436.1084M} {436, 1084}

\bibitem[\protect\citeauthoryear{{Mohan} \& {Rafferty}}{{Mohan} \&
  {Rafferty}}{2015}]{Mohan+Rafferty15}
{Mohan} N.,  {Rafferty} D.,  2015, {PyBDSM: Python Blob Detection and Source
  Measurement}, Astrophysics Source Code Library (\mn@eprint {ascl} {1502.007})

\bibitem[\protect\citeauthoryear{{Mullin}, {Riley}  \& {Hardcastle}}{{Mullin}
  et~al.}{2008}]{Mullin+08}
{Mullin} L.~M.,  {Riley} J.~M.,   {Hardcastle} M.~J.,  2008, \mn@doi [\mnras]
  {10.1111/j.1365-2966.2008.13534.x}, \href
  {http://adsabs.harvard.edu/abs/2008MNRAS.390..595M} {390, 595}

\bibitem[\protect\citeauthoryear{{Murgia} et~al.,}{{Murgia}
  et~al.}{2011}]{Murgia+11}
{Murgia} M.,  et~al., 2011, \mn@doi [\aap] {10.1051/0004-6361/201015302}, \href
  {http://adsabs.harvard.edu/abs/2011A%26A...526A.148M} {526, A148}

\bibitem[\protect\citeauthoryear{{Murphy}}{{Murphy}}{2009}]{Murphy09}
{Murphy} E.~J.,  2009, \mn@doi [\apj] {10.1088/0004-637X/706/1/482}, \href
  {http://adsabs.harvard.edu/abs/2009ApJ...706..482M} {706, 482}

\bibitem[\protect\citeauthoryear{{Offringa} et~al.,}{{Offringa}
  et~al.}{2014}]{Offringa+14}
{Offringa} A.~R.,  et~al., 2014, \mn@doi [\mnras] {10.1093/mnras/stu1368},
  \href {http://adsabs.harvard.edu/abs/2014MNRAS.444..606O} {444, 606}

\bibitem[\protect\citeauthoryear{{Orr{\`u}} et~al.,}{{Orr{\`u}}
  et~al.}{2015}]{Orru+15}
{Orr{\`u}} E.,  et~al., 2015, \mn@doi [\aap] {10.1051/0004-6361/201526501},
  \href {http://adsabs.harvard.edu/abs/2015A%26A...584A.112O} {584, A112}

\bibitem[\protect\citeauthoryear{{Pandey}, {van Zwieten}, {de Bruyn}  \&
  {Nijboer}}{{Pandey} et~al.}{2009}]{Pandey+09}
{Pandey} V.~N.,  {van Zwieten} J.~E.,  {de Bruyn} A.~G.,   {Nijboer} R.,  2009,
  in {Saikia} D.~J.,  {Green} D.~A.,  {Gupta} Y.,   {Venturi} T.,  eds,
  Astronomical Society of the Pacific Conference Series Vol. 407, The
  Low-Frequency Radio Universe. p.~384

\bibitem[\protect\citeauthoryear{{Parma}, {Murgia}, {de Ruiter}, {Fanti},
  {Mack}  \& {Govoni}}{{Parma} et~al.}{2007}]{Parma+07}
{Parma} P.,  {Murgia} M.,  {de Ruiter} H.~R.,  {Fanti} R.,  {Mack} K.-H.,
  {Govoni} F.,  2007, \mn@doi [\aap] {10.1051/0004-6361:20077592}, \href
  {http://adsabs.harvard.edu/abs/2007A%26A...470..875P} {470, 875}

\bibitem[\protect\citeauthoryear{{Pascale} et~al.,}{{Pascale}
  et~al.}{2011}]{Pascale+11}
{Pascale} E.,  et~al., 2011, \mn@doi [\mnras]
  {10.1111/j.1365-2966.2011.18756.x}, \href
  {http://adsabs.harvard.edu/abs/2011MNRAS.415..911P} {415, 911}

\bibitem[\protect\citeauthoryear{{Pilbratt} et~al.,}{{Pilbratt}
  et~al.}{2010}]{Pilbratt+10}
{Pilbratt} G.~L.,  et~al., 2010, \aap, 518, L1

\bibitem[\protect\citeauthoryear{{Poglitsch} et~al.,}{{Poglitsch}
  et~al.}{2010}]{Poglitsch+10}
{Poglitsch} A.,  et~al., 2010, \mn@doi [\aap] {10.1051/0004-6361/201014535},
  \href {http://adsabs.harvard.edu/abs/2010A%26A...518L...2P} {518, L2}

\bibitem[\protect\citeauthoryear{{Pracy} et~al.,}{{Pracy}
  et~al.}{2016}]{Pracy+16}
{Pracy} M.,  et~al., 2016, preprint, \href
  {http://adsabs.harvard.edu/abs/2016arXiv160404332P} {} (\mn@eprint {arXiv}
  {1604.04332})

\bibitem[\protect\citeauthoryear{{Prescott} et~al.,}{{Prescott}
  et~al.}{2016}]{Prescott+16}
{Prescott} M.,  et~al., 2016, \mn@doi [\mnras] {10.1093/mnras/stv3020}, \href
  {http://adsabs.harvard.edu/abs/2016MNRAS.457..730P} {457, 730}

\bibitem[\protect\citeauthoryear{{Rawlings}, {Eales}  \& {Lacy}}{{Rawlings}
  et~al.}{2001}]{Rawlings+01}
{Rawlings} S.,  {Eales} S.,   {Lacy} M.,  2001, \mn@doi [\mnras]
  {10.1046/j.1365-8711.2001.04151.x}, \href
  {http://adsabs.harvard.edu/abs/2001MNRAS.322..523R} {322, 523}

\bibitem[\protect\citeauthoryear{{Rigby} et~al.,}{{Rigby}
  et~al.}{2011}]{Rigby+11}
{Rigby} E.~E.,  et~al., 2011, \mn@doi [\mnras]
  {10.1111/j.1365-2966.2011.18864.x}, \href
  {http://adsabs.harvard.edu/abs/2011MNRAS.415.2336R} {415, 2336}

\bibitem[\protect\citeauthoryear{{R\"ottgering} et~al.,}{{R\"ottgering}
  et~al.}{2006}]{Rottgering+06}
{R\"ottgering} H.~J.~A.,  et~al., 2006, ArXiv Astrophysics e-prints, \href
  {http://adsabs.harvard.edu/abs/2006astro.ph.10596R} {}

\bibitem[\protect\citeauthoryear{{Saripalli}, {Subrahmanyan}, {Thorat},
  {Ekers}, {Hunstead}, {Johnston}  \& {Sadler}}{{Saripalli}
  et~al.}{2012}]{Saripalli+12}
{Saripalli} L.,  {Subrahmanyan} R.,  {Thorat} K.,  {Ekers} R.~D.,  {Hunstead}
  R.~W.,  {Johnston} H.~M.,   {Sadler} E.~M.,  2012, \mn@doi [\apjs]
  {10.1088/0067-0049/199/2/27}, \href
  {http://adsabs.harvard.edu/abs/2012ApJS..199...27S} {199, 27}

\bibitem[\protect\citeauthoryear{{Scaife} \& {Heald}}{{Scaife} \&
  {Heald}}{2012}]{Scaife+Heald12}
{Scaife} A.~M.~M.,  {Heald} G.~H.,  2012, \mn@doi [\mnras]
  {10.1111/j.1745-3933.2012.01251.x}, \href
  {http://adsabs.harvard.edu/abs/2012MNRAS.423L..30S} {423, L30}

\bibitem[\protect\citeauthoryear{{Schlafly} \& {Finkbeiner}}{{Schlafly} \&
  {Finkbeiner}}{2011}]{Schlafly+Finkbeiner11}
{Schlafly} E.~F.,  {Finkbeiner} D.~P.,  2011, \mn@doi [\apj]
  {10.1088/0004-637X/737/2/103}, \href
  {http://adsabs.harvard.edu/abs/2011ApJ...737..103S} {737, 103}

\bibitem[\protect\citeauthoryear{{Schoenmakers}, {de Bruyn}, {R{\"o}ttgering},
  {van der Laan}  \& {Kaiser}}{{Schoenmakers} et~al.}{2000}]{Schoenmakers+00}
{Schoenmakers} A.~P.,  {de Bruyn} A.~G.,  {R{\"o}ttgering} H.~J.~A.,  {van der
  Laan} H.,   {Kaiser} C.~R.,  2000, \mn@doi [\mnras]
  {10.1046/j.1365-8711.2000.03430.x}, \href
  {http://adsabs.harvard.edu/abs/2000MNRAS.315..371S} {315, 371}

\bibitem[\protect\citeauthoryear{{Schoenmakers}, {de Bruyn}, {R{\"o}ttgering}
  \& {van der Laan}}{{Schoenmakers} et~al.}{2001}]{Schoenmakers+01}
{Schoenmakers} A.~P.,  {de Bruyn} A.~G.,  {R{\"o}ttgering} H.~J.~A.,   {van der
  Laan} H.,  2001, \mn@doi [\aap] {10.1051/0004-6361:20010746}, \href
  {http://adsabs.harvard.edu/abs/2001A%26A...374..861S} {374, 861}

\bibitem[\protect\citeauthoryear{{Serjeant} et~al.,}{{Serjeant}
  et~al.}{2010}]{Serjeant+10}
{Serjeant} S.,  et~al., 2010, \mn@doi [\aap] {10.1051/0004-6361/201014565},
  \href {http://adsabs.harvard.edu/abs/2010A%26A...518L...7S} {518, L7}

\bibitem[\protect\citeauthoryear{{Seymour} et~al.,}{{Seymour}
  et~al.}{2008}]{Seymour+08}
{Seymour} N.,  et~al., 2008, \mn@doi [\mnras]
  {10.1111/j.1365-2966.2008.13166.x}, \href
  {http://adsabs.harvard.edu/abs/2008MNRAS.386.1695S} {386, 1695}

\bibitem[\protect\citeauthoryear{{Simpson} et~al.,}{{Simpson}
  et~al.}{2012}]{Simpson+12}
{Simpson} C.,  et~al., 2012, \mn@doi [\mnras]
  {10.1111/j.1365-2966.2012.20529.x}, \href
  {http://adsabs.harvard.edu/abs/2012MNRAS.421.3060S} {421, 3060}

\bibitem[\protect\citeauthoryear{{Smith}}{{Smith}}{2015}]{Smith15}
{Smith} D.~J.~B.,  2015, preprint, \href
  {http://adsabs.harvard.edu/abs/2015arXiv150605630S} {} (\mn@eprint {arXiv}
  {1506.05630})

\bibitem[\protect\citeauthoryear{{Smith} et~al.,}{{Smith}
  et~al.}{2013}]{Smith+13}
{Smith} D.~J.~B.,  et~al., 2013, \mn@doi [\mnras] {10.1093/mnras/stt1737},
  \href {http://adsabs.harvard.edu/abs/2013MNRAS.436.2435S} {436, 2435}

\bibitem[\protect\citeauthoryear{{Smith} et~al.,}{{Smith}
  et~al.}{2014}]{Smith+14}
{Smith} D.~J.~B.,  et~al., 2014, \mn@doi [\mnras] {10.1093/mnras/stu1830},
  \href {http://adsabs.harvard.edu/abs/2014MNRAS.445.2232S} {445, 2232}

\bibitem[\protect\citeauthoryear{{Smol{\v c}i{\'c}} et~al.,}{{Smol{\v c}i{\'c}}
  et~al.}{2009}]{Smolcic+09}
{Smol{\v c}i{\'c}} V.,  et~al., 2009, \mn@doi [\apj]
  {10.1088/0004-637X/690/1/610}, \href
  {http://adsabs.harvard.edu/abs/2009ApJ...690..610S} {690, 610}

\bibitem[\protect\citeauthoryear{{Sutherland} \& {Saunders}}{{Sutherland} \&
  {Saunders}}{1992}]{Sutherland+Saunders92}
{Sutherland} W.,  {Saunders} W.,  1992, \mn@doi [\mnras]
  {10.1093/mnras/259.3.413}, \href
  {http://adsabs.harvard.edu/abs/1992MNRAS.259..413S} {259, 413}

\bibitem[\protect\citeauthoryear{{Tasse}, {van der Tol}, {van Zwieten}, {van
  Diepen}  \& {Bhatnagar}}{{Tasse} et~al.}{2013}]{Tasse+13}
{Tasse} C.,  {van der Tol} S.,  {van Zwieten} J.,  {van Diepen} G.,
  {Bhatnagar} S.,  2013, \mn@doi [\aap] {10.1051/0004-6361/201220882}, \href
  {http://cdsads.u-strasbg.fr/abs/2013A%26A...553A.105T} {553, A105}

\bibitem[\protect\citeauthoryear{{Taylor}}{{Taylor}}{2005}]{Taylor05}
{Taylor} M.~B.,  2005, in {Shopbell} P.,  {Britton} M.,   {Ebert} R.,  eds,
  Astronomical Society of the Pacific Conference Series Vol. 347, Astronomical
  Data Analysis Software and Systems XIV. p.~29

\bibitem[\protect\citeauthoryear{{Turner} \& {Shabala}}{{Turner} \&
  {Shabala}}{2015}]{Turner+Shabala15}
{Turner} R.~J.,  {Shabala} S.~S.,  2015, \mn@doi [\apj]
  {10.1088/0004-637X/806/1/59}, \href
  {http://adsabs.harvard.edu/abs/2015ApJ...806...59T} {806, 59}

\bibitem[\protect\citeauthoryear{{Virdee} et~al.,}{{Virdee}
  et~al.}{2013}]{Virdee+13}
{Virdee} J.~S.,  et~al., 2013, \mn@doi [\mnras] {10.1093/mnras/stt488}, \href
  {http://adsabs.harvard.edu/abs/2013MNRAS.432..609V} {432, 609}

\bibitem[\protect\citeauthoryear{{Whittam} et~al.,}{{Whittam}
  et~al.}{2013}]{Whittam+13}
{Whittam} I.~H.,  et~al., 2013, \mn@doi [\mnras] {10.1093/mnras/sts478}, \href
  {http://adsabs.harvard.edu/abs/2013MNRAS.429.2080W} {429, 2080}

\bibitem[\protect\citeauthoryear{{Williams} et~al.,}{{Williams}
  et~al.}{2016}]{Williams+16}
{Williams} W.~L.,  et~al., 2016, preprint, \href
  {http://adsabs.harvard.edu/abs/2016arXiv160501531W} {} (\mn@eprint {arXiv}
  {1605.01531})

\bibitem[\protect\citeauthoryear{{Willott}, {Rawlings}, {Blundell}  \&
  {Lacy}}{{Willott} et~al.}{1999}]{Willott+99}
{Willott} C.~J.,  {Rawlings} S.,  {Blundell} K.~M.,   {Lacy} M.,  1999, \mnras,
  309, 1017

\bibitem[\protect\citeauthoryear{{Willott}, {Rawlings}, {Archibald}  \&
  {Dunlop}}{{Willott} et~al.}{2002}]{Willott+02}
{Willott} C.~J.,  {Rawlings} S.,  {Archibald} E.~N.,   {Dunlop} J.~S.,  2002,
  \mnras, 331, 435

\bibitem[\protect\citeauthoryear{{Wilman} et~al.,}{{Wilman}
  et~al.}{2008}]{Wilman+08}
{Wilman} R.~J.,  et~al., 2008, \mnras, 388, 1335

\bibitem[\protect\citeauthoryear{{Yun}, {Reddy}  \& {Condon}}{{Yun}
  et~al.}{2001}]{Yun+01}
{Yun} M.~S.,  {Reddy} N.~A.,   {Condon} J.~J.,  2001, \mn@doi [\apj]
  {10.1086/323145}, \href {http://adsabs.harvard.edu/abs/2001ApJ...554..803Y}
  {554, 803}

\bibitem[\protect\citeauthoryear{{Zwart}, {Jarvis}, {Deane}, {Bonfield},
  {Knowles}, {Madhanpall}, {Rahmani}  \& {Smith}}{{Zwart}
  et~al.}{2014}]{Zwart+14}
{Zwart} J.~T.~L.,  {Jarvis} M.~J.,  {Deane} R.~P.,  {Bonfield} D.~G.,
  {Knowles} K.,  {Madhanpall} N.,  {Rahmani} H.,   {Smith} D.~J.~B.,  2014,
  \mn@doi [\mnras] {10.1093/mnras/stu053}, \href
  {http://adsabs.harvard.edu/abs/2014MNRAS.439.1459Z} {439, 1459}

\bibitem[\protect\citeauthoryear{{de Jong}, {Klein}, {Wielebinski}  \&
  {Wunderlich}}{{de Jong} et~al.}{1985}]{deJong+85}
{de Jong} T.,  {Klein} U.,  {Wielebinski} R.,   {Wunderlich} E.,  1985, \aap,
  \href {http://adsabs.harvard.edu/abs/1985A%26A...147L...6D} {147, L6}

\bibitem[\protect\citeauthoryear{{van Haarlem} et~al.,}{{van Haarlem}
  et~al.}{2013}]{vanHaarlem+13}
{van Haarlem} M.~P.,  et~al., 2013, \mn@doi [\aap]
  {10.1051/0004-6361/201220873}, \href
  {http://adsabs.harvard.edu/abs/2013A%26A...556A...2V} {556, A2}

\bibitem[\protect\citeauthoryear{{van Weeren} et~al.,}{{van Weeren}
  et~al.}{2016}]{vanWeeren+16}
{van Weeren} R.~J.,  et~al., 2016, \mn@doi [\apjs] {10.3847/0067-0049/223/1/2},
  \href {http://adsabs.harvard.edu/abs/2016ApJS..223....2V} {223, 2}

\bibitem[\protect\citeauthoryear{{van der Kruit}}{{van der
  Kruit}}{1971}]{vanderKruit71}
{van der Kruit} P.~C.,  1971, \aap, \href
  {http://adsabs.harvard.edu/abs/1971A%26A....15..110V} {15, 110}

\makeatother
\end{thebibliography}

\appendix
\section{They might be giants}
\label{app:tmbg}

\begin{figure*}
  \includegraphics[width=.32\linewidth]{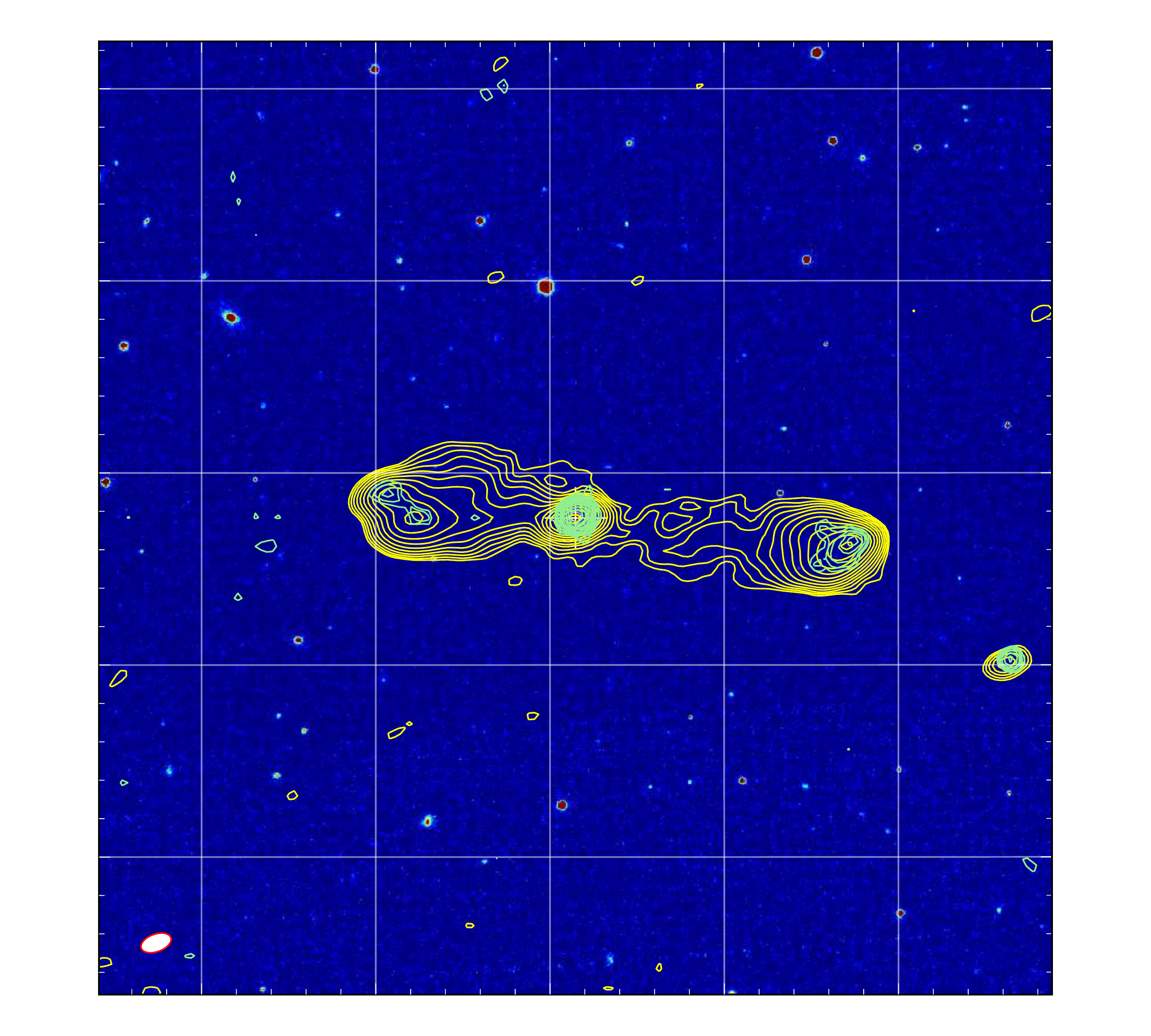}
  \includegraphics[width=.32\linewidth]{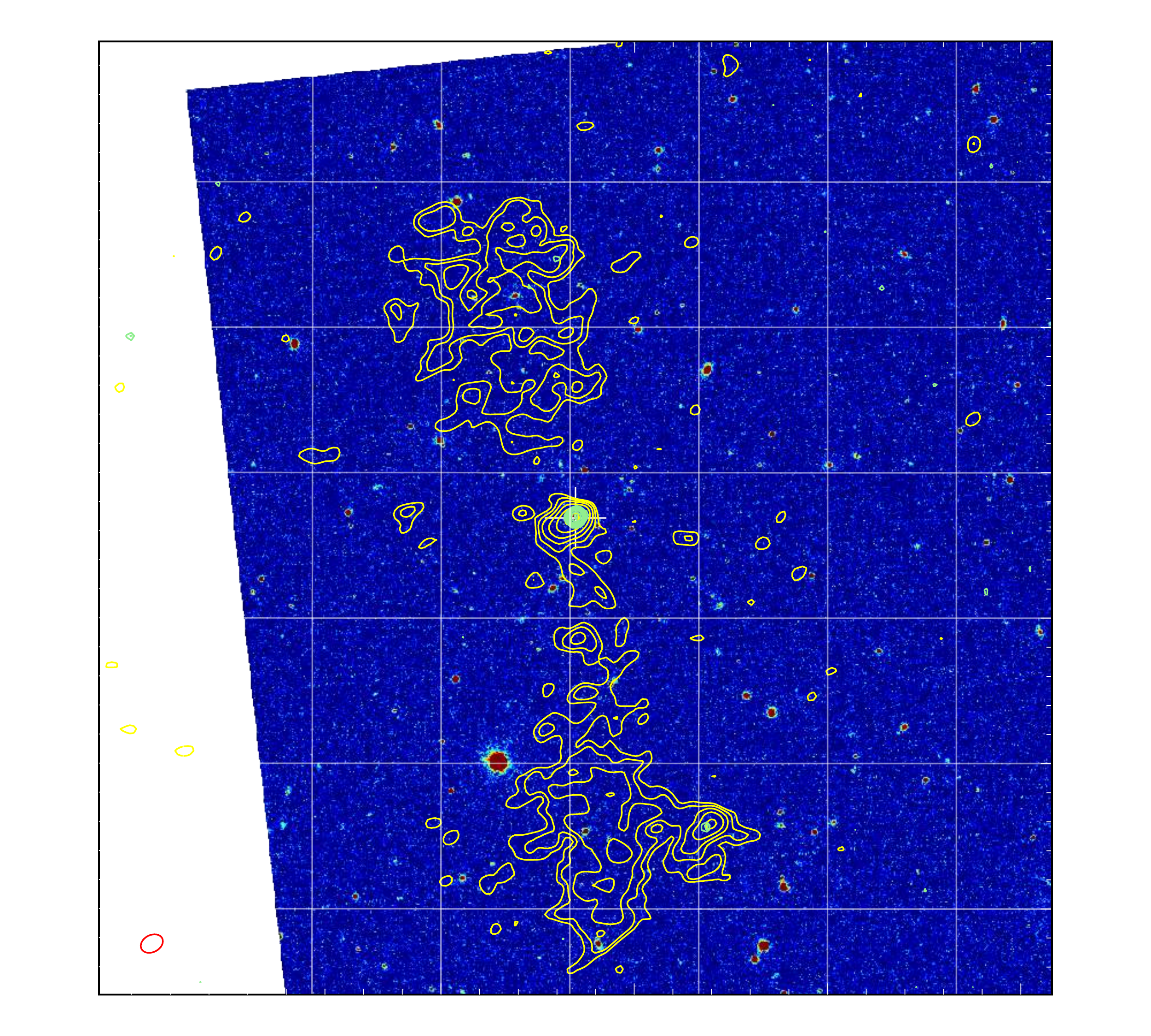}
  \includegraphics[width=.32\linewidth]{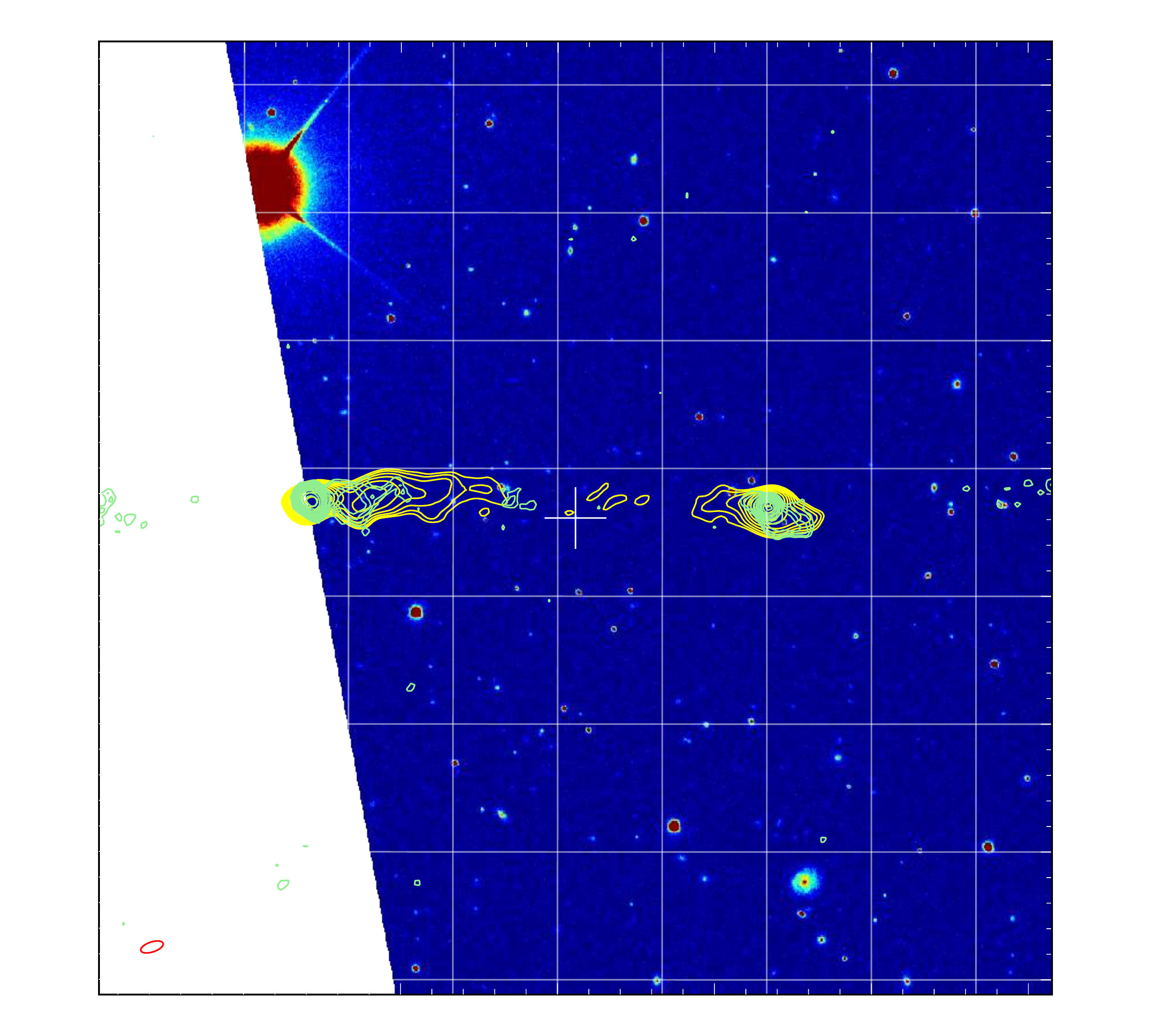}
  \includegraphics[width=.32\linewidth]{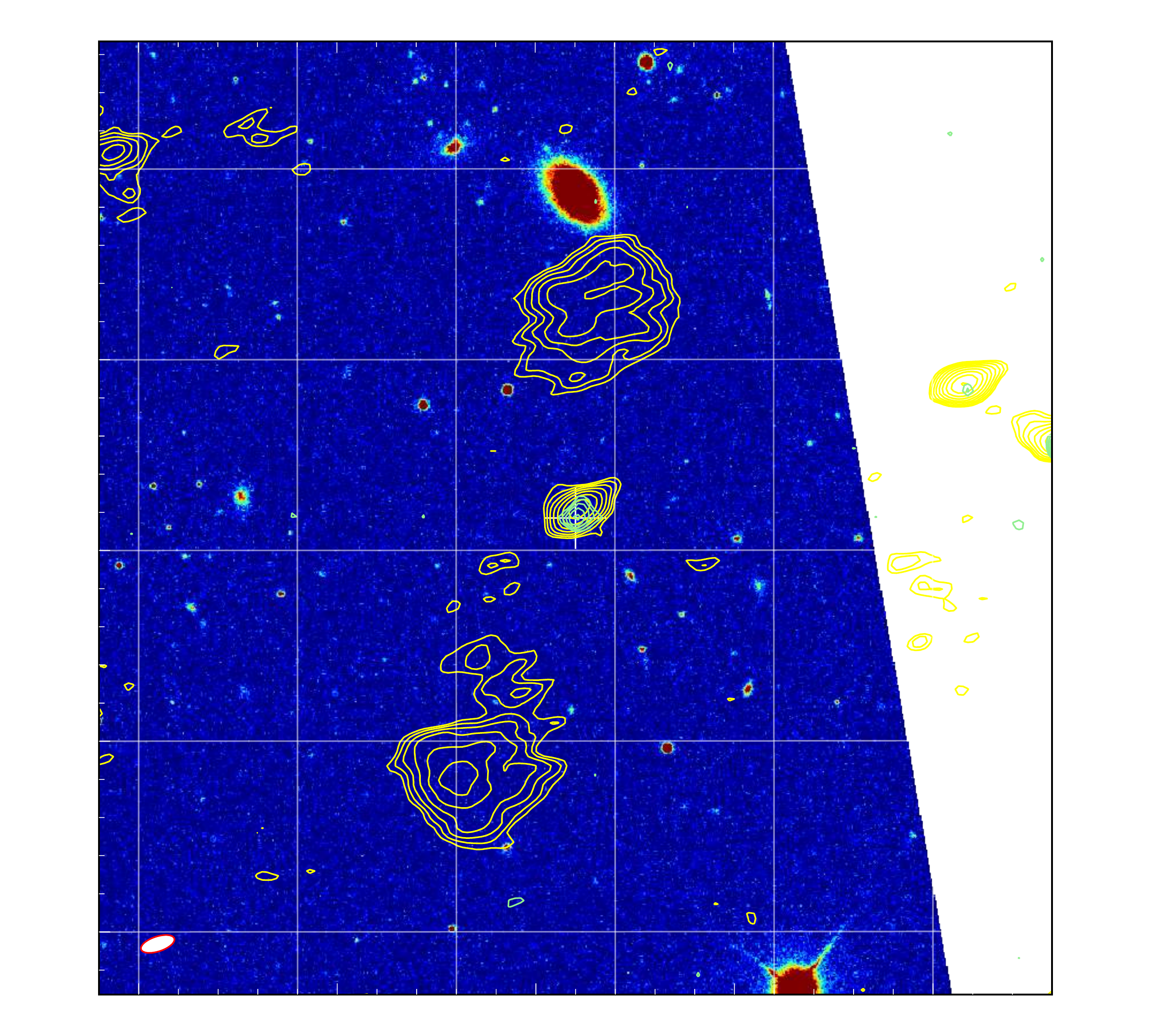}
  \includegraphics[width=.32\linewidth]{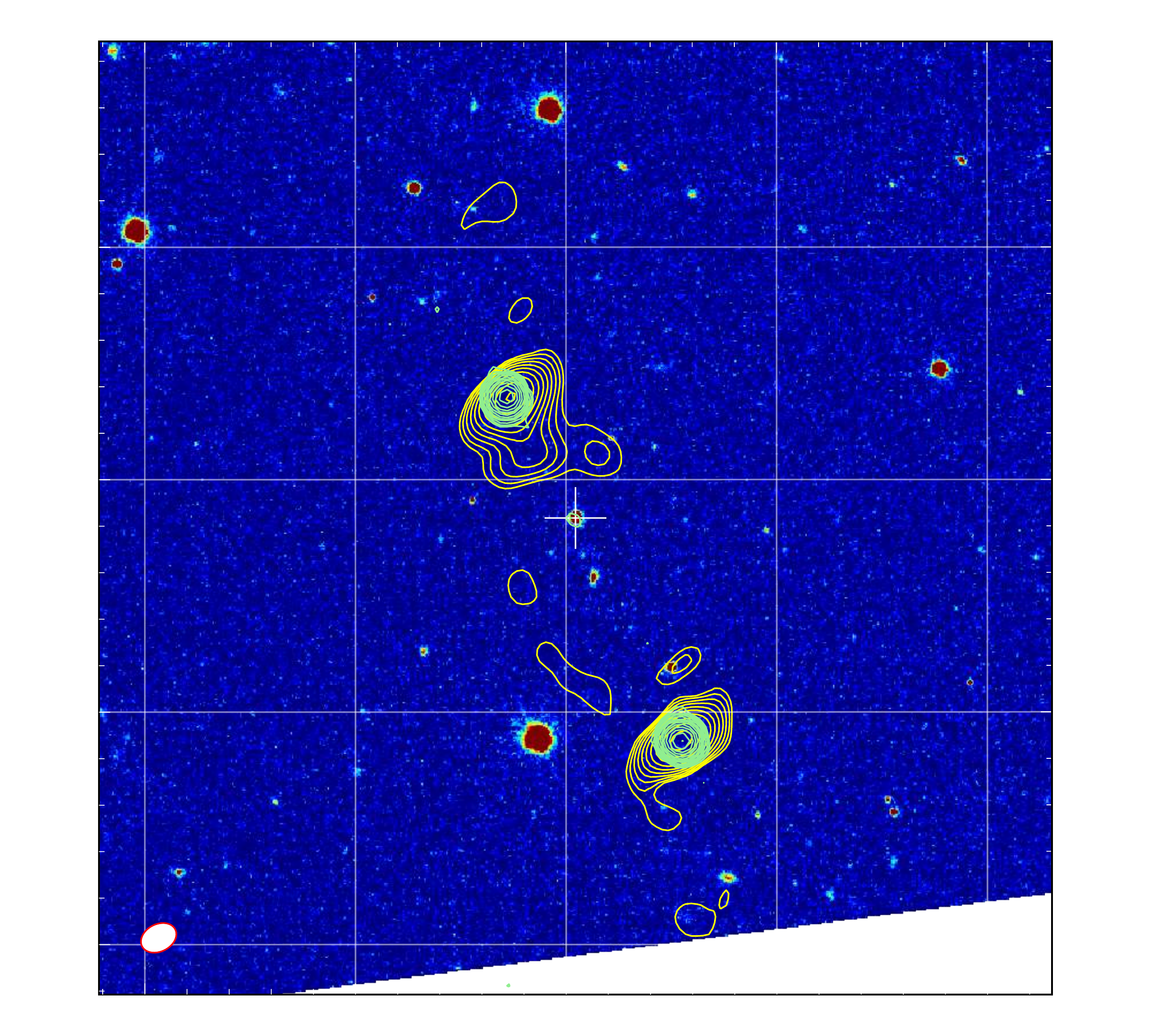}
  \includegraphics[width=.32\linewidth]{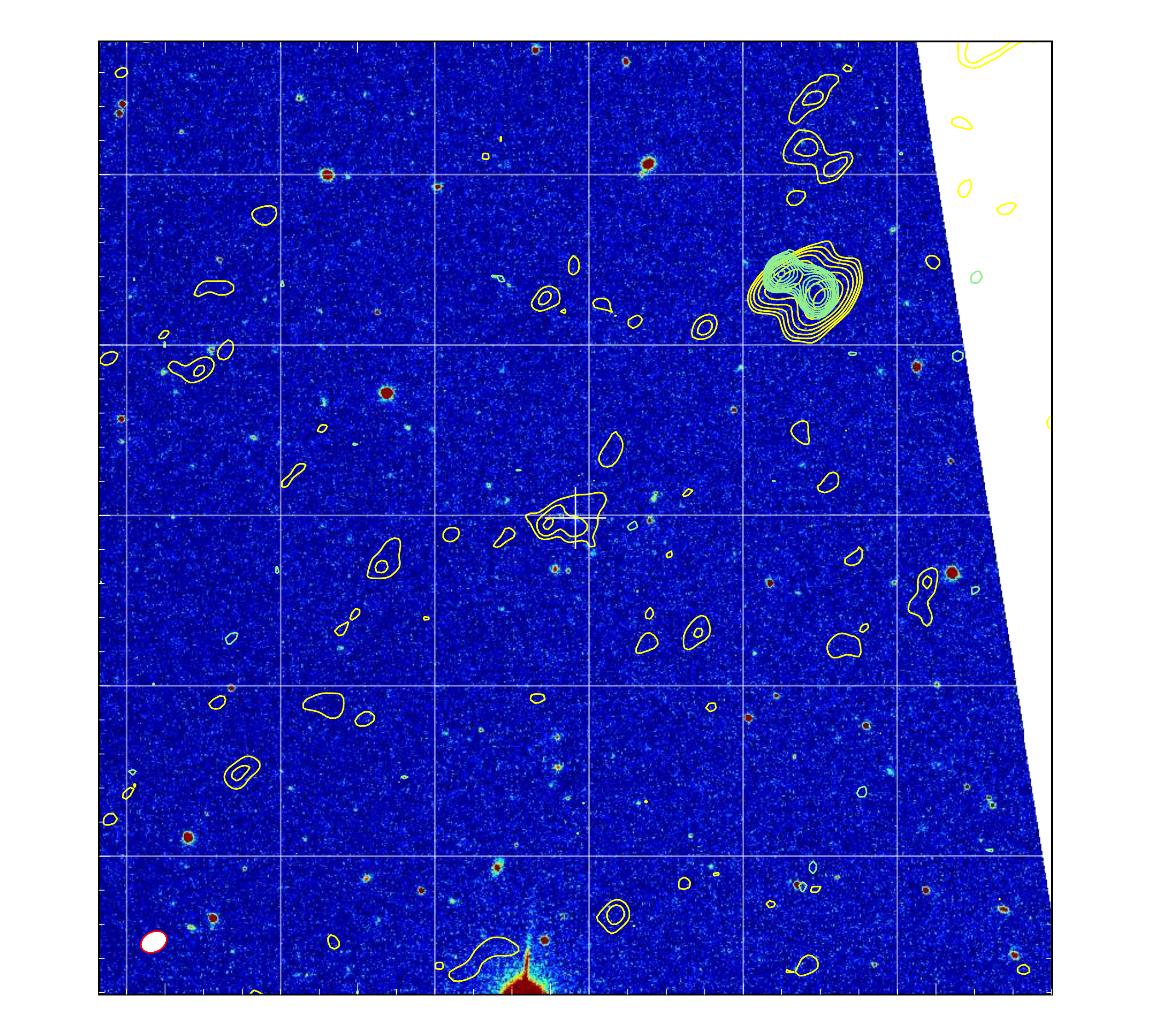}
  \includegraphics[width=.32\linewidth]{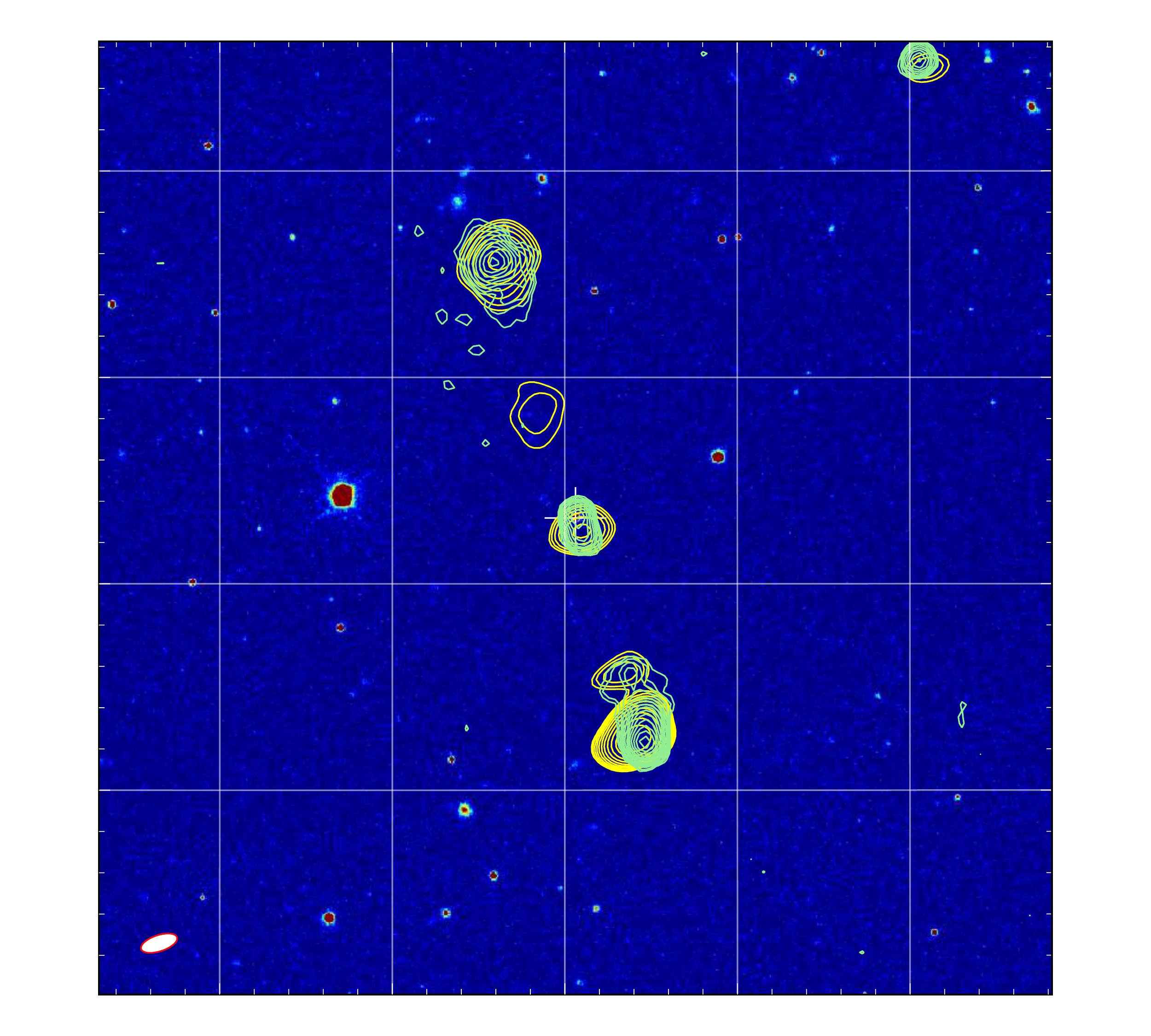}
\caption{`Postage stamp' images of the 7 optically identified objects
  at or close to 1 Mpc in physical size in the sample. Colours show
  SDSS $r$-band images: contours are LOFAR (yellow) and FIRST (green)
  contours starting at the local 3$\sigma$ value and increasing by
  steps of a factor 2. An ellipse in the bottom left-hand corner shows
  the LOFAR beam and a white cross the location of the optical
  identification. Grid lines are spaced by 1 arcmin in right ascension
  and declination: north is up. Top row, from left to right, are:
  SDSS J133127.82+250050.0 ($z = 0.8040$, $D = 0.9$
  Mpc), J134415.75+331719.1 ($z = 0.6863$, $D = 1.0$ Mpc), and
  J130451.41+245445.9 ($z = 0.6025$, $D = 1.0$ Mpc). Middle row:
  J132928.99+333810.1 ($z = 0.5404$, $D = 1.0$ Mpc), and
  J132735.32+350636.7 ($z = 0.5003$, $D = 1.4$ Mpc),
  J133744.35+251359.0 ($z = 0.6816$, $D = 1.2$ Mpc). Bottom:
  J131443.83+273741.3 ($z = 0.4179$, $D = 1.1$ Mpc).}
  \label{fig:tmbg}
  \end{figure*}

Fig. \ref{fig:tmbg} shows postage stamps of the 7 objects discussed in
Section \ref{sec:pdd} as candidate optically identified giants in the
field. Of these, only one (the one identified with SDSS
J133744.35+2513359.0) seems likely to be spurious. This is one of the two
non-composite sources of the seven and its large size is a result of
{\sc pybdsm} associating it with a large region of
low-surface-brightness emission. Further investigation with
low-resolution imaging would be required to say whether the extended
emission is really associated with the LOFAR source. All the remaining
6 objects are clear double radio sources. Most of the optical
identifications are unambiguous, and the source identified with SDSS
J132735.32+350636.7, which is less certain, is identified with the
brightest plausible galaxy in a crowded field and so is unlikely to be
at lower redshift. Three of the optical identifications
(SDSS J133127.82+250050.0, SDSS J134415.75+331719.1,
SDSS J130451.41+245445.9) are quasars with spectroscopic redshifts;
they are therefore likely substantially larger in physical size than
they appear. SDSS J131443.83+273741.3 is a radio galaxy with a
spectroscopic redshift; the other redshifts are photometric.

\section{Remnants and non-remnants}
\label{app:remnant}

\begin{figure*}
  \includegraphics[width=.32\linewidth]{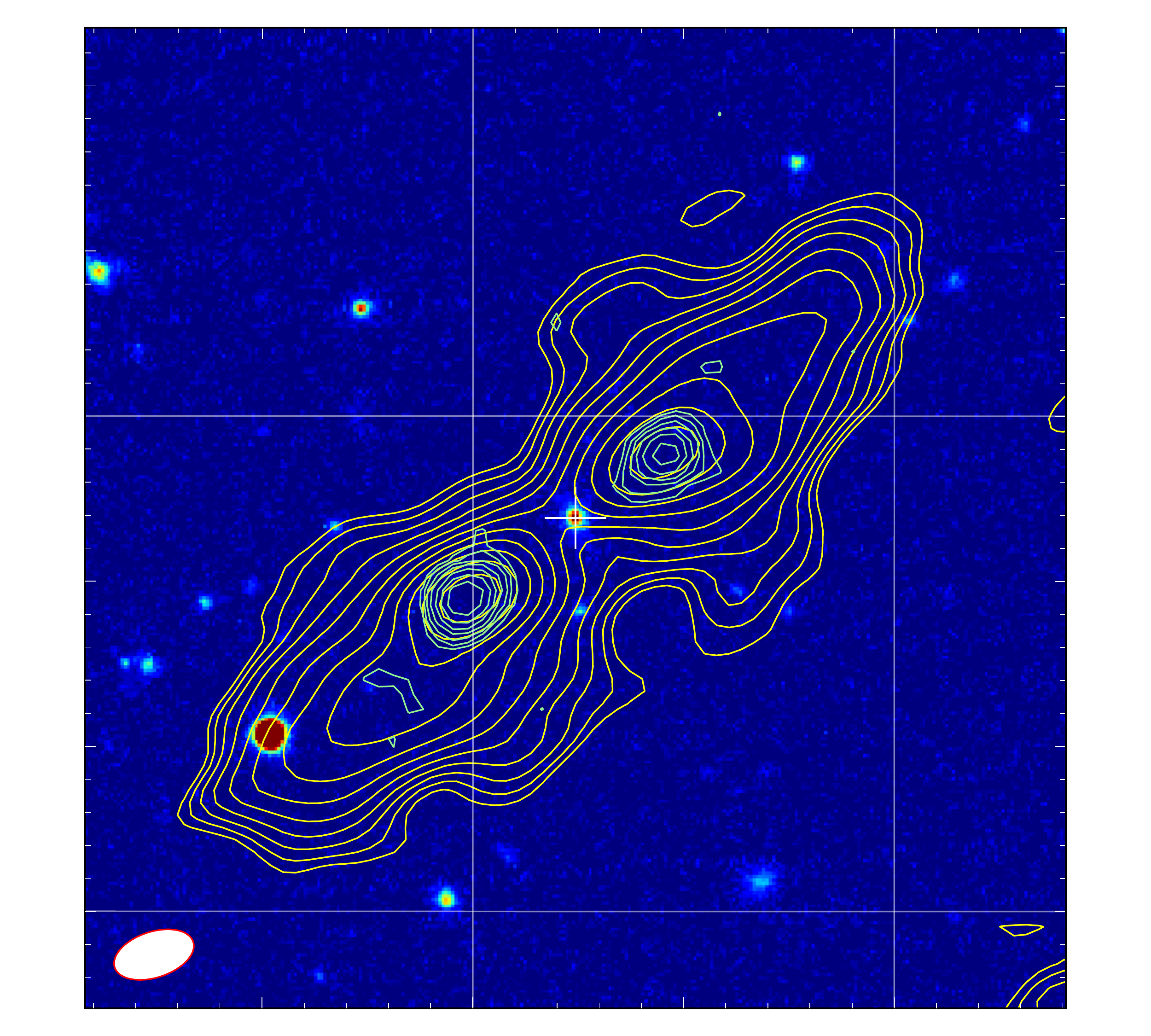}
  \includegraphics[width=.32\linewidth]{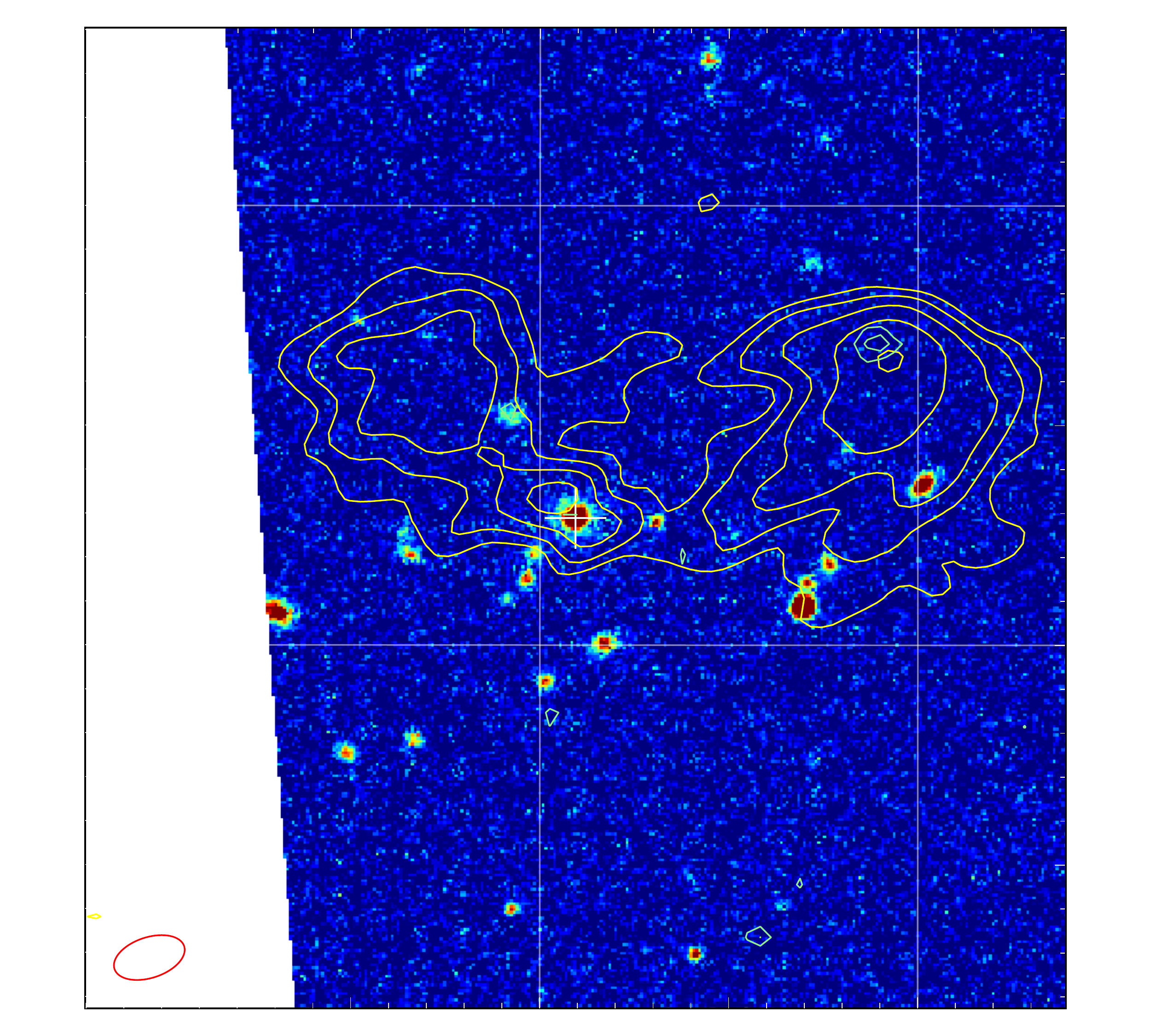}
  \includegraphics[width=.32\linewidth]{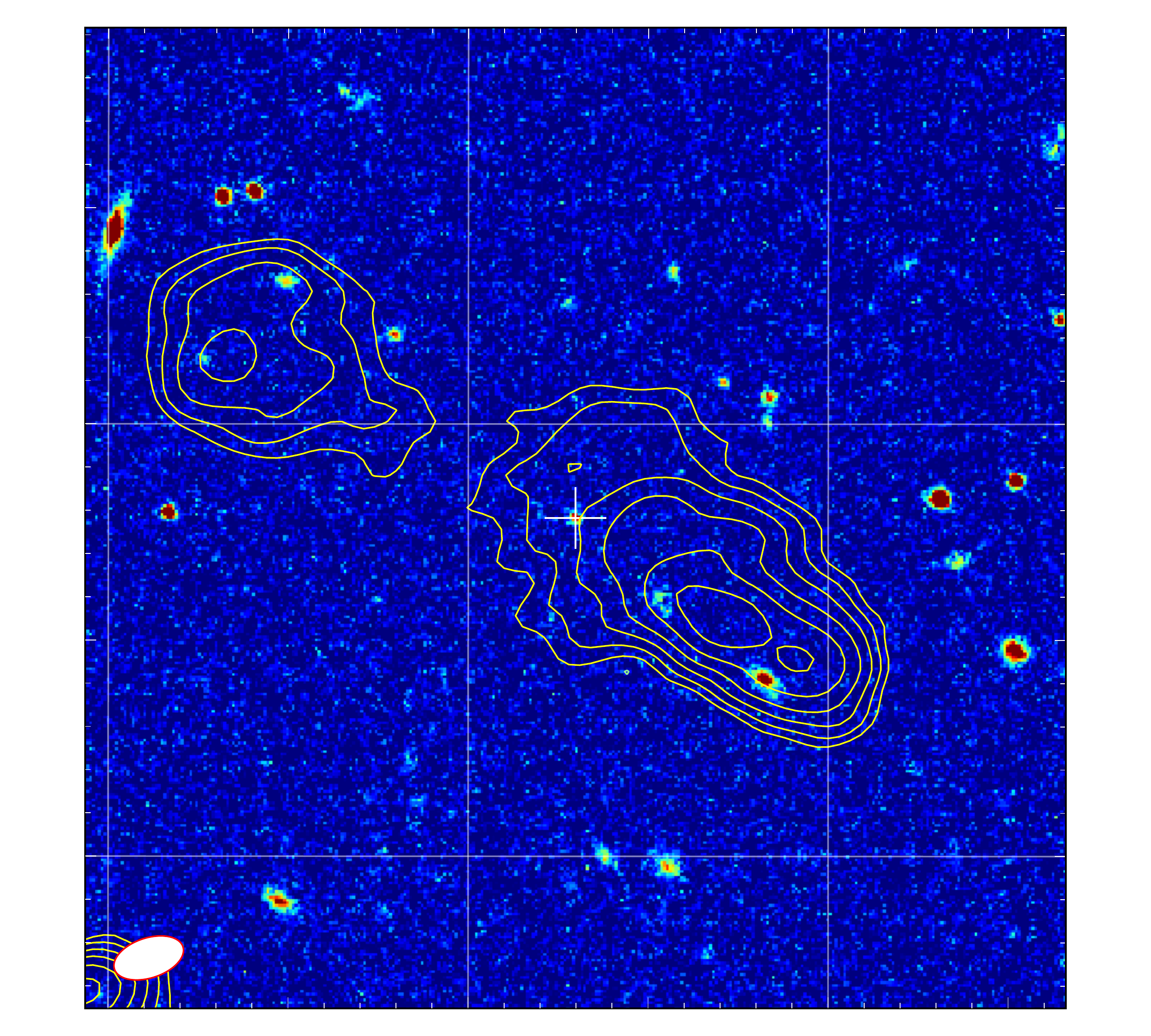}
  \includegraphics[width=.32\linewidth]{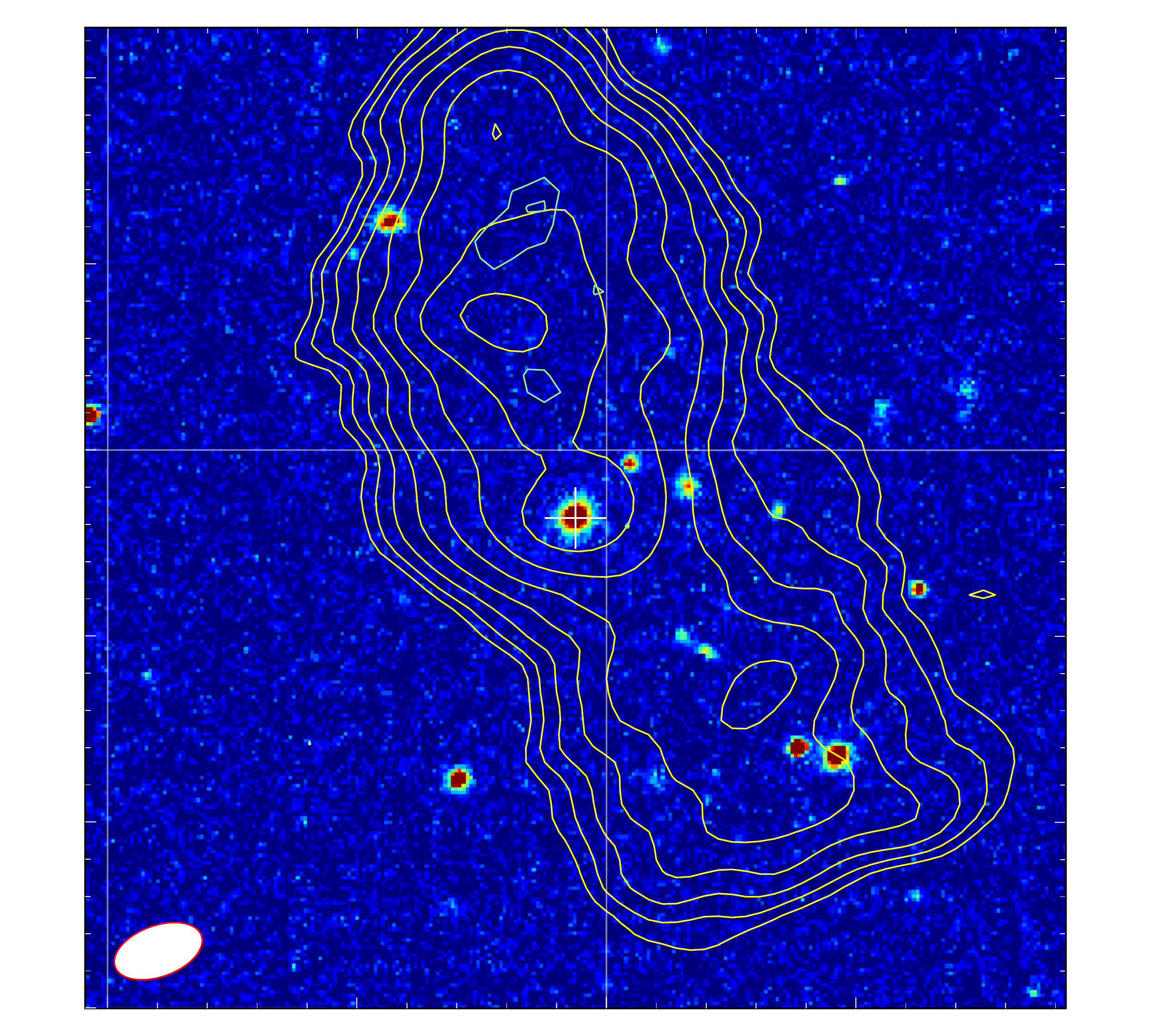}
  \includegraphics[width=.32\linewidth]{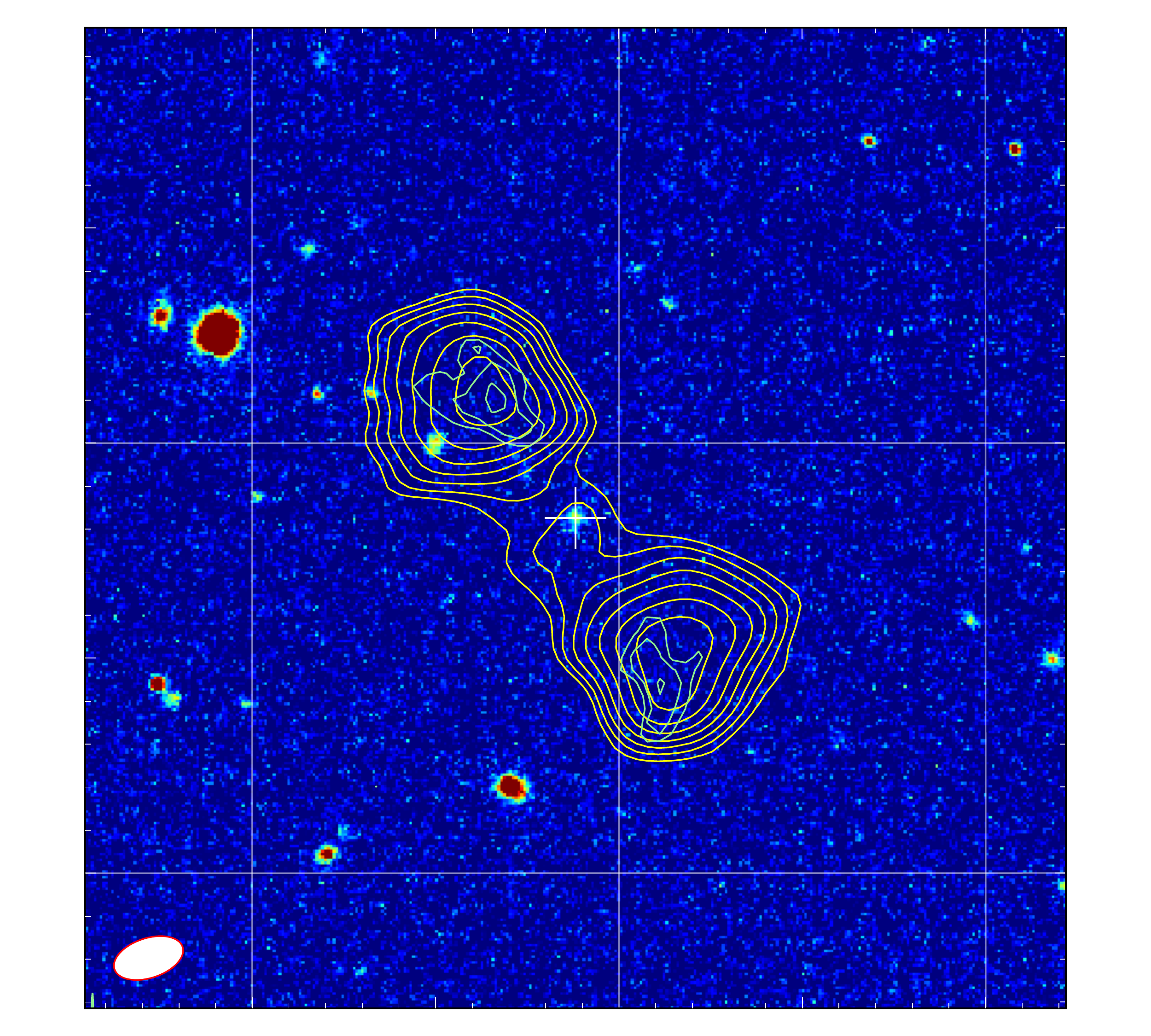}
  \includegraphics[width=.32\linewidth]{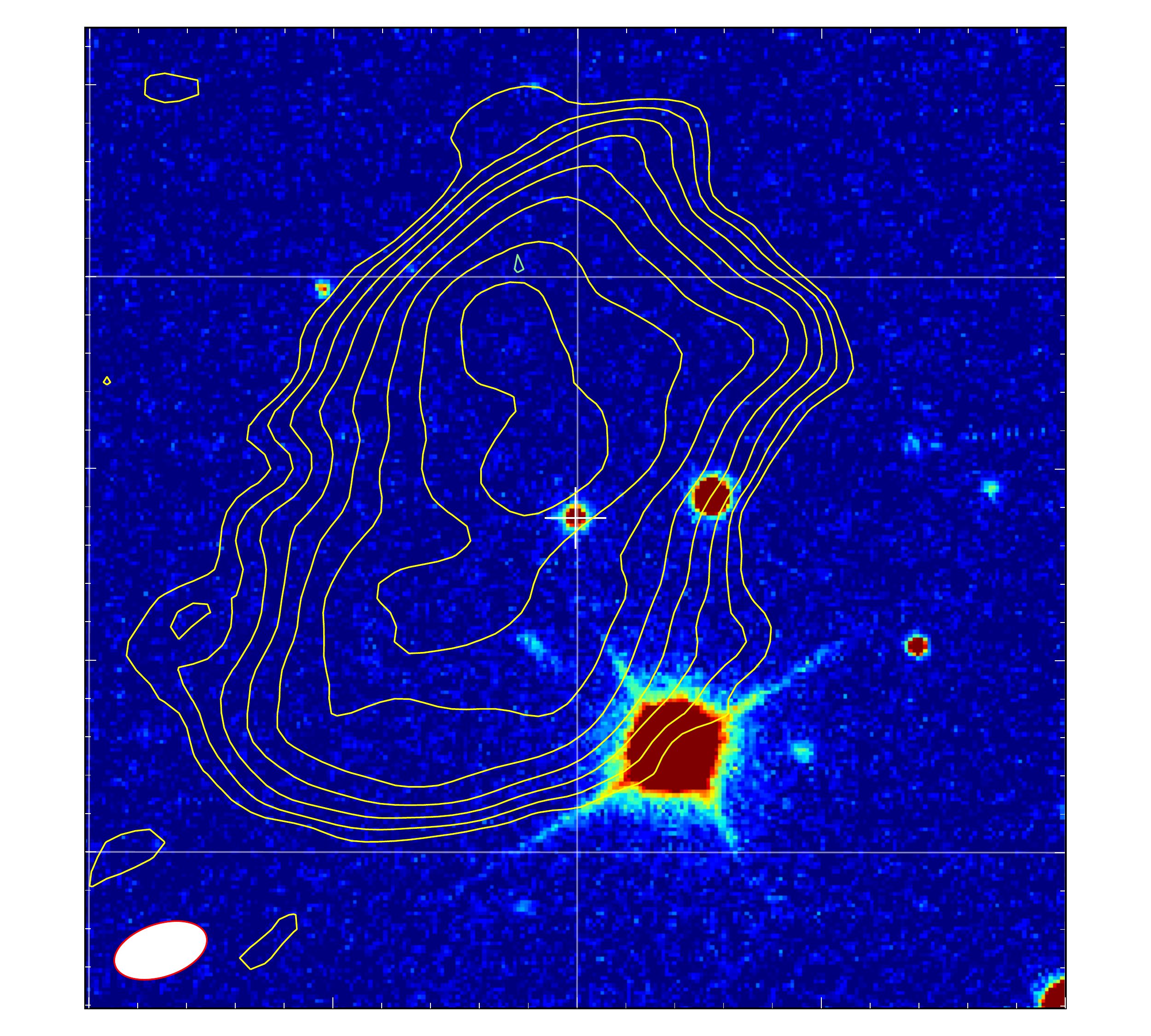}
  \includegraphics[width=.32\linewidth]{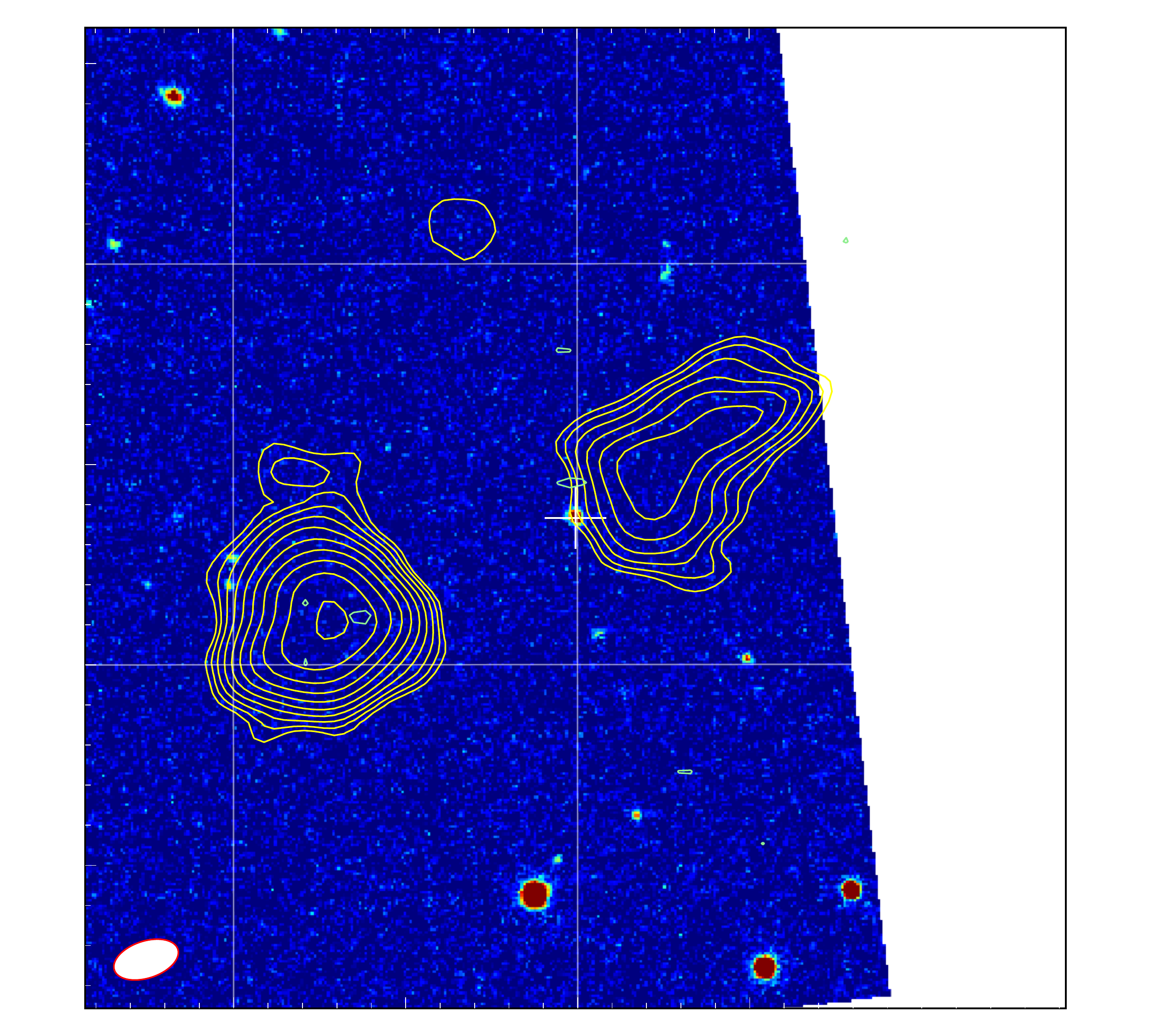}
  \includegraphics[width=.32\linewidth]{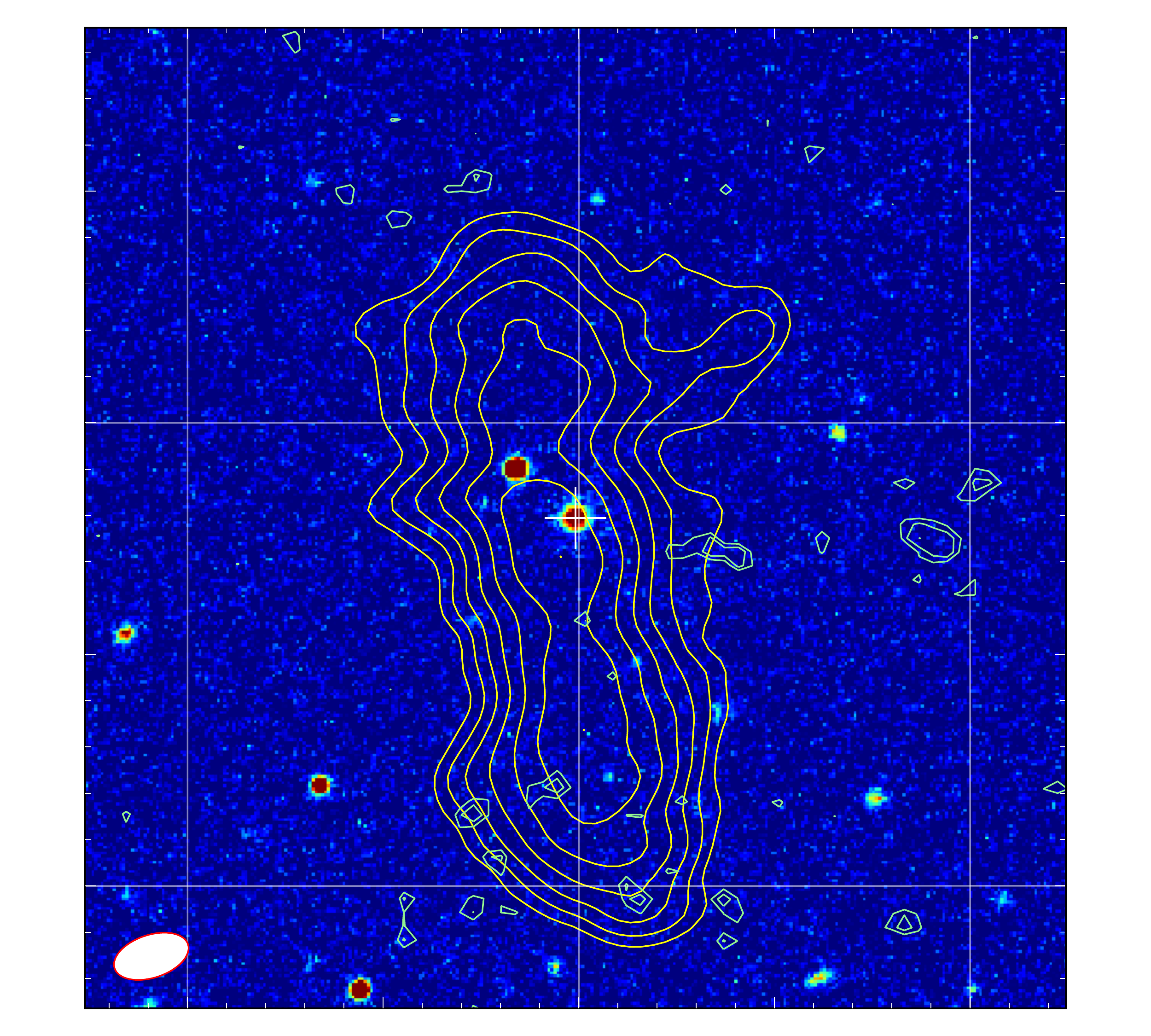}
  \includegraphics[width=.32\linewidth]{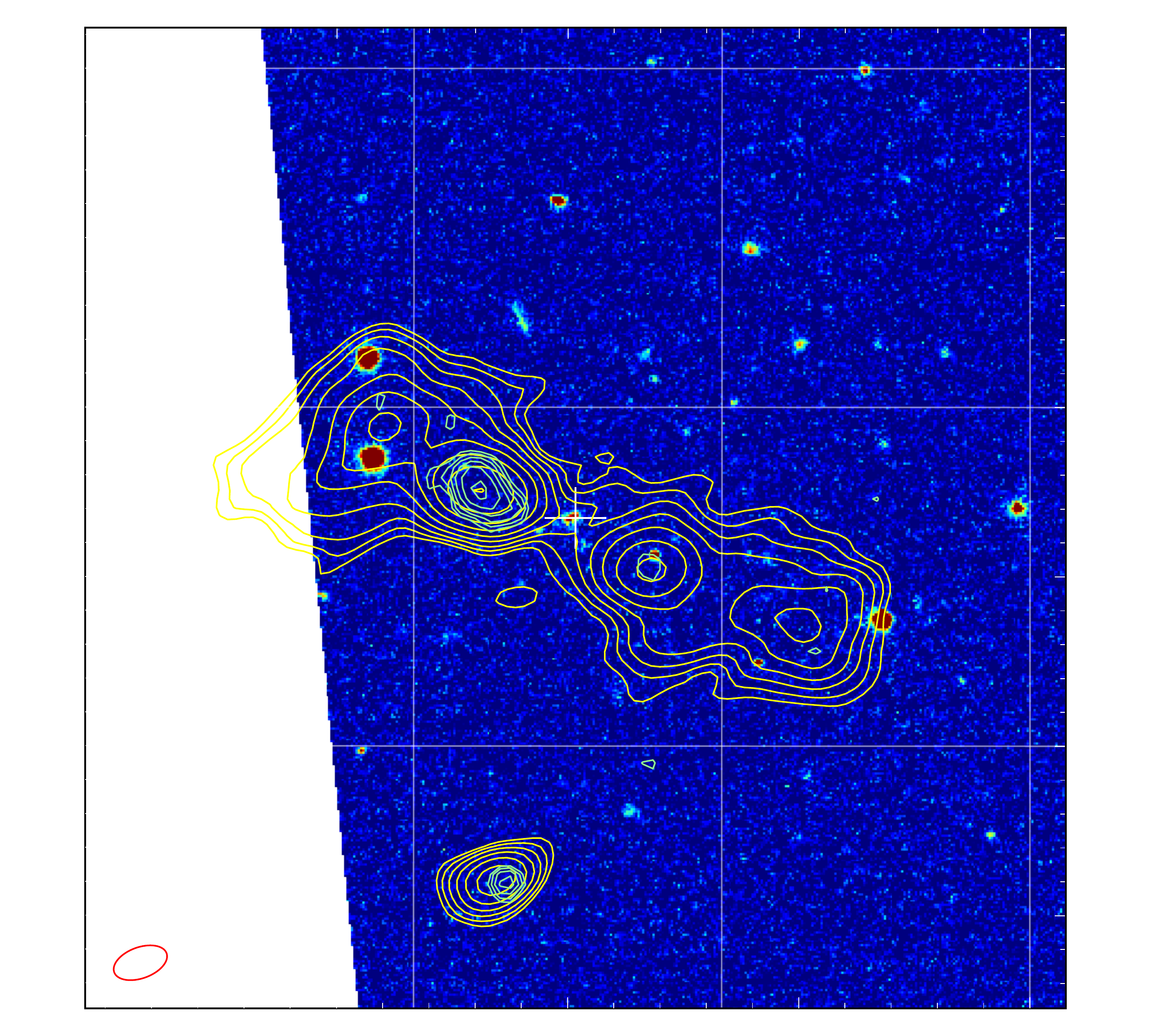}
  \includegraphics[width=.32\linewidth]{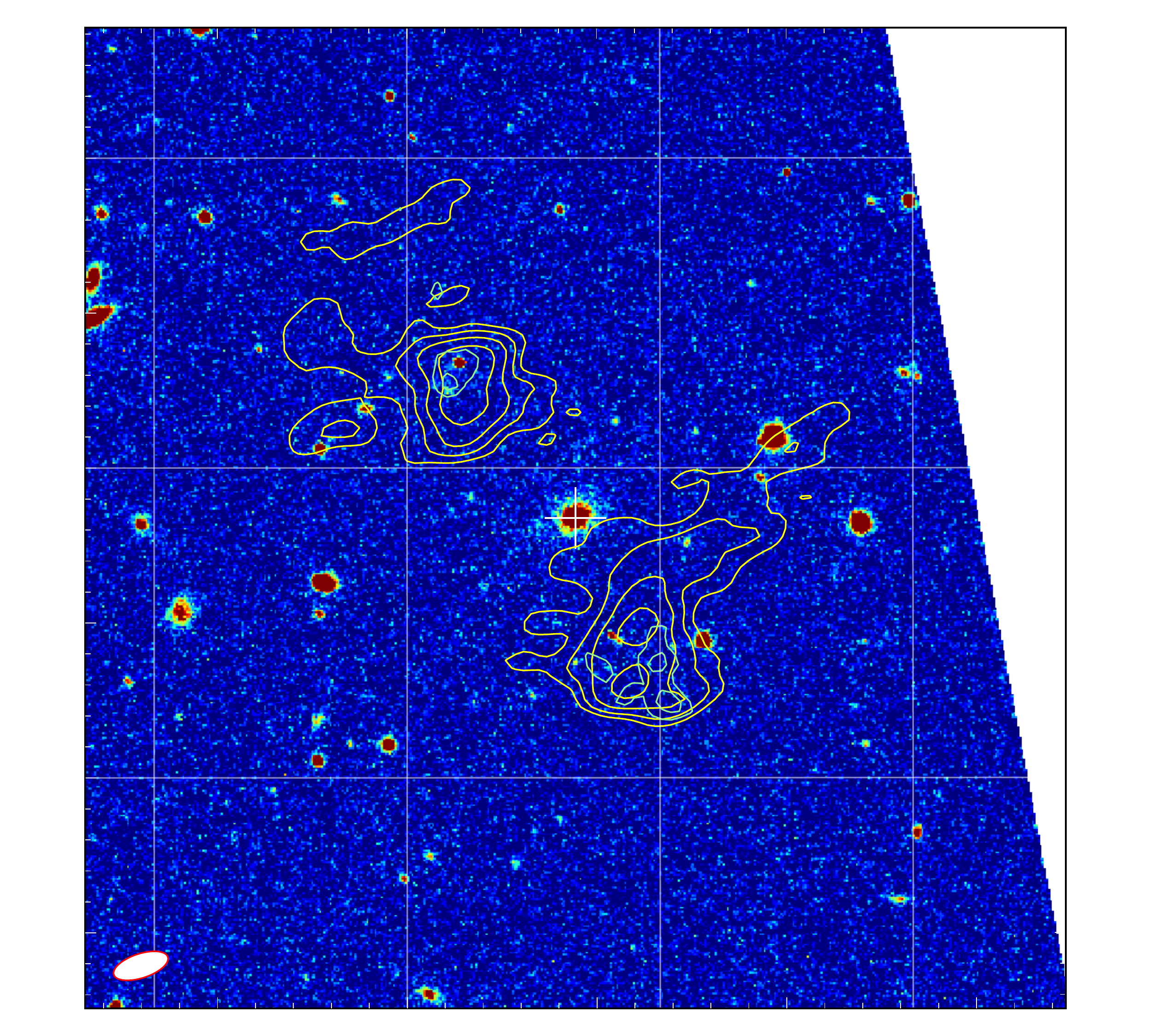}
  \includegraphics[width=.32\linewidth]{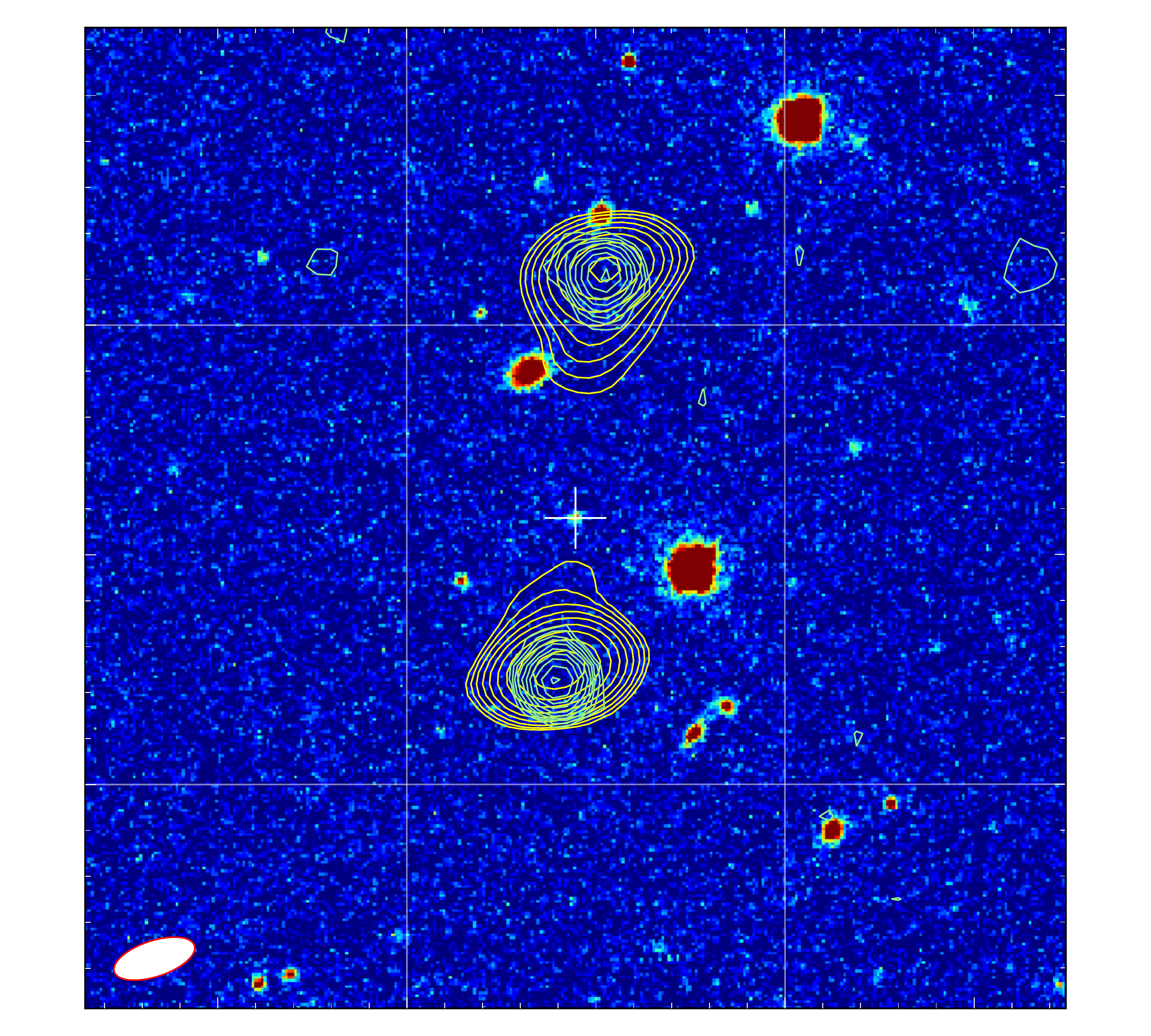}
  \includegraphics[width=.32\linewidth]{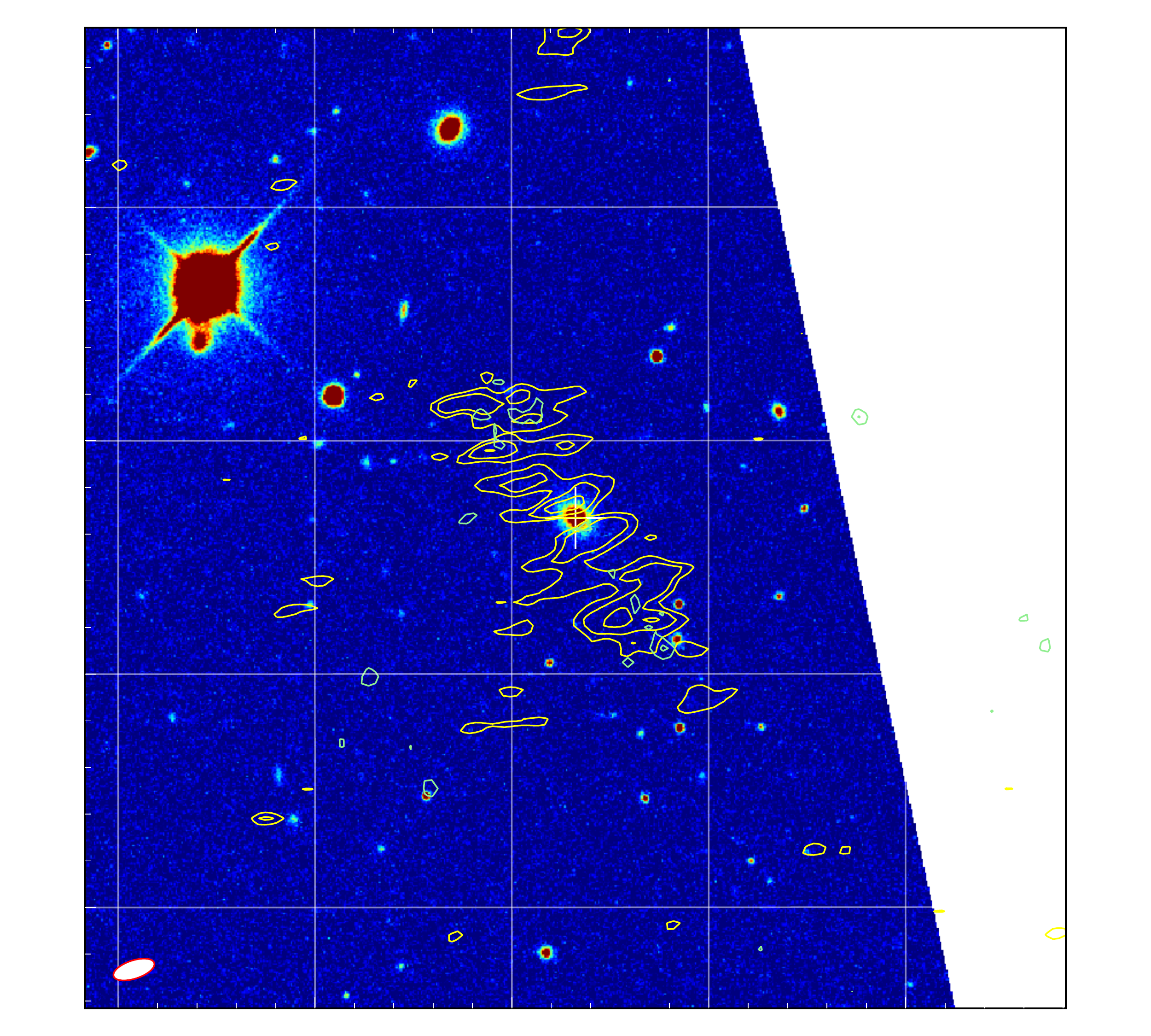}
  \caption{Postage stamps of example candidate remnant sources. Colours,
    contours and lines as in Fig. \ref{fig:tmbg}. From left to right, top to
  bottom: SDSS J125147.03+314047.6, J125311.62+304017.3,
J125930.81+333646.9, J130004.25+263652.7,
J130332.47+312949.5, J130532.02+315634.8,
J130916.02+305121.9, J131040.03+322047.6,
J131405.90+243240.3, J133057.34+351650.2, J133422.21+343634.8, and J134802.70+322940.1.}
  \label{fig:remnant}
  \end{figure*}

Candidate remnant sources (Fig.\ \ref{fig:remnant}) show a wide range
of morphologies. Many of them have no compact emission in FIRST at
all, though some have hotspots; as noted in the main text, the
presence of hotspots does not in itself indicate that a source is not
a recent remnant. Sources where double lobes cannot be distinguished
are quite common (e.g. SDSS J130532.02+315634.8), as are sources like
SDSS J134802.70+322940.1 where LOFAR surface brightness sensitivity at
full resolution limits our ability to image the source structure.
These may be examples of the `relaxed double' class \citep{Leahy93}.
There are also a number of objects like SDSS J125147.03+314047.6 and
SDSS J131405.90+243240.3, whose `inner hotspots' may indicate a
restarting source rather than a remnant radio galaxy. LOFAR provides a
good opportunity for the study of restarting (double-double) sources
\citep[e.g.][]{Orru+15}.

\begin{figure*}
  \includegraphics[width=.32\linewidth]{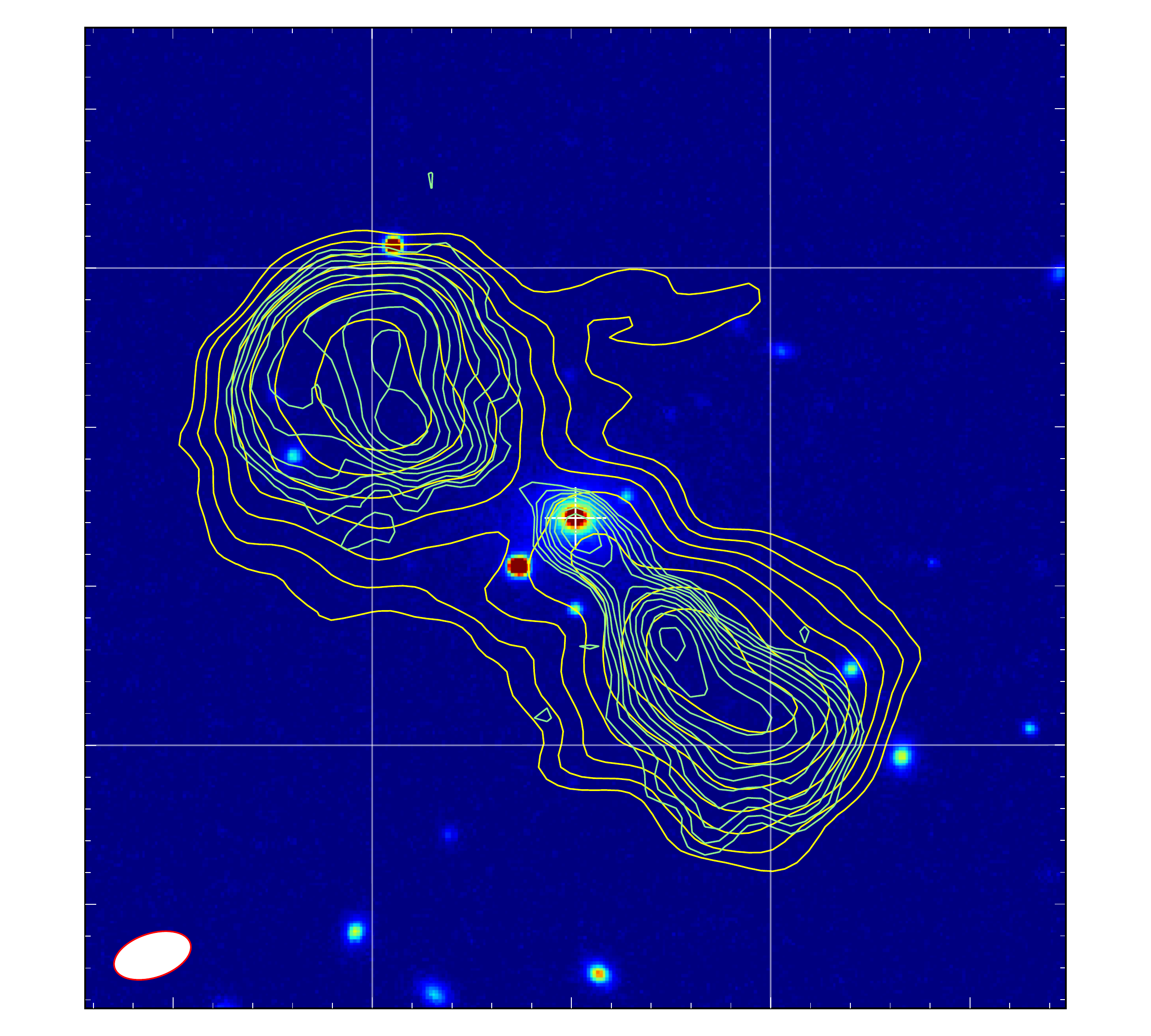}
  \includegraphics[width=.32\linewidth]{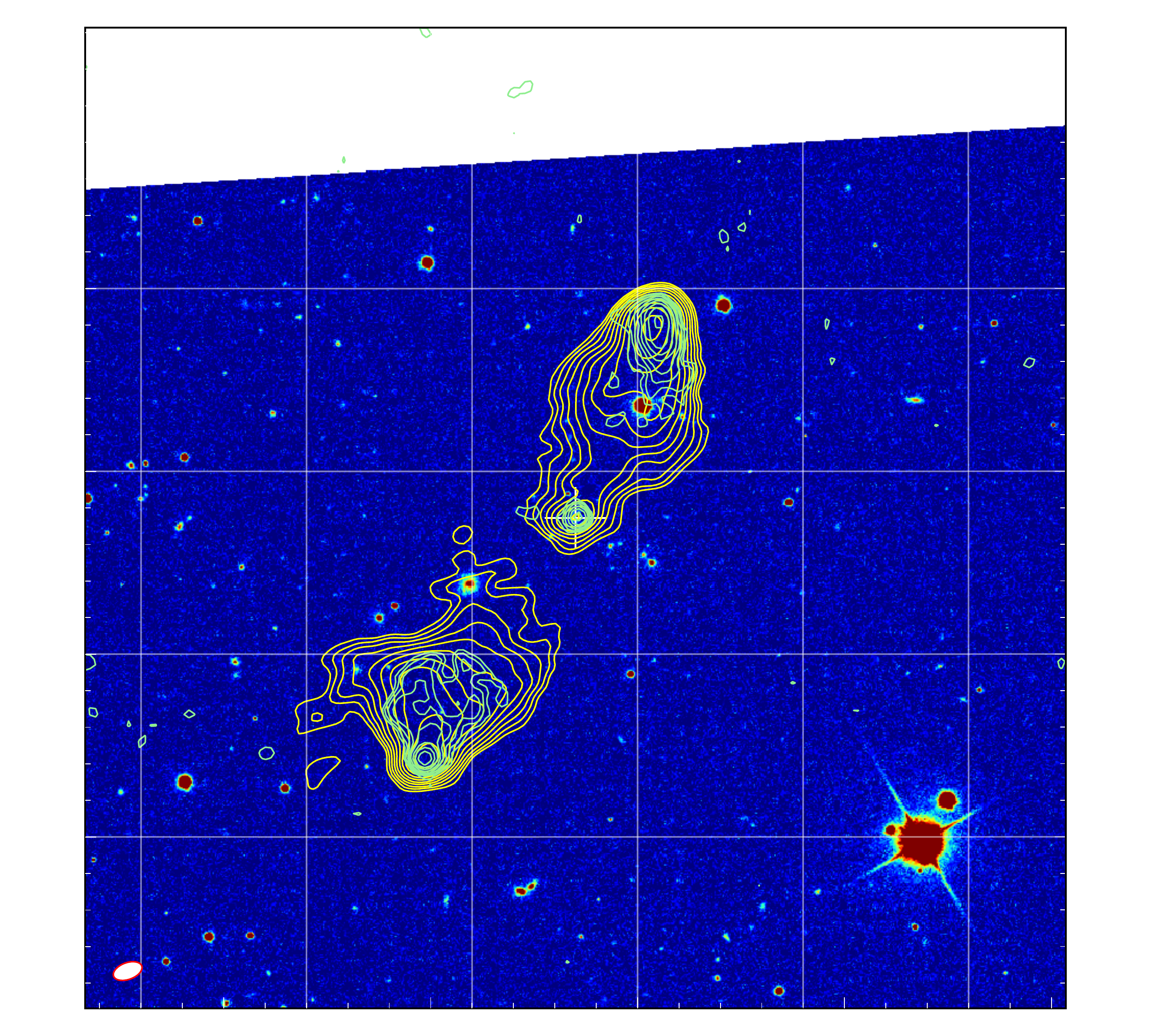}
  \includegraphics[width=.32\linewidth]{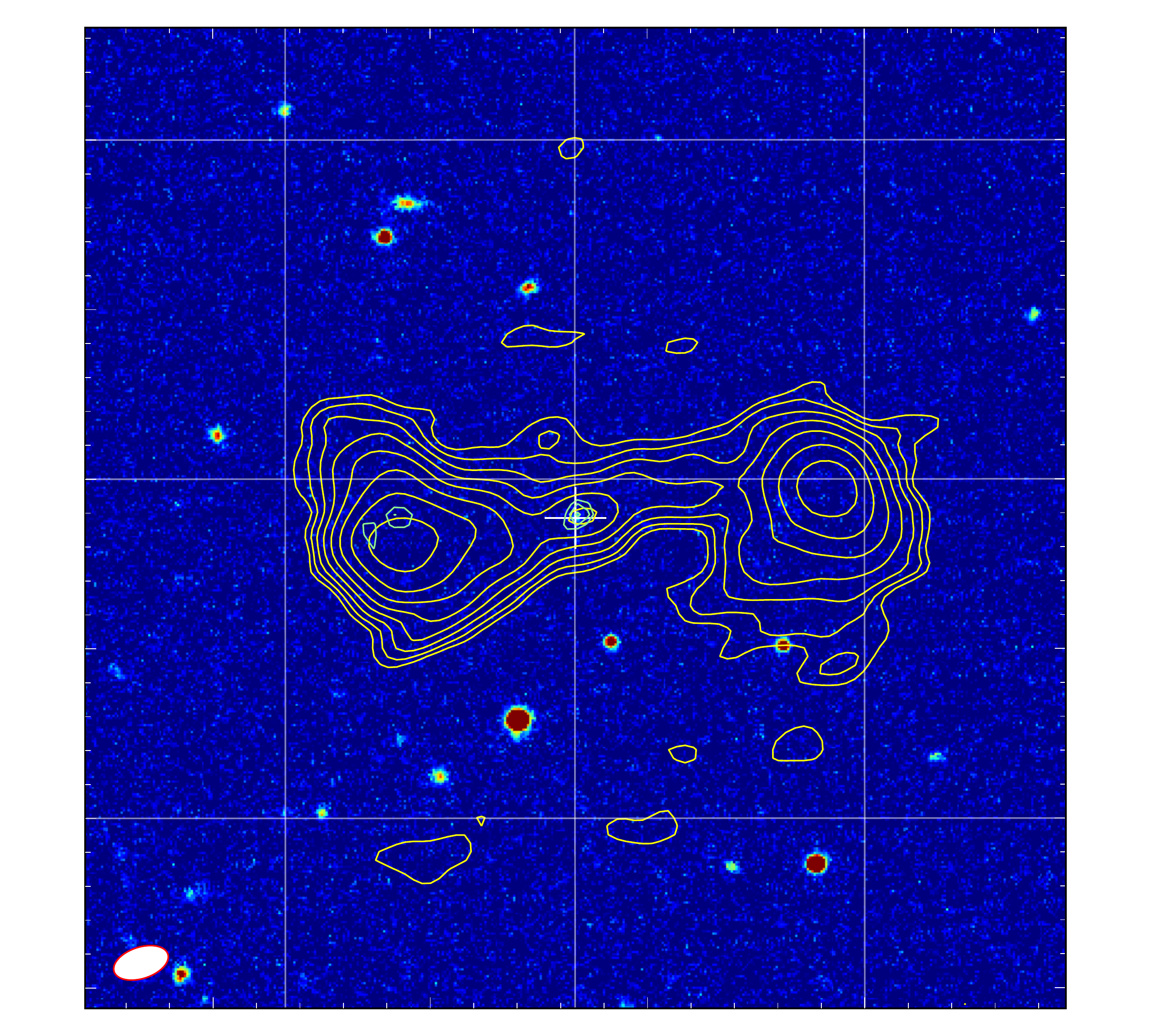}
  \includegraphics[width=.32\linewidth]{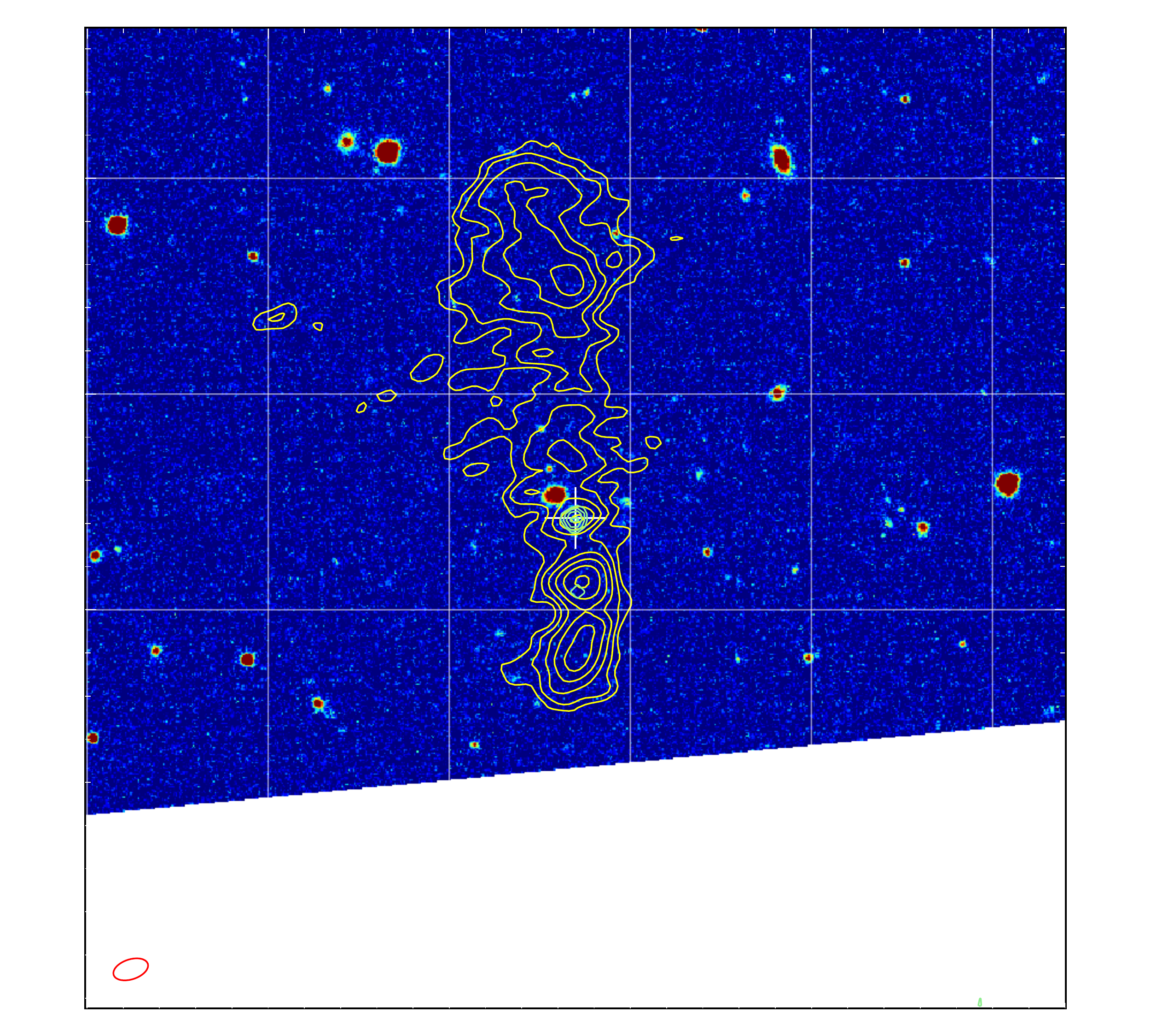}
  \includegraphics[width=.32\linewidth]{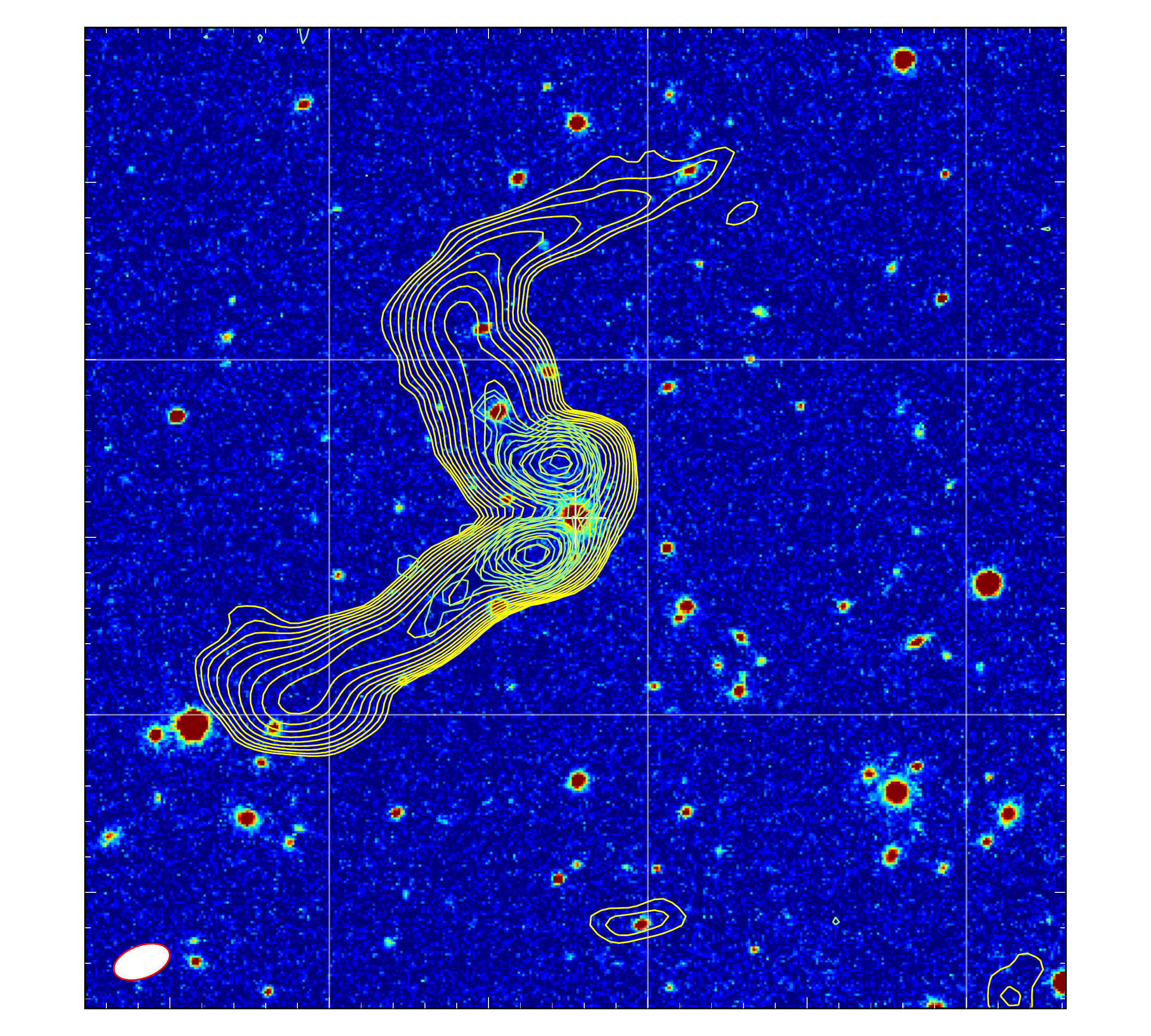}
  \includegraphics[width=.32\linewidth]{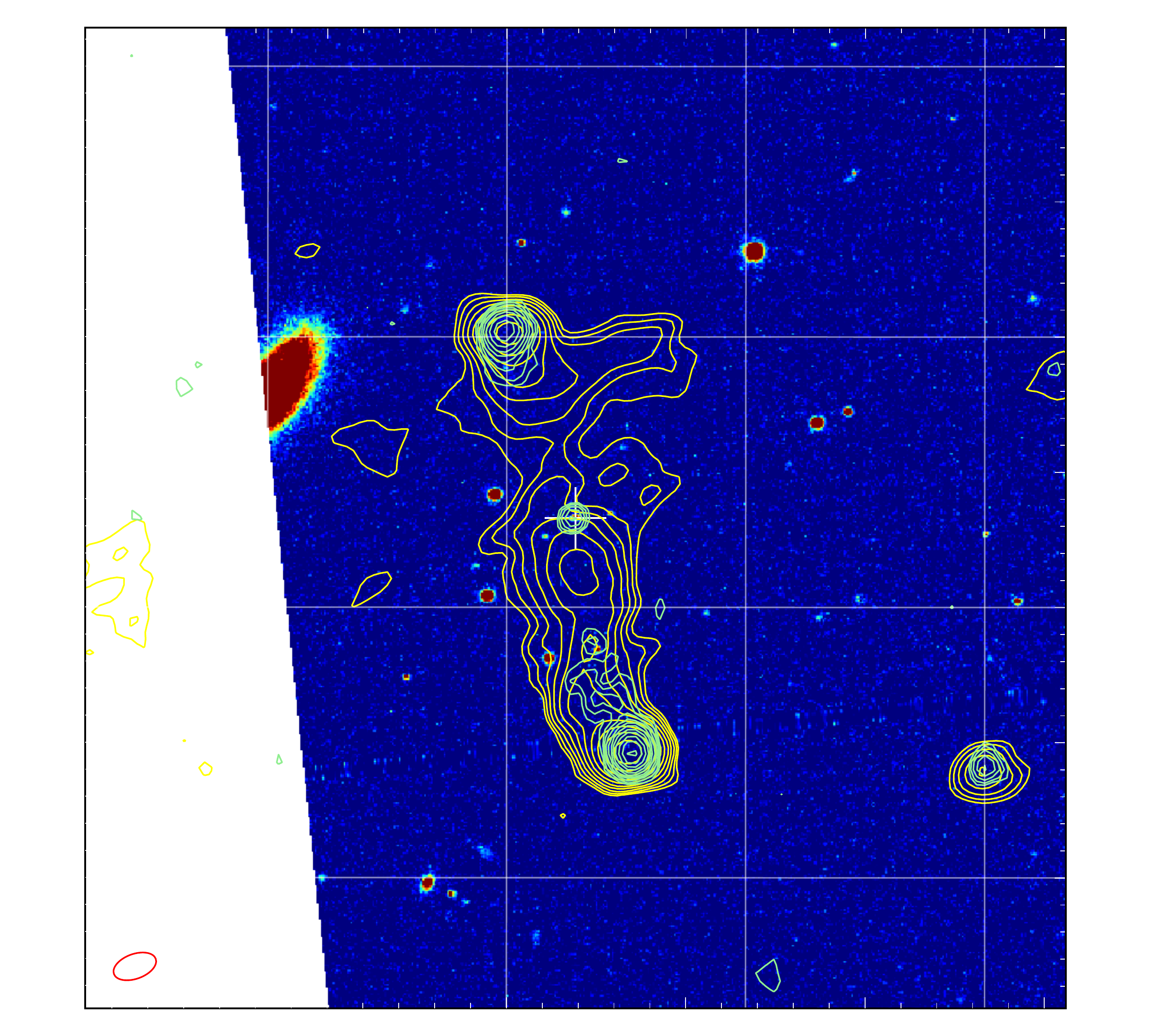}
  \includegraphics[width=.32\linewidth]{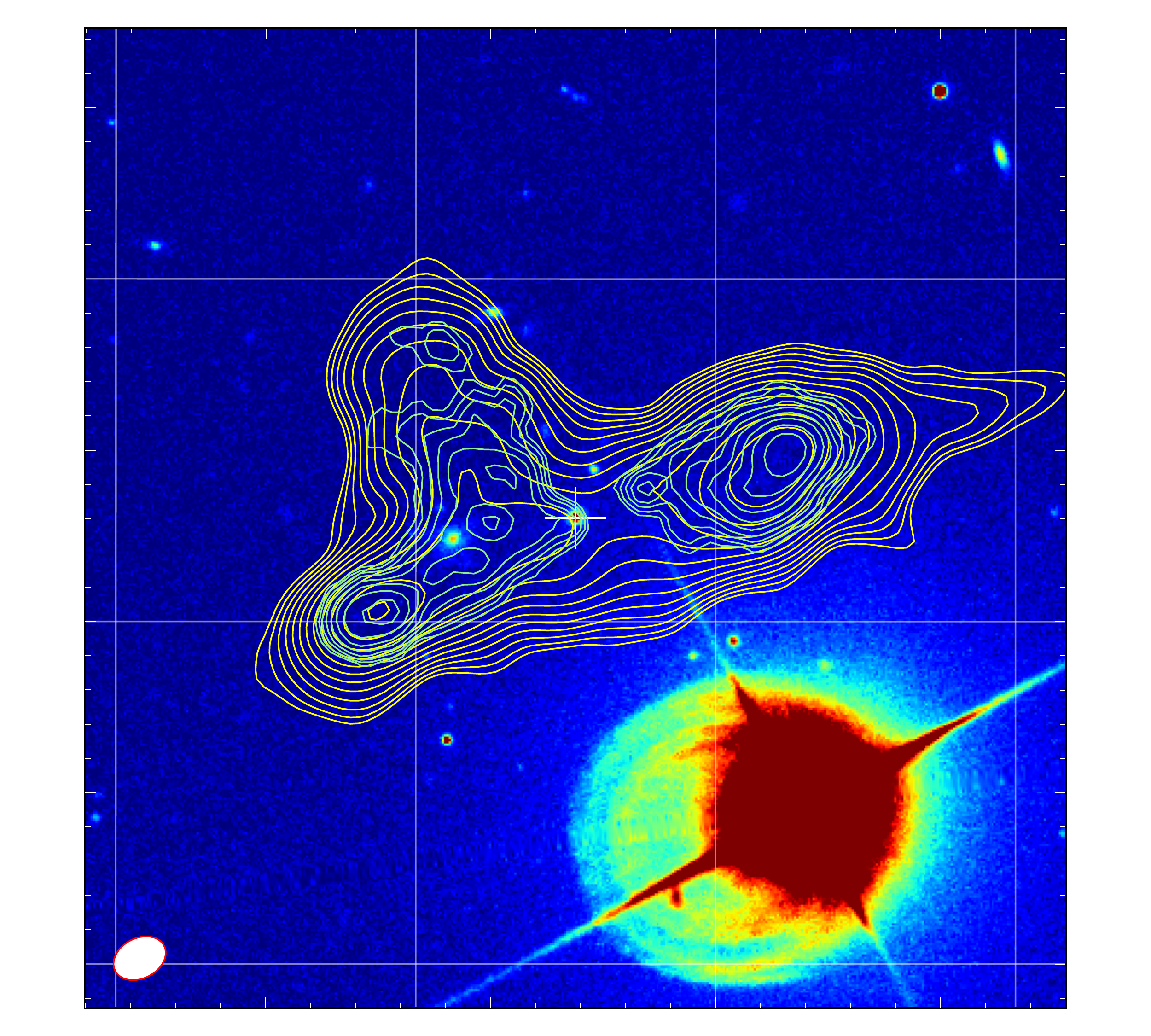}
  \includegraphics[width=.32\linewidth]{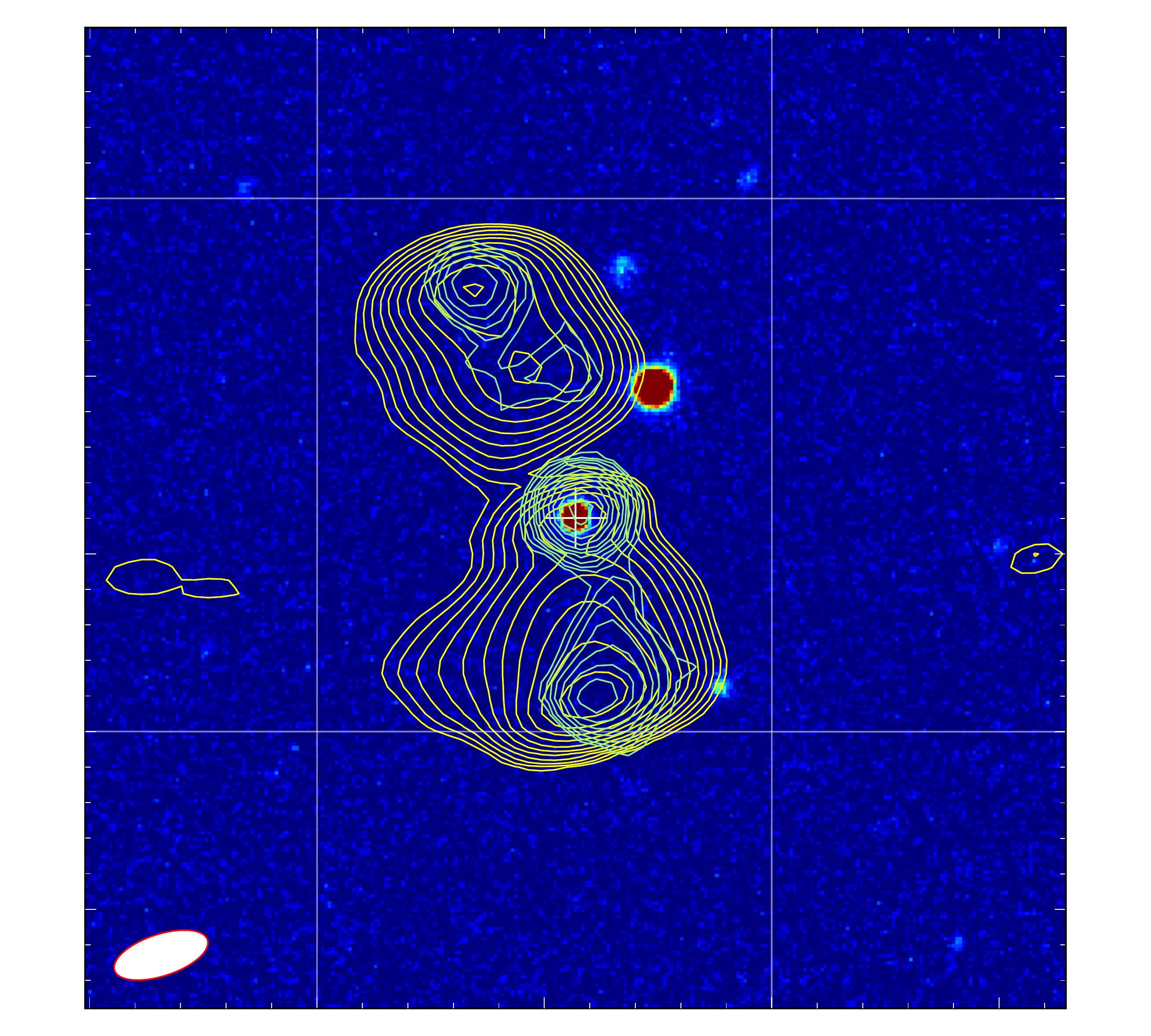}
  \includegraphics[width=.32\linewidth]{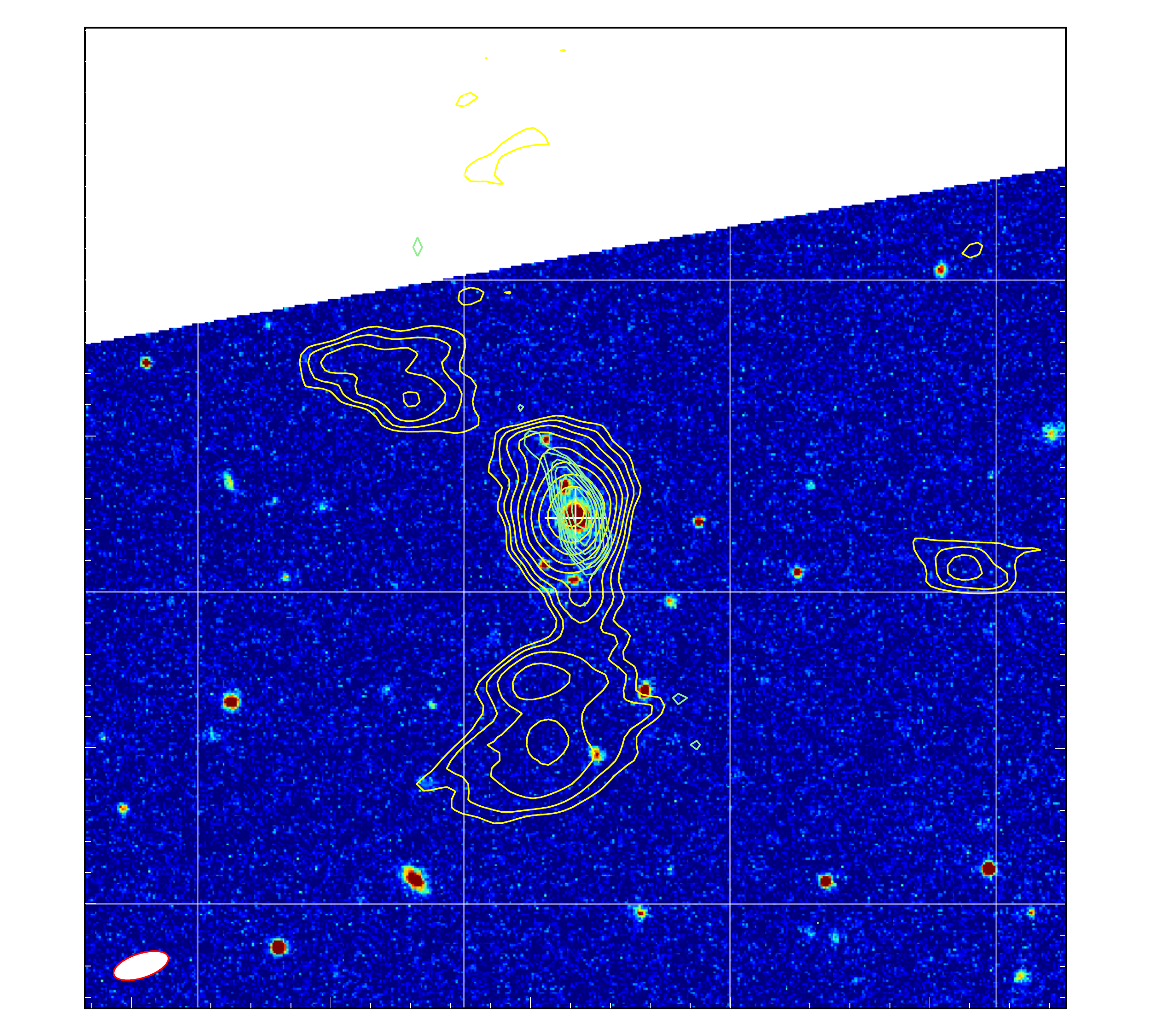}
  \includegraphics[width=.32\linewidth]{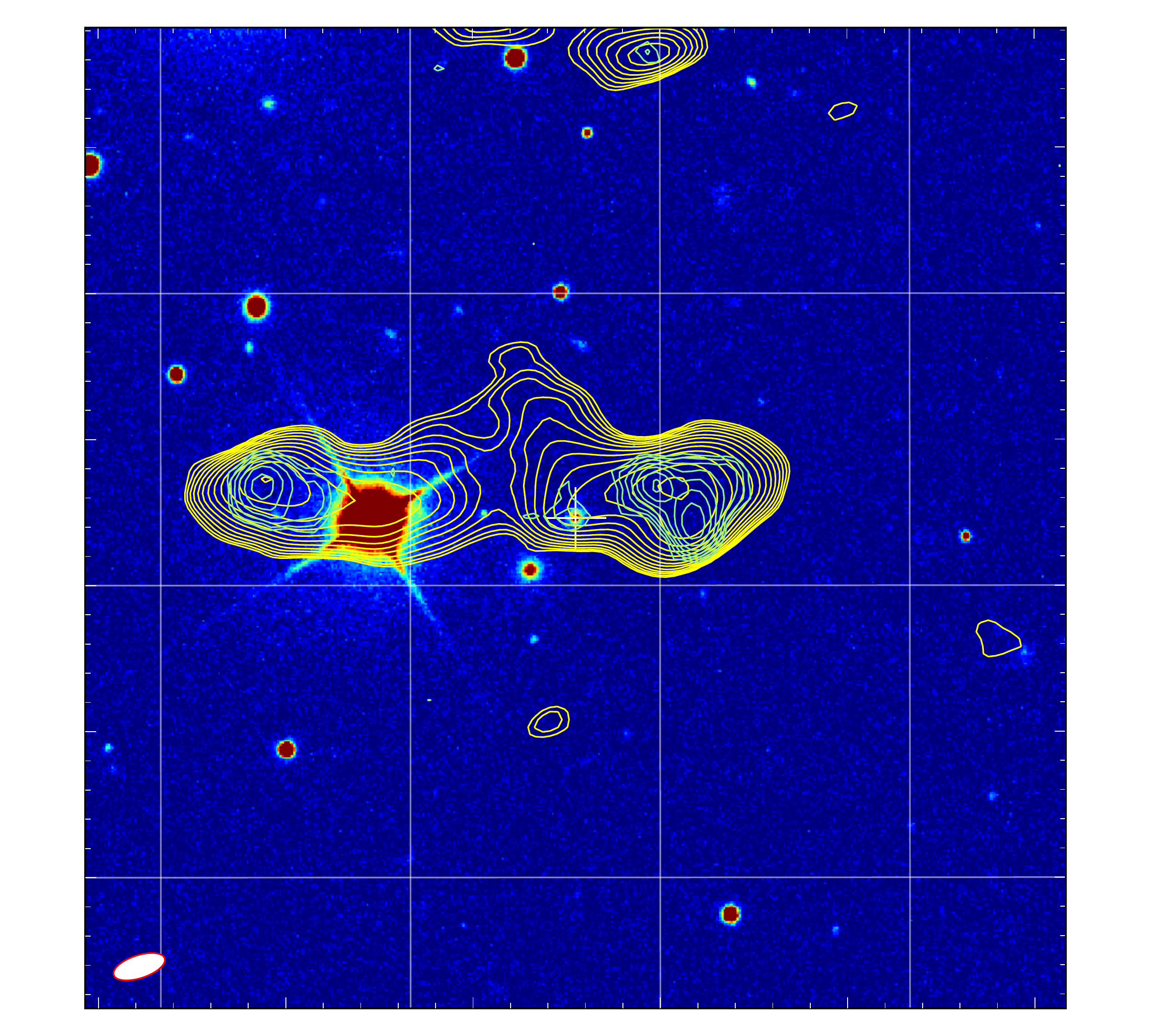}
  \includegraphics[width=.32\linewidth]{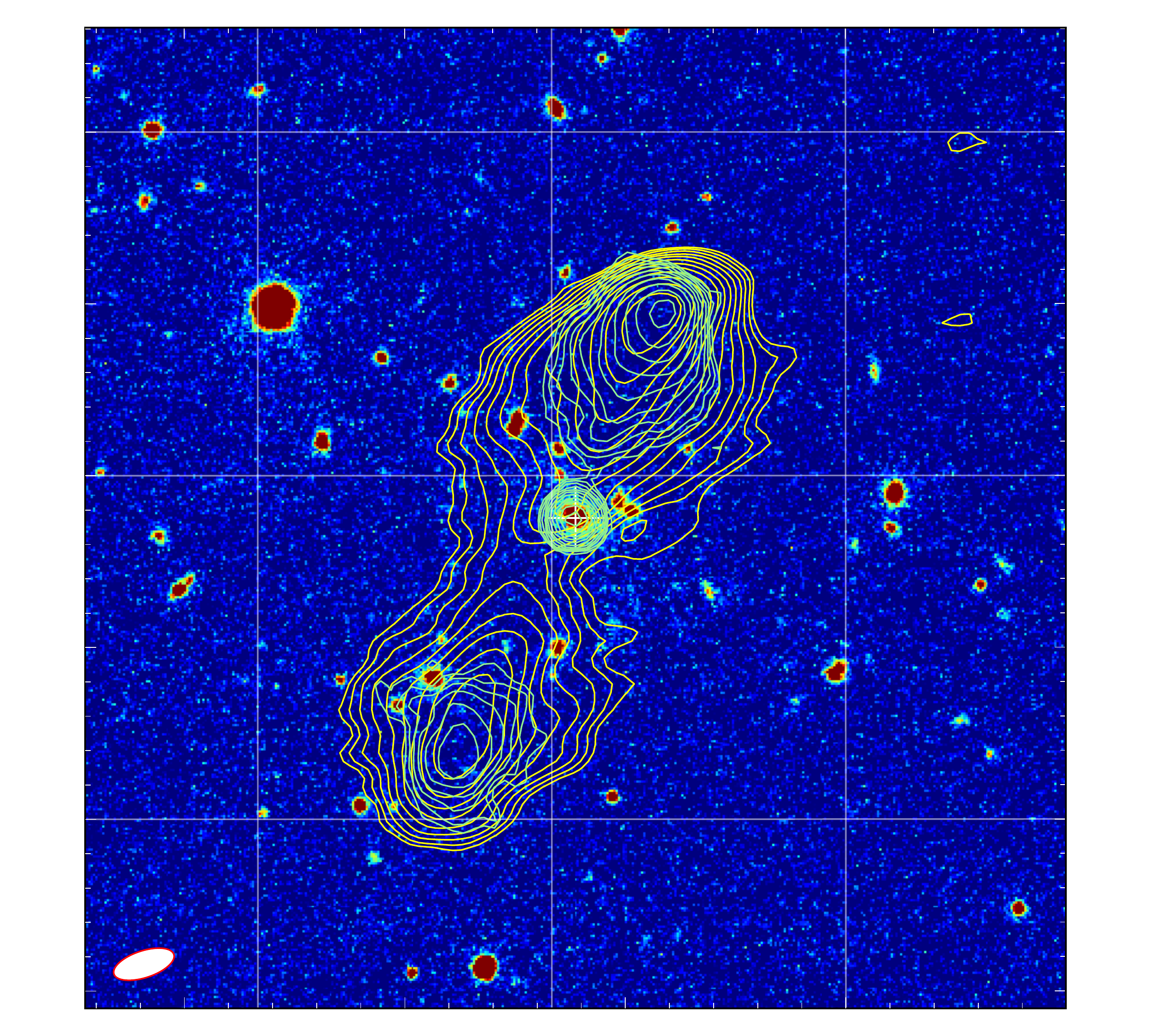}
  \includegraphics[width=.32\linewidth]{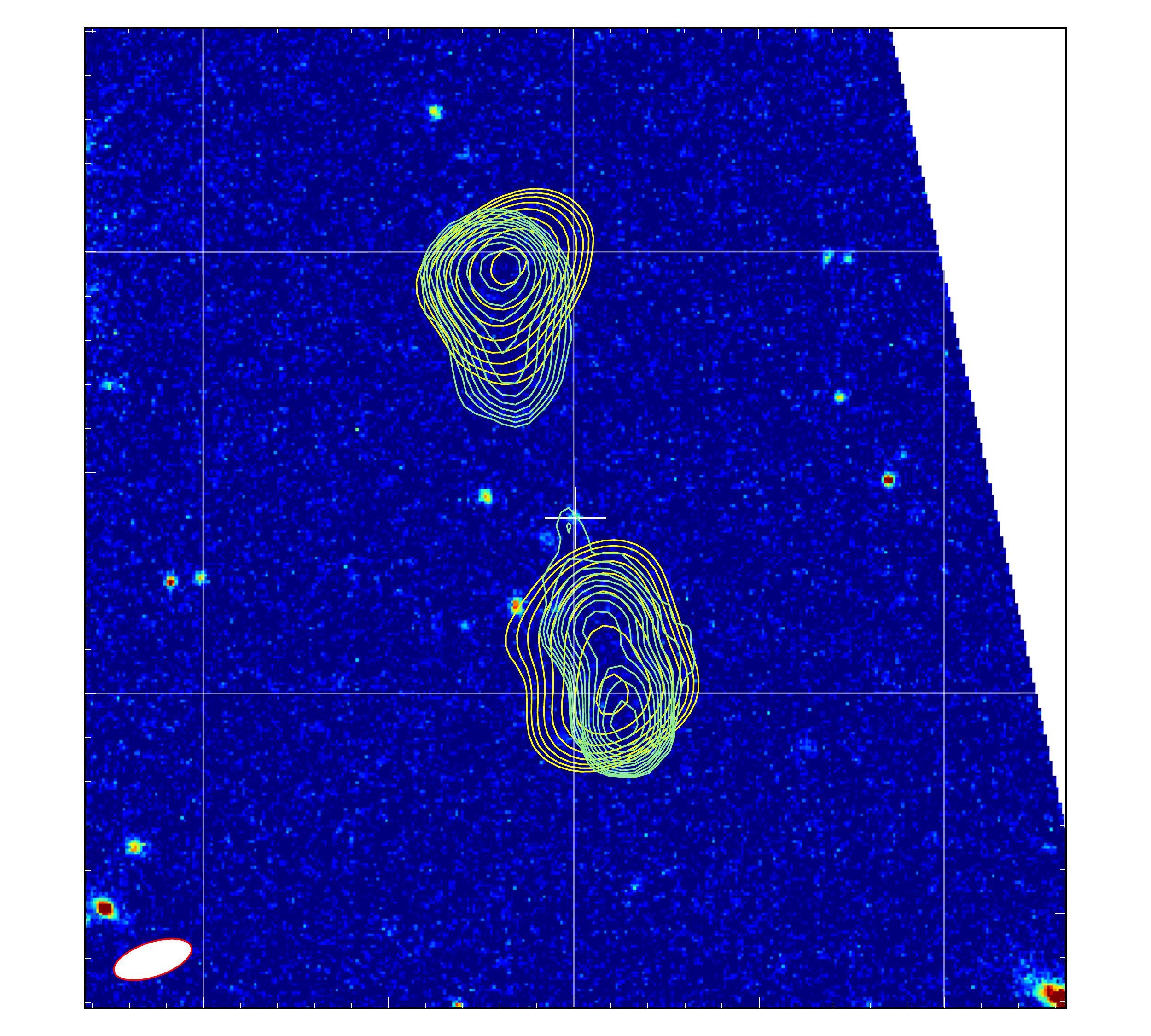}
  \caption{Postage stamps of example sources with FIRST cores. Colours,
    contours and lines as in Fig. \ref{fig:tmbg}. From left to right, top to
    bottom:
SDSS J124541.96+332428.5, 
J125541.50+250744.7, 
J125715.99+312153.1, 
J130057.21+325625.5, 
J130856.91+261333.2, 
J131246.85+275219.8, 
J132713.87+285318.1, 
J133449.73+312824.0, 
J133738.33+312514.2, 
J133837.35+311413.8, 
J134251.68+311052.6, 
and J134747.98+325823.8.}
  \label{fig:cored}
  \end{figure*}

For comparison, and to illustrate the range of source structures seen,
we plot some examples of sources from the 80-mJy subsample not classed as
remnants in Fig.\ \ref{fig:cored}. As can be seen, core strength in
these objects ranges from a dominant component in FIRST (e.g. in the
quasar SDSS J133449.73+312824.0) to a weak extension of lobe emission
over the optical ID (SDSS J134747.98+325823.8). Tailed morphology is
seen in many of these (e.g. the wide-angle tail
SDSS J130856.91+261333.2): we have not seen a convincing example of a
tailed remnant candidate.

\end{document}